\documentclass[prb,reprint,aps,superscriptaddress,longbibliography]{revtex4-2}
\usepackage[colorlinks=true,linkcolor=blue,citecolor=blue,urlcolor=blue]{hyperref}
\usepackage{graphicx}
\usepackage{bm}% bold math
\usepackage{braket}
\usepackage{amsmath}
\usepackage{amssymb}
\usepackage{graphics}
\usepackage{siunitx}
\usepackage{ulem}
\usepackage[dvipsnames]{xcolor}
\usepackage{enumitem}
\setlist{nolistsep}
\begin{document}

\title{Dirac-point spectroscopy of flat-band systems with the quantum twisting microscope}

\author{Nemin Wei}
\affiliation{\mbox{Department of Physics and Yale Quantum Institute, Yale University, New Haven, Connecticut 06520, USA}}

\author{Felix von Oppen}
\affiliation{\mbox{Dahlem Center for Complex Quantum Systems and Fachbereich Physik, Freie Universit\"at Berlin, 14195 Berlin, Germany}}

\author{Leonid I.\ Glazman}
\affiliation{\mbox{Department of Physics and Yale Quantum Institute, Yale University, New Haven, Connecticut 06520, USA}}

\date{\today}

\begin{abstract}
    Motivated by the recent development of the quantum twisting microscope, we formulate a theory of elastic momentum-resolved tunneling across a planar tunnel junction between a monolayer graphene layer situated on a tip and a twisting graphene-based sample. We elucidate features in the dependence of the tunnel current on bias and twist angle, which reflect  the sample band structure and allow the tip to probe the momentum-and energy-resolved single-particle excitations of the sample. While the strongest features originate from the Fermi edge of the tip, we argue that features associated with the tip Dirac points provide a more immediate and precise map of the sample band structure. We specifically compute the low-temperature tunneling spectrum of magic angle twisted bilayer graphene (MATBG) rotated relative to the tip by nearly commensurate angles, highlighting the potential of Dirac-point spectroscopy to measure single-particle spectral functions of flat bands along specific lines in reciprocal space. Furthermore, our analysis of tunneling matrix elements suggests a method to extract the ratio of the intra-and inter-sublattice tunneling parameters $w_0/w_1$ of MATBG from the differential tunneling conductance. Finally, we discuss signatures of $C_{3z}$ symmetry breaking in the tunneling spectrum using strained MATBG as an example. Our work establishes a general theoretical framework for Dirac-point spectroscopy of flat-band systems using the quantum twisting microscope.  
\end{abstract}

\maketitle

%{\color{blue} Option: split in 2 separate works: Theory for scanning TBG bands with QTM,\\ and\\ QTM signatures of spontaneously broken symmetries in normal-state TBG. For the first work:\\ 1. Advantages of scanning with Dirac point vs. scanning with the Fermi edge. Inefficiency of scanning with ``dull'' Dirac point (additional advantage, reduces the number of resonances in $d^2I/dV^2$). Advantages compared to ARPES (scanning states above FL).\\ 2. Theory of tunneling, continuum model. TBG band energies and wave functions information contained in the resonances.\\ 3. Application to TBG in the absence of symmetry breaking. Small-angle and ``Umklapp'' results: (i) symmetries in the angle-bias traces; (ii) effects of interference; (iii) information about $|\psi_{A,B}|^2$ and $w_{0,1}$ from the tunneling data.\\ 4. Effects of strain\\ }

\section{Introduction}
\label{sec:intro}

Electron tunneling has been an indispensable tool for probing quantum systems \cite{wolf2011principles}. Scanning tunneling microscopy and spectroscopy provide topographic images and local-density-of-states maps of materials. Tunneling between parallel two-dimensional electron systems can reveal complementary information including Fermi surfaces \cite{eisenstein1991probing}, quantum lifetimes \cite{murphy1995lifetime}, and energy dispersions \cite{jang2017full} of the electrons.

The quantum twisting microscope (QTM) is emerging as a versatile instrument for tunneling spectroscopy of two-dimensional materials \cite{inbar2023quantum}. It uses a Van der Waals tip to probe local properties of other Van der Waals samples by forming a twistable finite-area tunnel junction. Distinct from the scanning tunneling microscope, tunneling across the finite-area junction obeys energy and (in-plane) momentum conservation. This enables the QTM to reveal momentum-resolved energy dispersions by measuring the tunneling current as a function of bias voltage and twist angle between tip and sample. This powerful capability of the QTM was harnessed in Ref.\ \cite{inbar2023quantum} to reveal the electronic band structure of both graphene and large-angle twisted bilayer graphene. More recent experimental \cite{birkbeck2024measuring} and theoretical \cite{xiao2024theory} progress established inelastic tunneling spectroscopy using a QTM at cryogenic temperatures as a tool to measure phonon dispersions and electron-phonon coupling constants. Additionally, theoretical proposals suggest that the QTM could be used to probe two-dimensional superconductors \cite{xiao2023probing}, spin liquids \cite{peri2024probing}, and magnetically ordered phases \cite{pichler2024probing}.

\begin{figure}[b]
    \centering \includegraphics[width=.9\linewidth]{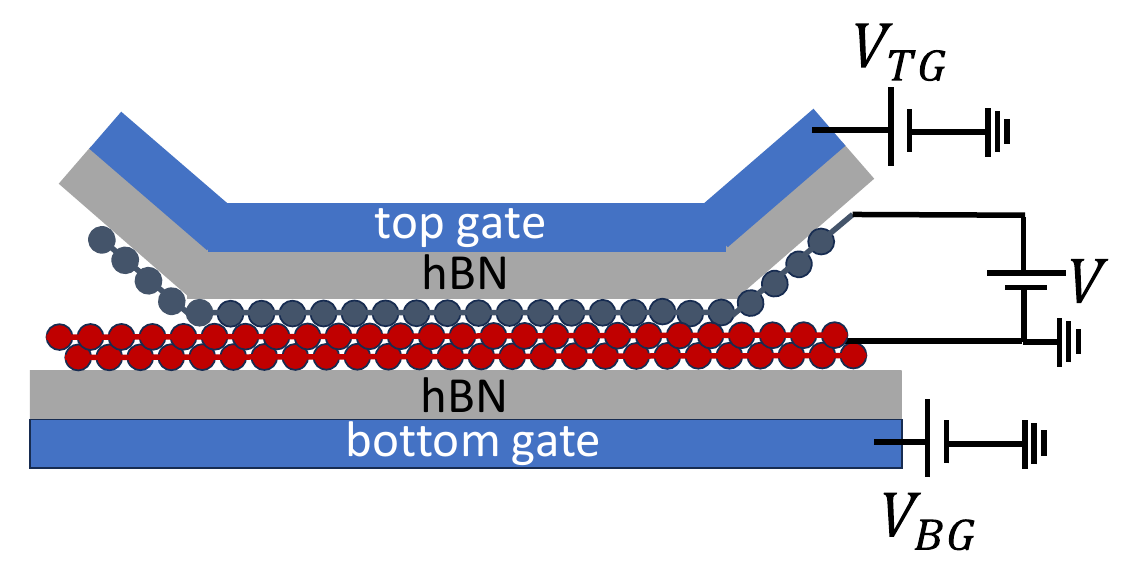} \caption{Schematic setup of a quantum twisting microscope with monolayer graphene as the tip layer (blue) and twisted bilayer graphene (red) as the sample layer. Probing elastic tunneling near the twist angle $\theta = 0$ between tip and sample may require introducing an additional decoupling layer as described in Ref.\ \cite{inbar2023quantum}. The electron densities in tip and sample layers can be tuned by top and bottom gates (light blue; gate voltages $V_{TG}$ and $V_{BG}$, respectively). Gate dielectrics are shown in gray (e.g., hBN). The tunneling current between tip and sample layers is measured as a function of the bias voltage $V$ and the twist angle $\theta$.}
    \label{fig:setup}
\end{figure}

Since the initial discovery of correlated insulating and unconventional superconducting phases in magic-angle twisted bilayer graphene (MATBG) \cite{cao2018unconventional,cao2018correlated}, extensive research has advanced our understanding of these phases by observing a cascade of phase transitions \cite{zondiner2020cascade,wong2020cascade}, intervalley coherent order \cite{nuckolls2023quantum, kim2023imaging}, and highly anisotropic or nodal superconducting gaps \cite{oh2021evidence,kim2022evidence,tanaka2024kinetic,banerjee2024superfluid}, among many other remarkable phenomena. Nevertheless, measuring electronic band structures of nearly-flat bands and their evolution with varying electron density in MATBG or other moir\'e graphene systems \cite{choi2021interaction,bocarsly2024imaging}, remains challenging. While angle-resolved photoemission spectroscopy (ARPES) on micrometer-scale MATBG devices showed evidence for flat bands \cite{li2024evolution, chen2023strong, lisi2021observation}, its energy resolution and temperature constraints have been hampering further studies of band structures and correlated phases.

These advances motivate us to explore the possibility of studying the excitation spectrum of moir\'e graphene structures such as twisted bilayer graphene (TBG) by means of momentum-resolved tunneling of electrons from monolayer graphene (MLG) in a QTM, see Fig.\ \ref{fig:setup} for a schematic illustration. In this setup, adjusting the bias voltage $V$ in coordination with gate voltages $V_{TG}$ and $V_{BG}$ allows for control of the band offset while maintaining the electron density of the sample at a target level. Twisting the junction displaces the tip Dirac points along arcs in reciprocal space. We present a theory of elastic momentum-conserving tunneling between MLG and moir\'e graphene samples in such a QTM device, and of its use for probing the sample band structure.

%\textcolor{red}{The dominant features in the current due to elastic tunneling between tip and sample originate from two sources.} The strongest signals arise from the Fermi edge of the tip. While these features encode information on the sample band structure, extracting this information is complicated by the finite extent of the tip's Fermi line in momentum space. 
The dominant features in the current due to elastic tunneling between tip and sample originate from two sources, Fermi edges and singularities in the density of states. The strongest signals which encode information on the sample band structure arise from the Fermi edge of the tip. However, extracting the band structure information is complicated by the finite extent of the tip's Fermi line in momentum space. In particular, this is the case, when the sample Brillouin zone is small due to a large moir\'e period. More immediate imaging of the sample band structure uses the singularity in the density of states of the tip, which is associated with a Dirac point. This leads to singularities in the tunneling current at characteristic bias voltages, where the energy of a tip Dirac point aligns with the band energy of the sample at the same momentum. We show that these features directly map the momentum-resolved energy dispersion of the sample through the twist-angle dependence of the characteristic bias voltages.

We develop a general formalism for this Dirac-point spectroscopy of flat bands and apply it to tunneling between MLG and the flat bands of MATBG near a few commensurate angles (Sec.~\ref{sec:normal}). We illustrate how the tunneling spectra reveal the one-particle spectral function of flat bands along the trajectories of the tip Dirac points, which traverse all high-symmetry points in the mini-Brillouin zone (mBZ) of MATBG. We find that the ratio of intra-and inter-sublattice tunneling parameters, a key parameter of low-energy effective models of MATBG, can be inferred from the differential tunneling conductances measured when the MLG Dirac point is aligned with the two Dirac points in the same valley of MATBG. Further analysis on strained MATBG indicates that Dirac point spectroscopy also provides a direct probe of the breaking of
the $C_{3z}$ symmetry, which triples the singularities in the tunneling spectra (Sec.~\ref{sec:strain}). 

%We present a theory of elastic momentum-conserving interlayer tunneling between van der Waals materials in a QTM device (Sec.~\ref{sec:main}). Our theory focuses on singularities of the tunneling current arising from a singularity of the tip density of states. We elucidate how this type of tunneling-current singularity reveals sample band structures (or one-particle spectral functions) along specific lines in reciprocal space. We subsequently apply this formalism to tunneling between MLG and the flat bands of MATBG near a few commensurate angles. %By analyzing tunneling matrix element effects in the normal state, we find that the ratio of intra- and intersublattice tunneling parameters, a key parameter of low-energy effective models of MATBG, can be inferred from the differential tunneling conductances measured when the MLG Dirac point is aligned with the two Dirac points in the same valley of MATBG. 
%Next, we explore QTM spectroscopic properties of low-temperature collective orders in MATBG, such as flavor symmetry breaking and superconductivity. In particular, we propose smoking-gun signatures of time-reversal and moir\'e-translation invariant intervalley coherent orders as well as nematic superconducting orders in samples without strain that explicitly break the three-fold rotation symmetry of MATBG.

\section{Dirac-point spectroscopy}\label{sec:main}

\subsection{General formalism}
\label{sec:general}

Given the high DOS associated with narrow bands, we simplify the consideration of electrostatics of the QTM junction by assuming that the chemical potential of the studied structure is parked in a band and remains unaffected by the bias between tip and sample. This simplification allows for an explicit relation between the second derivative of the current, $d^2I/dV^2$, and the quasiparticle spectrum of the moir\'e material. %We will illustrate the relation  assuming the absence of spontaneous symmetry breaking. 
To ensure that the chemical potentials $\mu_T$ and $\mu_S$ of tip and sample are independent of bias voltage $V$, one may in practice need to adjust the gate voltages $V_{TG}$ and $V_{BG}$, see Fig.~\ref{fig:setup}.

\begin{figure}[b]
    \centering
    \includegraphics[width=.9\linewidth]{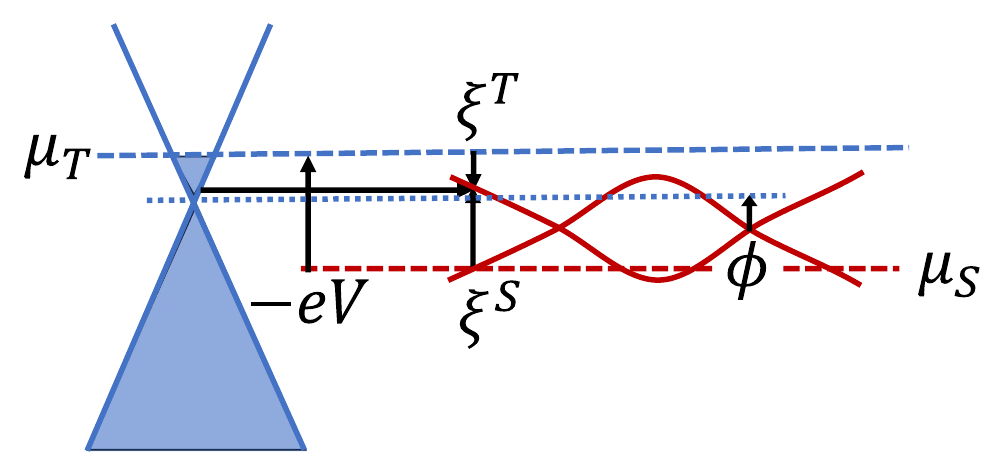}
    \caption{ Schematic band diagram for a given bias $-eV$ between MLG tip (blue) and TBG sample (red). The Dirac points of tip (blue dotted line) and sample are offset in energy by the electrostatic potential $\phi$. The chemical potentials $\mu_{T/S}$ of tip and sample are measured from their respective Dirac points, so that $-eV=\mu_T-\mu_S+\phi$. The gate voltages are adjusted such that the tip chemical potential $\mu_T$ coincides with the gap between the flat and dispersive bands of the sample. A representative elastic-tunneling process near the tip Dirac point is indicated by the hori\-zontal arrow (black). The momenta of the dispersions of tip and sample are offset relative to each other in the figure, and do not reflect the twist of the reciprocal spaces, cf.\ Fig.~\ref{fig:k_trajectory}. The energy arguments of the spectral functions of tip and sample are measured from the respective chemical potentials. An energy $\omega$ measured from the sample chemical potential corresponds to $\omega + eV$ relative to the tip chemical potential, cf.\ Eq.\ (\ref{eq:i_basic}).}
    \label{fig:energydiagram}
\end{figure}

The elastic momentum-conserving tunneling current between tip and sample reads \cite{bistritzer2010transport}
\begin{widetext}
    \begin{equation}\label{eq:i_basic}
    I=\frac{2\pi e}{\hbar} \int d \omega\left[f(\omega)-f(\omega+e V)\right] \sum_{\bm k\bm k^{\prime}} \sum_{\lambda\lambda^{\prime}} |\langle \bm k^{\prime}\lambda^{\prime} T| H_{\text{tun}}|  \bm k\lambda S\rangle|^2 A_{\lambda^{\prime}}^{T}\left(\bm k^{\prime}, \omega+e V\right)A_{\lambda}^{S} \left(\bm k, \omega\right),%\delta_{\bm k+\bm{G_t},\bm k'+\bm{G_b}},
\end{equation}
\end{widetext}
%The spectral function of non-interacting bands are diagonal in the energy band basis. 
where $f(\omega)=[\exp(\beta\omega)+1]^{-1}$ is the Fermi-Dirac distribution function and the energy $\omega$ is measured relative to the Fermi level of the sample (see Fig.\ \ref{fig:energydiagram}). We denote the spectral functions of tip ($T$) and sample ($S$) for Bloch states $|\bm k \lambda T/S\rangle$ as $A_{\lambda}^{T,S}\left(\bm k, \omega\right)$, with $\bm k$ being the wave vector and $\lambda$ the band label. $\bm k,\bm k'$ are restricted to one Brillouin zone of the sample and tip, respectively. The tunneling Hamiltonian $H_{\text{tun}}$ conserves momentum,
\begin{equation}\label{eq:ht}
    \langle \bm k^{\prime}\lambda^{\prime} T| H_{\text{tun}}|  \bm k\lambda S\rangle = \sum_{\bm {Q},\bm{Q}'}T_{\lambda^{\prime}\lambda}(\bm k + \bm{Q})\delta_{\bm k^{\prime}+\bm{Q}^{\prime},\bm k+\bm{Q}}
\end{equation}
with reciprocal lattice vectors $\bm{Q}$ of the sample and $\bm{Q}'$ of the tip. Here, we use that for a given twist angle between the tip and sample and in view of the momentum-conservation condition, $\bm{k}+\bm{Q}$ uniquely determines the wave vectors $\bm{k}$, $\bm{Q}$, $\bm{k}'$, and $\bm{Q}'$, so that we can consider $T_{\lambda'\lambda}$ to be a function of $\bm{k}+\bm{Q}$ only.  
%\textcolor{brown}{N: Owing to the kronecker delta constraint, the four variables $\bm k, \bm Q, \bm k', \bm Q'$ are not independent and the tunneling matrix elements $ T_{\lambda'\lambda}$ is a function of a single vector $\bm p$ in the 2D reciprocal space because any 2D wave vector $\bm p$ admits a unique decomposition $\bm p = \bm k' + \bm Q' = \bm k + \bm Q$ when the tip and sample lattices are fixed. } 
Each term on the right hand side of Eq.~\eqref{eq:ht} contributes to the square of the tunneling matrix element separately without interference,
\begin{equation}\label{eq:ht_squared}
    |\langle \bm k^{\prime}\lambda^{\prime} T| H_{\text{tun}}|  \bm k\lambda S\rangle|^2 = \sum_{\bm {Q},\bm{Q}'}|T_{\lambda^{\prime}\lambda}(\bm k + \bm{Q})|^2\delta_{\bm k'+\bm{Q}',\bm k+\bm{Q}}.
\end{equation}
We denote the tip and sample band dispersions, measured relative to the respective chemical potentials $\mu_T$ and $\mu_S$, as $\xi_{\bm k' \lambda'}^{T}$ and $\xi_{\bm k\lambda}^{S}$ (cf.\ Fig.\ \ref{fig:energydiagram}). For the free-fermion spectral functions $A_{\lambda}^{T,S}(\bm k, \omega)=\delta(\omega-\xi_{\bm k \lambda}^{T,S})$, Eq.~\eqref{eq:i_basic} reduces to Fermi's golden rule,
\begin{align}\label{eq:fermigolden}
    I =\frac{2\pi e }{\hbar}\sum_{\lambda\lambda'}\sum_{\bm k}\sum_{\bm Q} &(f(\xi_{\bm p\lambda}^{S})-f(\xi_{\bm p\lambda'}^{T}))|T_{\lambda'\lambda}(\bm p)|^2\notag\\
    &\times\delta(\xi_{\bm p\lambda}^{S}-\xi_{\bm p\lambda'}^{T}+eV)\Big |_{\bm p =\bm k +\bm{Q}}.
\end{align}
%
%Among all $\bm{Q}$'s we will only keep $\bm{Q}_n$ which can make $\xi_{\bm k\lambda}^{T}$ and $\xi_{\bm k+\bm{Q}_n\lambda'}^{S}$ simultaneously close to charge neutrality and accessible by electrical gating in experiment.
$\xi_{\bm k'\lambda'}^{T}=\xi_{\bm k'+\bm {Q}'\lambda'}^{T}$ and $\xi_{\bm k\lambda}^{S}=\xi_{\bm k+\bm {Q}\lambda}^{S}$ are considered as periodic in the extended Brillouin zone. To simplify notations, we define the band energy difference
\begin{equation}\label{eq:deltaepsilon}   
\delta\epsilon_{\bm p}^{\lambda'\lambda} = \xi_{\bm p\lambda}^{S} + \mu_S - \xi_{\bm{p}\lambda'}^{T} - \mu_{T}   
\end{equation}
of a particular pair of bands $\lambda'$
and $\lambda$, which is independent of the chemical potentials. We also define the electrostatic potential difference \begin{equation}\label{eq:phi}
    \phi = -eV-\mu_{T}+\mu_{S}
\end{equation}
between tip and sample, which determines the band offset in energy (Fig.\ \ref{fig:energydiagram}). 

Using these definitions, Eq.~\eqref{eq:fermigolden} can be rewritten as a line integral over the equal-energy contours $\delta\epsilon_{\bm p}^{\lambda'\lambda}=\phi$ in the extended Brillouin zone, 
\begin{align}
    I &= \frac{e \Omega}{2\pi\hbar} \sum_{\lambda\lambda'}\int_{\delta\epsilon_{\bm p}^{\lambda'\lambda}=\phi} \frac{d p_{\scriptscriptstyle\parallel}}{|\partial_{\bm p}\delta\epsilon_{\bm p}^{\lambda'\lambda}|}
    \nonumber\\
    &\qquad \times(f(\xi_{\bm p\lambda}^{S})-f(\xi_{\bm {p}\lambda'}^{T}))|T_{\lambda'\lambda}(\bm p)|^2, \label{eq:i_contour}
\end{align}
where $\Omega$ is the contact area between tip and sample. %We denote the equal-energy contours as $C_n(\phi)$, where the index $n$ encompasses the band indices as well as an index enumerating disconnected pieces originating from the same pair of bands

\begin{figure}
    \centering
    \includegraphics[width=1\linewidth]{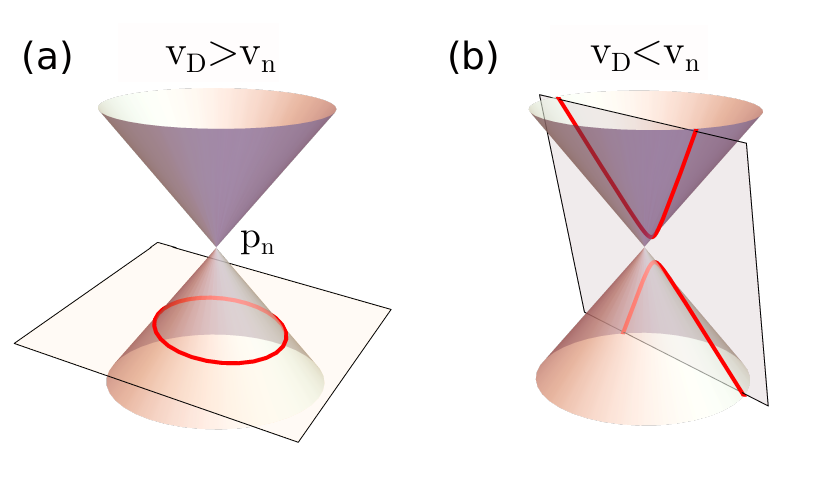}
    \caption{Schematics of the intersections (red) between a Dirac cone at velocity $v_D$ and a plane representing a probed band with linearized energy dispersion relative to the Dirac point $\bm p_n$, $\bm v_n\cdot\bm (\bm p - \bm p_n) + \phi-\phi_n$, see Eq.~\eqref{eq:resonance} for $\phi_n$. Elastic momentum-conserving tunneling occurs at the band intersections. (a) For $v_D > v_n$, the intersections are ellipses and tunneling current $I\propto |\phi-\phi_n|$ becomes singular as the Dirac point crosses the probed band at the characteristic band offset $\phi=\phi_n$. (b) For $v_D<v_n$, the intersections are hyperbolas and $I(\phi)$ is an analytic function at $\phi=\phi_n$.}
    \label{fig:intersections}
\end{figure}

%The Dirac point of the graphene tip leads to contours $C_n(\phi)$, which shrink to a point (denoted ${\bm p}_n$) at some values of $\phi$, see Fig.~\ref{fig:intersections}a.
The Dirac cones of the graphene tip lead to equal-energy contours $C_n(\phi)$ which shrink to the Dirac points ${\bm p}_n$ at some values of $\phi$, see Fig.~\ref{fig:intersections}a. The index $n$ enumerates disconnected contours (e.g., due to a valley degeneracy) for a fixed band index $\lambda$, which we leave implicit to shorten notation.
These points $\bm p_n$ are associated with singularities in the joint density of states 
\begin{equation}
    \nu_n(\phi) \equiv \frac{1}{(2\pi)^2}\int_{C_n(\phi)} \frac{d p_{\scriptscriptstyle\parallel}}{|\partial_{\bm p}\delta\epsilon_{\bm p}^{\lambda'\lambda}|}.
\end{equation}
Along $C_n(\phi)$, $\lambda'$ is uniquely specified by $\phi$ and $\lambda$, as indicated by Fig.~\ref{fig:intersections}a.
At the Dirac point of the tip, we have $\xi^T_{{\bm p}_n\lambda'}=-\mu_T$, so that the singularity occurs at 
\begin{equation}
\label{eq:resonance}
 \phi_n=\xi^S_{{\bm p}_n\lambda}+\mu_S,   
\end{equation}
see Eq.\ (\ref{eq:deltaepsilon}).

%At a given angle, $\nu_n(\phi)$ exhibits a singularity associated with the integration contour $C_n(\phi)$ shrinking to a point, see Fig.~\ref{fig:intersections}a. This happens at a specific ${\bm p}_n$ corresponding to the location of the $K$ point of the tip (with respect to  the $\Gamma$ point common to the tip and sample), and specific energy associated with the neutrality point of the tip, $\xi^T_{{\bm p}_n}=-\mu_T$. That results in $\delta\epsilon_{{\bm p}_n}=\xi^S_{{\bm p}_n}+\mu_S$, see Eq. (5); here $n$ enumerates such singular points $(\bm p_n, \delta\epsilon_n)$and the contours $C_n(\phi)$ near them.

The singularities in $\nu_n(\phi)$  lead to singularities in the tunneling current at bias voltages
\begin{equation}\label{eq:Vn}
    V_n = \frac{1}{e} \left(\xi_{\bm p_n\lambda'}^{T} - \xi_{\bm{p}_n\lambda}^{S} \right) .
\end{equation}
In the vicinity of these bias voltages, we can approximate the current as 
\begin{align}
     I \approx \frac{2\pi e \Omega}{\hbar}\sum_{n}(f(\xi_{\bm p_n\lambda}^{S})-f(\xi_{\bm {p}_n\lambda'}^{T}))|T|_n^2 \nu_n(\phi) + \ldots , \label{eq:i_approximate}
\end{align}
where we introduced the normalized value  
\begin{equation}\label{eq:t_averaged}
    |T|_n^2 = \lim_{\phi\rightarrow  \phi_{n}} \frac{1}{(2\pi)^2\nu_n(\phi)} \oint_{C_n(\phi)} \frac{d p_{\scriptscriptstyle\parallel}}{|\partial_{\bm p}\delta\epsilon_{\bm p}^{\lambda'\lambda}|} |T_{\lambda'\lambda}(\bm p)|^2
\end{equation}
of the square of the  tunneling matrix element.
Analytically, the singularities take the from (see App.~\ref{app_sec:dos})
\begin{equation}\label{eq:dos}
    \nu_n(\phi) = \frac{|\phi-\phi_n|}{2\pi\hbar^2 v_D^2 (1-v_n^2/v_D^2)^{\frac{3}{2}}}\,,
\end{equation}
after combining the contributions of the $\lambda'=\pm$ bands of MLG. This assumes that the  Dirac velocity $v_{D}$ of the tip is larger than the group velocity $v_n$ of the sample's energy band. In this case, the equal-energy contours approach ellipses for $\phi$ close to $\phi_n$, see Fig.~\ref{fig:intersections}a. Importantly, there is no singularity in $\nu_n(\phi)$ in the opposite case $v_D<v_n$. 
Mathematically, this arises from a cancellation between the two hyperbolic intersections in Fig.\ \ref{fig:intersections}b, see App.~\ref{app_sec:dos}. For flat-band samples, this asymmetry ensures that Dirac points of the tip scan the sample band structure, but Dirac points of the sample would not scan the tip band structure.

Combining these results, we find that the singularity in the current manifests as a Dirac $\delta$-function in $d^2I/dV^2$ \footnote{In Eq.~\eqref{eq:d2IdV2_dp}, $d\phi/dV =-e$ if $\mu_{T,S}$ are fixed},
\begin{align}\label{eq:d2IdV2_dp}
    \frac{d^2I}{dV^2} &= \frac{2 e \Omega}{\hbar^3 v_D^2}\left(\frac{d\phi}{d V}\right)^2(f(-eV-\mu_T)-f(-\mu_T)) \notag\\
    &\times\sum_{n} \left(1-\frac{v_{n}^2}{v_D^2}\right)^{-\frac{3}{2}}|T|_{n}^2\delta(eV+\mu_T+\xi_{\bm p_n\lambda}^S). 
\end{align}
As the twist angle is varied, the Dirac point ${\bm p}_n$ of the tip scans along certain lines in the Brillouin zone of the sample, while the sample bias varies the band offset between tip and sample. The sample dispersion can then be extracted from the resonance condition
\begin{equation}\label{eq:vn}
   -eV_{n}=\xi_{\bm p_n\lambda}^S+\mu_T.
\end{equation}
This provides the basis for Dirac-point spectroscopy of flat-band systems with a QTM. 

Several comments are in order. (i) 
In principle, the bias voltage enters the tunnel current in Eq.\ (\ref{eq:i_approximate}) through the Fermi functions %(see Sec.\ \ref{sec:DiracFermi} below for a more detailed discussion) 
and the joint density of states. In deriving Eq.\ \eqref{eq:d2IdV2_dp}, we only retained the  derivative of the joint density of states. This is the leading contribution when the derivatives of the Fermi-Dirac functions are suppressed by a judicious choice of $\mu_T$. 
(ii) One convenient choice sets $\mu_T$ such that it coincides with the Dirac point of the tip. Then, we have $f(-\mu_T) = 1/2$. This ensures that the difference of Fermi functions is a nonzero constant for both empty and occupied states of the sample band, allowing one to scan the entire band with a single setting of $\mu_T$. However, it requires $\mu_T$ to be at the Dirac point with an accuracy better than the thermal energy. (iii) Alternatively, one can choose $\mu_T$ such that it coincides with a band gap between the flat and dispersive bands of the sample. Unlike for the previous choice, this requires measurements at two values of $\mu_T$ to reveal the entire sample band structure. When $\mu_T>0$ is chosen in the band gap above the sample flat band, the tunneling current will be nonzero only for empty states of the flat band. Conversely, when $\mu_T<0$ is chosen in the band gap below the sample flat band, the tunneling current will be nonzero only for occupied states. 
%We note in passing that
%{\color{blue} As we mentioned in the Introduction,} scanning the flat band as a function of filling is of much interest since band deformations arising from Hartree-Fock effects are believed to play an important role in the physics of twisted bilayer graphene. 
(iv) We note that in the absence of $C_{3z}$ symmetry breaking of the tip MLG, the location of the Dirac point is fixed by symmetry, providing an important robustness to Dirac-point spectroscopy.  
(v) The singularity is enhanced as $v_n\rightarrow v_D$ and the bands of tip and sample are nested near $\bm p_n$. (vi) Despite an infinite number of $\bm p_n$ values in the reciprocal space, only a few of them need consideration due to practical constraints. First, tunneling matrix elements $|T|_n^2$ decrease rapidly with increasing $|\bm p_n|$. Second, the characteristic bias voltage $V_{n}$ and band offset $\phi_n$ of the singularity should fall within the range accessible by electrical tuning.

%To derive the tunneling current Eq.\ \eqref{eq:i_approximate} from Eq.~\eqref{eq:i_contour}, we assumed momentum-independent Fermi-Dirac functions. 
When deriving the tunneling current Eq.\ \eqref{eq:i_approximate} from Eq.~\eqref{eq:i_contour}, we disregarded the energy dependence of the Fermi-Dirac functions in the integrand of Eq.~\eqref{eq:i_contour}. In the $T\to 0$ limit, however, a Fermi-Dirac distribution becomes a step function of energy. This non-analytical behavior leads to a singular dependence of $d^2I/dV^2$ on bias. The singularity is associated with the Fermi edge in one of the electrodes ``scanning'' the electron spectrum in the other.
%Strictly speaking, when the Fermi level of the tip or sample intersects the bands of the other, that approximation becomes invalid close to the Fermi level where the Fermi-Dirac functions of tunneling electrons depend sensitively on bias voltage through $\mu_{T,S}$ and the band offset $\phi$ according to Eq.~\eqref{eq:i_contour}. The bias-dependence of the Fermi-Dirac functions can lead to Fermi edge singularities in tunneling current 
(see Sec.\ \ref{sec:DiracFermi} below for a more detailed discussion). 
We have discussed choices of $\mu_T$ to suppress the singularity induced by the Fermi edge of the tip. For the Fermi edge in a flat band of the sample, we find that the associated singularity can be strongly smeared by temperature and quasiparticle broadening of the flat band, see App.~\ref{app_sec:fs}. Thus, the Dirac-point singularity can become the main feature in the tunneling spectrum within a range of bias voltage.

%\textcolor{red}{F: fine by me} An important advantage of Dirac-point spectroscopy of flat bands is the asymmetry in response depending on the ratio of the band velocities of tip and sample, cf.\ Fig.\ \ref{fig:intersections}. This ensures that singularities of the tip band structure scan the sample band structure, while the reverse does not happen. Mathematically, this emerges from a cancellation between the two hyperbolic intersections in Fig.\ \ref{fig:intersections}b. Strictly speaking, this cancellation only occurs when the chemical potential of the tip is within a band gap of the sample between the flat and a dispersive band. It does not occur when $\mu_T$ is located at the charge-neutrality point of the tip. DETAILS; MORE IN APPENDIX/OTHER SECTION. 

%QTM is able to control the in-plane rotation angle of MLG relative to the sample and by doing so rotate the Dirac points $\bm p_n$ in reciprocal space continuously, leading to lines of $d^2I/dV^2$ singularity as the twist angle varies. The angle-dependent bias voltages at the singularity $V_{n}=-(\xi_{\bm p_n\lambda}^S+\mu_T)/e$ measures the sample band structures along the $\bm p_n$ trajectories.

\subsection{Application to twisted graphene samples}\label{sec:commensuration}

We specify this general discussion to a QTM junction consisting of a MLG tip and a graphene-based sample. This allows us to express the tunneling matrix element [Eq.\ \eqref{eq:t_averaged}] in terms of the wave function of the sample. 

\begin{figure}
    \centering
    \includegraphics[width=1\linewidth]{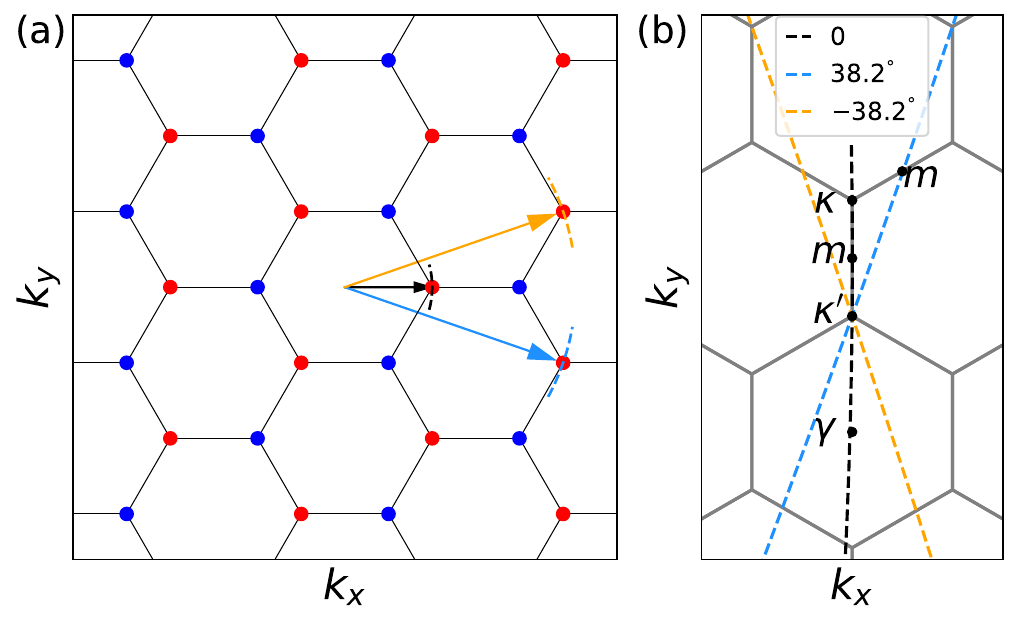}
    \caption{(a) The reciprocal lattice of a graphene layer in the sample with $\pm K$ valleys marked as red and blue dots, respectively. As the MLG tip is twisted relative to this graphene layer by commensurate twist angles $\theta_c=0,\pm 38.2^{\circ}$, the Dirac points of two layers overlap at $\bm K + \bm G_{1}$ marked by the black, blue, and orange arrows, respectively.  The dashed arcs are trajectories of the tip Dirac point $\bm p_1$ near commensuration. (b) The three dashed arcs in (a) are plotted in the mBZ of a $1.05^{\circ}$-TBG. They cross the Dirac points $\kappa'(\kappa)$ of the TBG top (bottom) layer when the MLG is commensurate with the corresponding layer. %By scanning around $\theta_c=0$ and $\pm 38.2^{\circ}$, one can move the Dirac point (nearly) cross all higher symmetry points of TBG.
    }
    \label{fig:k_trajectory}
\end{figure}

The tip weakly couples to the topmost graphene layer of the sample with a relative twist angle $\theta$. The two inequivalent corners of the first Brillouin zone of the tip are rotated to $\pm\bm K_{\theta}= \pm O(\theta)\bm K$, where $\pm\bm K$ denotes the corresponding points in the Brillouin zone of the top layer of the sample and  
\begin{equation}
    O(\theta) = \begin{pmatrix}
    \cos\theta & \sin\theta \\
    -\sin\theta & \cos\theta \\
    \end{pmatrix}.
\end{equation}
is an in-plane clockwise rotation matrix. The angle misalignment largely forbids elastic momentum-conserving interlayer tunneling between low-energy states close to the Dirac points $\pm\bm K_\theta$ and $\pm\bm K$ except near commensurate twist angles $\theta_c$, where elastic tunneling is enabled via umklapp processes. Specifically, we will consider twist angles near $\theta_c=0$, where the first Brillouin zones of the tip and topmost sample layer coincide, as well as $\theta_c = 2\arctan{\sqrt{3}/5}\approx 38.2^{\circ}$ (and symmetry-related configurations), where corners of the third Brillouin zones coincide, see Fig.\ \ref{fig:k_trajectory}a and Ref.\ \cite{inbar2023quantum}.

At the commensurate angles $\theta_c$, there exist reciprocal lattice vectors $\bm{G}_n'$ and $\bm{G}_{n}$ of the tip and the topmost graphene layer of the sample, respectively, such that the Dirac points of both layers overlap at (I) $\bm{K}_{\theta_c}+\bm{G}_n'=\bm{K}+\bm{G}_{n}$ or (II) $\bm{K}_{\theta_c}+\bm{G}_n'=-\bm{K}+\bm{G}_n$. A valley in one layer couples with either the same (I) or the opposite (II) valley in the other layer. If $\theta_c$ corresponds to configuration II, $\theta_c-\pi/3$ will correspond to configuration I. Since all twist angles which differ by multiples of $\pi/3$ are physically identical owing to the sixfold rotational symmetry $C_{6z}$ of MLG, we can limit our discussion to configuration I with $|\theta_c|<\pi/3$.
For later convenience, $n=1,2,3$ labels the three shortest $|\bm K + \bm G_n|$. 
At $\theta_c=0$, the black arrow in Fig.~\ref{fig:k_trajectory}a depicts the vector $\bm K + \bm G_1$ with $\bm G_1=0$, while the $\bm K+\bm G_{2,3}$ (not shown) connect the origin with the other two $K$-valley corners of the first Brillouin zone. The blue arrow represents $\bm K+\bm G_{1}$ for $\theta_c\approx 38.2^{\circ}$. When the reci\-procal lattice in Fig.~\ref{fig:k_trajectory}a is rotated clockwise by the angle between the blue and orange arrows ({\it i.e.}, at $\theta_c = 2\arctan{\sqrt{3}/5}\approx 38.2^{\circ}$), the rotated and original lattices are commensurate and coincide at the Dirac point indicated by the blue arrow. 
%Thus, the blue arrow represents $\bm K+\bm G_{1}$ for $\theta_c\approx 38.2^{\circ}$. 
Likewise, the orange arrow represents $\bm K+\bm G_{1}$ for $\theta_c\approx -38.2^{\circ}$.
%{For $\theta_c=0$, $\bm K + \bm G_1$ corresponds to the black arrow for $\bm G_1=0$, while $\bm K+\bm G_{2,3}$ are the other two $K$ valley corners of the first Brillouin zone. For $0<|\theta_c|<\pi/3$, the smallest $|\bm K + \bm G_1|$ is $\sqrt{7}|\bm K|$ (\textit{e.g.}, the orange and blue arrows in Fig.~\ref{fig:k_trajectory}a). When the reciprocal lattice in Fig.~\ref{fig:k_trajectory}a is rotated clockwise by $\theta_c = 2\arcsin{1/2\sqrt{7}}\approx 38.2^{\circ}$, the angle between the orange and blue arrows, the rotated and original lattices are commensurate and coincide at the Dirac point indicated by the blue arrow. Thus, the blue arrow represents $\bm K+\bm G_{1}$ for $\theta_c=38.2^{\circ}$. Likewise, the orange arrow represents $\bm K+\bm G_{1}$ for $\theta_c=-38.2^{\circ}$.}

Near commensuration, the long-wavelength modulation of the local stacking configuration (moir\'e superlattice) admits a continuum description of the interlayer tunneling Hamiltonian \cite{bistritzer2011moire}. We adopt the $\bm k\cdot \bm p$ approximation of the tip and sample Hamiltonians in the $\tau K$ valley ($\tau=\pm $) with respect to a common reference wave vector $\tau(\bm K+\bm G_1)$ and define continuum Dirac fields $\Phi_{\tau}^{\dagger}=(\Phi_{\tau A}^{\dagger}, \Phi_{\tau B}^{\dagger})$ for the MLG tip and $\Psi_{\tau l}^{\dagger}=(\Psi_{\tau l A}^{\dagger}, \Psi_{\tau l B}^{\dagger})$ for layer $l$ of the sample. % and require $\Psi^{T}(\bm r)$ and $\Psi_l^{S}(\bm r)$ to be spatially uniform for Bloch wave functions of quasimomentum $\bm K+\bm G_1$.
The most relevant interlayer tunneling Hamiltonian conserves the valley quantum number $\tau$, $H_{\text{tun}}=H_{\text{tun}}^{\tau=+} + H_{\text{tun}}^{\tau=-}$, and is local in real space. The minimal expression of $H_{\text{tun}}^{\tau}$ that conserves quasimomentum up to reciprocal lattice vectors of the two interfacial graphene layers and respects $C_{3z}$ symmetry is \cite{balents2019general},
\begin{equation}\label{eq:H_T}
    H_{\text{tun}}^{\tau} = \int d^2r \sum_{n=1}^3 \Phi_{\tau}^{\dagger}(\bm r) \hat{T}_{\tau n} e^{-i\tau\delta\bm G_n\cdot\bm r} \Psi_{\tau t}(\bm r) + h.c..
\end{equation}
Here, the $\delta\bm G_1 = 0$ term describes the spatially averaged  tunneling amplitude. $C_{3z}$ symmetry necessitates two additional tunneling terms with wave vectors 
\begin{equation}\label{eq:DeltaG}
\delta\bm G_n = \bm G_n'-\bm G_1' - (\bm G_n - \bm G_1).
\end{equation}
The $\delta\bm G_{2,3}$ are two primitive reciprocal lattice vectors of the moir\'e superlattice formed by the two graphene layers. $C_{2z}\mathcal{T}$ and $C_{3z}$ symmetries constrain the tunneling matrix to (see App.~\ref{app_sec:symmetry} and Ref.~\cite{scheer2022magic,vafek2023continuum,kang2023pesudomagnetic})
\begin{equation}\label{eq:tn_general}
    \hat{T}_{\tau n} = w_0 e^{i\chi\tau\sigma^z}+w_1\sigma^x e^{i\frac{2\pi (n-1)}{3}\tau\sigma^{z}},
\end{equation}
where $\sigma^i$ are Pauli matrices for the sublattice pseudospin. For $\theta\approx 0$, $\chi=0$ and the intra-and inter-sublattice tunneling parameters $w_0$ and $w_1$ are equal, $w_0=w_1\equiv t_{\theta_c=0}$, in the absence of lattice relaxation. For $\theta\approx \pm 38.2^{\circ}$, Ref.~\cite{scheer2022magic} finds that $\chi\approx 0$ and $w_0\approx w_1\equiv t_{38.2^{\circ}}$, see explanations in App.~\ref{app_sec:tightbinding}. %Based on the tight-binding model of $2p_z$ orbitals in graphene, 

% For simplicity, we focus on $\theta_c=0$ and briefly comment on $\theta_c=\pm 38.2^{\circ}$ cases in the end of the section. At small twist angles, the envelope functions of the Bloch states read that
We now derive tunneling matrix elements between the Bloch states $|\bm k'\lambda' T\rangle$ and $|\bm k\lambda S\rangle$ of tip and sample. The wave vectors $\bm k, \bm k'$ are measured relative to the graphene $\Gamma$ point and the wave functions are given by
\begin{equation} \label{eq:psi_mlg}
    \langle 0 |\Phi_{\tau}(\bm r)|\bm k'\lambda' T\rangle = \frac{1}{\sqrt{2\Omega}}
    \begin{pmatrix}
        \lambda'\\
        e^{i\tau\theta_{\bm k'}}\\
    \end{pmatrix} e^{i(\bm k' + \tau\bm G_1' -\tau\bm K - \tau\bm G_1)\cdot \bm r} , \notag
\end{equation}
\begin{equation}\label{eq:psi_s}
    \langle 0 |\Psi_{\tau l}(\bm r)|\bm k\lambda S\rangle = \frac{1}{\sqrt{\Omega}}\sum_{\bm g} \begin{pmatrix}
        \psi_{lA}^{\lambda}(\bm k + \bm g) \\
        \psi_{lB}^{\lambda}(\bm k + \bm g)\\
    \end{pmatrix}
     e^{i(\bm k + \bm g - \tau\bm K)\cdot \bm r}, %\notag
\end{equation}
where $|0\rangle$ is the vacuum. For the MLG tip, $\bm k'$ is restricted to be near the tip Dirac points $\tau\bm {K}_\theta$, $\theta_{\bm k'}= \arg [\tau (k_x'+ik_y')/(K_{\theta,x}+i K_{\theta,y}) - 1]$, and $\lambda'=\pm 1$ denotes the conduction and valence bands of MLG. In the sample, $\bm g$ represents all reciprocal lattice vectors which keep $\bm k+\bm g$ in the $\pm K$ valleys, where the low-energy continuum model is valid.
For instance, when the sample forms a long-period moir\'e superlattice, the wave vector $\bm k$ is defined in a mBZ and $\bm g$ contains the moir\'e reciprocal lattice vectors that are much shorter than $|\bm K|$. %, \textit{e.g.,} the moir\'e reciprocal lattice vectors.
The tunneling matrix elements read
\begin{align}
    &\langle \bm k'\lambda' T| H_{\text{tun}} |  \bm k\lambda S\rangle = \notag\\
    & \sum_{n=1}^{3}\sum_{\bm {g}} T_{\lambda^{\prime}\lambda}(\bm k+\bm g + \tau\bm{G}_{n})\delta_{\bm k'+\tau \bm{G}_{n}', \bm k+\bm {g}+\tau \bm{G}_{n}}, \label{eq:ht_tbg} \\
    & T_{\lambda^{\prime}\lambda}(\bm p) =  \left(\frac{\lambda'}{\sqrt{2}}, \frac{e^{i\tau\theta_{\bm p-\tau\bm G_n'}}}{\sqrt{2}}\right)^{*}\hat{T}_{\tau n}\psi_{t}^{\lambda}(\bm p- \tau \bm G_n). \label{eq:T_tbg}
\end{align}
%Here, $T_n$ is given by Eq.~\eqref{eq:tn}. 
In Eq.~\eqref{eq:T_tbg}, $\tau, n$ implicitly depend on $\bm p$ via the relation $\bm p = \bm k'+\tau\bm{G}_{n}'$. Equation~\eqref{eq:ht_tbg} is consistent with the general expression Eq.~\eqref{eq:ht}, where $\bm Q'=\tau \bm G_n'$ and $\bm Q$ are linear combinations of reciprocal lattice vectors in each layer of the sample (or equivalently, reciprocal lattice vectors $\tau\bm{G}_n$ of the topmost layer and moir\'e reciprocal lattice vectors $\bm g$). This equation indicates that the tip Dirac points $\tau \bm K_{\theta}$ couple with three wave vectors $\bm K_{\tau n}(\theta)  \equiv \bm p_{\tau n} - \tau \bm{G}_{n} (n=1,2,3)$ in the sample, where $\bm p_{\tau n} = \tau(\bm K_{\theta} + \bm G_{n}')$ are six Dirac points in the extended Brillouin zone of the MLG tip with six-fold rotation symmetry about the $\Gamma$ point. In valley $\tau$, the $\bm{K}_{\tau n}$ are $C_{3z}$ symmetric about $\tau\bm {K}$. 
% \textcolor{red}{
% Likewise, tunneling in the opposite valley occurs near the three Dirac points $\bm p_{-n} = -\bm p_n$ which are equivalent to $\bm K_{-n} = -\bm K_{n}$ in the sample.}
%

When probing a flat band of the graphene-based sample, the group velocity of the probed band is much smaller than the Dirac velocity in the tip, $\delta\epsilon_{\bm p}^{\lambda'\lambda}\approx - \xi_{\bm p\lambda}^{T} + \mathrm{const}$ and $|\partial_{\bm p}\delta\epsilon_{\bm p}^{\lambda'\lambda}/\hbar| \approx v_D $. The intersections between the tip and sample bands (Fig.~\ref{fig:intersections}) can be approximated as circles centered around $\bm p_{\tau n}$. Equation \eqref{eq:t_averaged} then reduces to 
\begin{align}
    |T|_{\tau n}^2(\theta,\lambda) &= \lim_{\bm k'\rightarrow \tau\bm K_{\theta}}\int \frac{d\theta_{\bm k'}}{2\pi} |T_{\lambda^{\prime}\lambda}(\bm k' + \tau\bm G_{n}')|^2 \notag\\
    & = t_{\theta_c}^2\left|e^{i\tau [\chi+\frac{2\pi (n-1)}{3}]} \psi_{tA}^{\lambda}+\psi_{tB}^{\lambda} \right|_{\bm{K}_{\tau n}(\theta)}^2. \label{eq:M_tbg}
    %&\equiv\overline{|T|_{\tau n}^2}(\theta,\lambda). \notag
\end{align}
Interestingly, the result equals the average of the tunneling matrix elements squared at the Dirac point, $|T|_{\tau n}^2(\theta,\lambda)=\sum_{\lambda'=\pm}|T_{\lambda'\lambda}(\bm p_{\tau n})|^2/2$. As shown in App.~\ref{sec:me_bloch}, corrections due to a finite dispersion of the probed band only rescale the right hand side of Eq.~\eqref{eq:M_tbg} by a velocity-dependent factor. In the limit $v_n\ll v_D$, the second derivative of the current,  Eq.~\eqref{eq:d2IdV2_dp}, becomes
\begin{align}\label{eq:d2IdV2_dp_flat}
    &\frac{d^2I}{dV^2} = \frac{2 \Omega e t_{\theta_c}^2}{\hbar^3 v_D^2}\left(\frac{d\phi}{dV}\right)^2 (f_{-eV-\mu_T} - f_{-\mu_{T}}) \sum_{n=1}^{3}\sum_{\tau=\pm}\notag\\
    &\sum_{\lambda}\left|e^{i\tau [\chi+\frac{2\pi (n-1)}{3}]} \psi_{tA}^{\lambda}+\psi_{tB}^{\lambda} \right|_{\bm{K}_{\tau n}}^2 \delta(eV+\mu_T+\xi_{\bm{K}_{\tau n}\lambda}^S). 
\end{align}
Notice that even for a single band $\lambda$, $d^2I/dV^2$ can contain multiple Dirac-$\delta$ singularities at different bias voltages because there are \textit{six} independent tunneling channels around $\bm p_{\tau n}$. The number of singularities is at most six but can be reduced by symmetries. Time-reversal (or inversion) symmetry ensures the valley degeneracy $\xi_{\bm K_{n}\lambda} = \xi_{\bm K_{-n}\lambda}$ and equal contributions to the tunneling current from the two valleys. 
%$C_{3z}$ symmetry imposes $\xi_{\bm K_{\tau n}}$ and $|T|^2_{\tau n} = |T|^2_{\tau n=1}$ due to $\psi_t^{\lambda}(\bm K_{\tau(n+1)}) = e^{i\frac{2\pi}{3}\tau\sigma^z}\psi_t^{\lambda}(\bm K_{\tau n})$.
$C_{3z}$ symmetry requires $\xi_{\bm K_{\tau n}\lambda}$ and $|T|^2_{\tau n}$  to be $n-$independent. When both symmetries are intact, we find that for $\theta\approx 0$ and $\pm 38.2^{\circ}$,
\begin{align}\label{eq:d2IdV2_dp_c3}
    \frac{d^2I}{dV^2} =& \frac{6 N_f \Omega e t_{\theta_c}^2}{\hbar^3 v_D^2}\left(\frac{d\phi}{dV}\right)^2 (f_{-eV-\mu_T} - f_{-\mu_{T}}) \notag\\
    &\times \sum_{\lambda}\left|\psi_{tA}^{\lambda}+\psi_{tB}^{\lambda} \right|_{\bm{K}_1}^2 \delta(eV+\mu_T+\xi_{\bm{K}_1\lambda}^S). 
\end{align}
%The singularities trace out the sample band structures along $\bm{K}_1 = \bm{K}_\theta\approx -\theta\hat{z}\times\bm{K}$.
%
% For $\theta\approx \pm 38.2^{\circ}$,
% \begin{align}\label{eq:d2IdV2_dp_c3_38.2}
%     &\frac{d^2I}{dV^2} = \frac{6 N_f \Omega e t_{38.2^{\circ}}^2}{\hbar^3 v_D^2}\left(\frac{d\phi}{dV}\right)^2 (f_{-eV-\mu_T} - f_{-\mu_{T}}) \notag\\
%     &\times \sum_{\lambda}\left|e^{\mp \frac{2\pi i}{3}}\psi_{tA}^{\lambda}+\psi_{tB}^{\lambda} \right|_{\bm{K}_1}^2 \delta(eV+\mu_T+\xi_{\bm{K}_1\lambda}^S). 
% \end{align}
%
When the system has SU(2) spin symmetry, $N_f=4$ accounts for the spin-valley degeneracy and $\lambda$ labels the bands of a single flavor in Eq.~\eqref{eq:d2IdV2_dp_c3}. Otherwise, $N_f=2$ accounts for the valley degeneracy and $\lambda$ includes all bands.
%The singularities trace out the sample band structures along $\bm{K}_1 = \bm{K}_\theta\approx -\theta\hat{z}\times\bm{K}$.

\section{Single-particle TBG model}\label{sec:normal}
We specify to a bilayer graphene sample at a small twist angle $\theta_{\text{TBG}}$ for the remaining parts of the paper.
%In the following sections, we study the elastic momentum-conserving tunneling between the MLG tip and 
% \begin{equation}
%     H_0 = -ie^{-i\frac{l\theta}{4}\sigma^z}(-i\bm \sigma\cdot \nabla)e^{i\frac{l\theta}{4}\sigma^z} + 
% \end{equation}
% $h_{\frac{l\theta}{2}}$.
We choose a coordinate system in which the $K$ valley Dirac points in the top and bottom layers of TBG are located at $\bm {K_{t/b}} = O(\pm \theta_{\text{TBG}}/2)(4\pi/3a_0,0)^{T}$, where $a_0$ is the lattice constant of graphene. In the mBZ, these two points are labeled as $\bm\kappa'$ and $\bm\kappa$, respectively, as depicted in Fig.~\ref{fig:k_trajectory}b.

Equations~\eqref{eq:d2IdV2_dp} and~\eqref{eq:d2IdV2_dp_flat} show that the tip scans the sample band structure simultaneously along three lines in the $K$ valley, $\bm K_{n}(\theta)  \equiv \bm p_{n}(\theta) - \bm{G}_{n}$ for $n=1,2,3$, and three lines in the opposite valley, $\bm K_{-n}=-\bm K_n$. According to the preceding section,
\begin{equation}\label{eq:pn}
    \bm p_n(\theta)= \bm{K}_{\theta} + \bm{G}_n' = O(\delta\theta)(\bm{K_t} + \bm{G}_n),
\end{equation}
\begin{equation}\label{eq:kn}
    \bm K_n(\theta)  = \bm {K_t} + \left[O(\delta\theta)-\hat{1}\right] (\bm {K_t}+\bm G_n),
\end{equation}
where $\delta\theta=\theta-\theta_c$ and $\bm{K_t} + \bm{G}_{n}$ are the three shortest overlapping Dirac momenta at commensurate angle $\theta_c$ between the tip and the top layer of TBG. The trajectories $\bm K_{1}(\theta)$ near $\theta_c=0,\pm 38.2^{\circ}$ are plotted as dashed lines in Fig.~\ref{fig:k_trajectory}b and can be approximated by
\begin{equation}\label{eq:k_1}
    \bm K_1 \approx \bm {K_t} -|\bm p_1|\delta\theta \left(\sin \frac{\theta_c}{2}, \cos \frac{\theta_c}{2}\right)^{T},
\end{equation}
where $|\bm p_1|=|\bm K|\ (\sqrt{7}|\bm K|)$ near $\theta_c=0\ (\pm 38.2^{\circ})$. The trajectories $\bm K_{2,3}$ can be obtained by rotating these lines about $\bm {K_t}$ clockwise by $\mp 2\pi/3$. The $\bm K_n(\theta)$ coincides with $\bm\kappa'$ in the mBZ when the MLG tip is commensurate with the top layer of TBG at $\delta\theta=0$, and $\bm\kappa$ when it is commensurate with the bottom layer at $\delta\theta=-\theta_{\text{TBG}}$. 

%In the remaining part of the section, we discuss the numerical calculations of tunneling current. 
The $K$ valley Dirac Hamiltonian of a graphene layer rotated clockwise by $\theta$ relative to the $x$ axis reads
\begin{equation}\label{eq:h}
    h_{\bm p}(\theta)=\hbar v_D (O(\theta)^{T}\bm p - \bm{K})\cdot\bm \sigma.
    % \begin{pmatrix}
    % 0 & (q_x-iq_y)e^{-i(\theta)} \\
    % (q_x+iq_y)e^{i\theta} & 0 \\
    % \end{pmatrix},
\end{equation}
Written in the plane-wave basis, the Bistritzer-MacDonald Hamiltonian 
of TBG reads
% \begin{equation}\label{eq:H_bm}
% \begin{split}
%     H_0 = &\sum_{\bm p, \alpha,\beta}\sum_{l=t,b}h_{\bm p-\bm K_l}(\theta_l)_{\alpha\beta} |\bm p, \alpha,l\rangle\langle \bm p, \beta,l| \\
%     &+\sum_{\bm p,\alpha,\beta} \sum_{n=1}^{3}(\tilde{T}_n)_{\alpha\beta} |\bm p+\bm {g}_n, \alpha,t\rangle\langle \bm p, \beta,b| + h.c.,
% \end{split}
% \end{equation}
\begin{align}\label{eq:H_bm}
\hat{H}_{ij}^0(\bm k) =
\begin{pmatrix}
    h_{\bm k + \bm g_i}(\frac{\theta_{\text{TBG}}}{2})\delta_{i,j} & \sum\limits_{n=1}^{3}\tilde{T}_n \delta_{\bm g_i+\bm g_n, \bm g_j}\\
    \sum\limits_{n=1}^{3}\tilde{T}_n^{\dagger} \delta_{\bm g_i, \bm g_j+\bm g_n} & h_{\bm k + \bm g_i}(-\frac{\theta_{\text{TBG}}}{2})\delta_{i,j}
\end{pmatrix}.
\end{align}
The moir\'e reciprocal lattice vectors are $\bm {g}_1=0$ and $\bm {g}_{2,3}= k_M(\pm\sqrt{3}/2,3/2)^{T}$ with $k_M =|\bm {K_b}-\bm {K_t}| = 2|\bm K|\sin (\theta_{\text{TBG}}/2)$. For simplicity, we use momentum-independent interlayer tunneling matrices $\tilde{T}_n$ of the same form as Eq.~\eqref{eq:tn_general}. Diagonalizing Eq.~\eqref{eq:H_bm} yields band energies $\xi_{\bm k\lambda}^{S}+\mu_{S}$ and eigenstates $\psi_{l\beta}^{\lambda}(\bm k + \bm g_i)$  defined in Eq.~\eqref{eq:psi_s}. These quantities may then be used to evaluate the second derivative of the current by two methods. One method exploits the approximate analytic expression Eq.~\eqref{eq:d2IdV2_dp_c3}. The other  computes the current as a function of bias voltage numerically after inserting the matrix elements Eqs.~\eqref{eq:ht_tbg} and~\eqref{eq:T_tbg} into Eq.~\eqref{eq:i_basic}, see App.~\ref{sec:numerics}.

Both methods are adopted and their results are compared in this section. While assuming that quasiparticles in MLG have infinite lifetime and $A_{\lambda}^{T}(\bm{k},\omega) = \delta(\omega - \xi_{\bm k, \lambda}^{T})$, we improve the stability of our numerical calculations by using a Lorentzian spectral function 
\begin{equation}\label{eq:As}
    A_{\lambda}^{S}(\bm{k},\omega) = \frac{1}{\pi}\frac{\gamma_S}{(\omega-\xi_{\bm k\lambda}^{S})^2+\gamma_S^2}
\end{equation}
for MATBG with a constant broadening parameter $\gamma_S=0.6$\si{meV}
and by taking a finite temperature $T=1$\si{\kelvin}. Unless otherwise stated, we focus on the flat bands and 
neglect remote bands of MATBG in our simulations to simplify the analysis. Simulations including the remote bands are shown at the end of this section. Below, we evaluate the Hamiltonian in Eq.~\eqref{eq:H_bm} for $\theta_{\text{TBG}}=1.05^{\circ}$, $v_D=10^6$\si{.\meter\per\second}, and intra-and inter-sublattice tunneling parameters $w_0=88$\si{meV} and $w_1=110$\si{meV}, respectively. $w_0<w_1$ accounts partially for the lattice relaxation and corrugation effects. The ratio $w_{0}/w_1\approx 0.8$ is suggested by tight-binding model estimates, but should be checked experimentally \cite{li2024evolution,li2024infrare}. Nonetheless, our qualitative results are not specific to the particular choice of model parameters.

\subsection{Dirac-point singularity \textit{vs.} Fermi-edge singularity}
\label{sec:DiracFermi}
\begin{figure}
    \centering
    \includegraphics[width=0.95\linewidth]{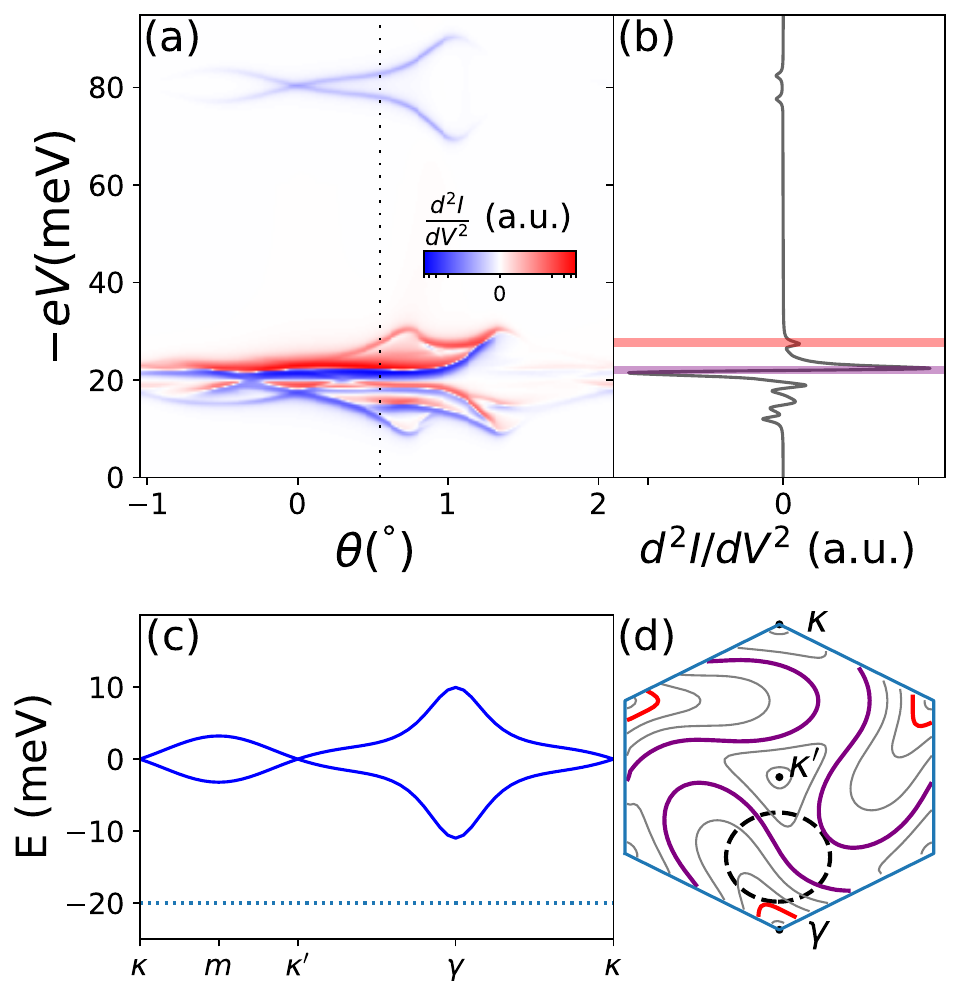}
    \caption{(a) Simulation of $d^2I/dV^2$ map for the tunnel junction between $1.05^{\circ}$-TBG with unoccupied flat bands and an electron-doped MLG tip. The chemical potentials of the TBG and MLG are $\mu_S=-20$\si{meV} and $\mu_T=60$\si{meV}, respectively. The map exhibits two types of $d^2I/dV^2$ singularities. Weaker ones induced by the Dirac points of the tip at bias voltages close to $\mu_T-\mu_S=80$\si{meV} and strong ones induced by the Fermi edge of the tip at smaller bias voltages. (b) A linecut along the dashed line in (a). (c) Band structure of $1.05^{\circ}$-TBG along the high symmetry line $\bm{\kappa}-\bm m - \bm \kappa' - \bm \gamma -\bm \kappa$. The dashed line marks the Fermi level. (d) Equal-energy contours of the conduction flat band of TBG are shown as solid lines. The dashed circle represents the Fermi circle of the tip. Tangency of this Fermi circle to two contours (of different energies) highlighted in red and purple contributes the two $d^2I/dV^2$ singularities in (b) at the bias voltages marked by the two horizontal lines in the corresponding colors. The crowding of Fermi-edge traces in (a) arises from the combination of the flatness of the TBG bands and the finite size of the MLG Fermi circle. The absence of the latter factor for scanning with a MLG Dirac point results in the two well-resolved lines in the top part of (a), which directly trace the sample band structure.}
    \label{fig:theta_c_0_N_fermi_edge}
\end{figure}

Figure \ref{fig:theta_c_0_N_fermi_edge}a presents a numerical simulation of the second derivative of the tunneling current between a hole-doped MATBG sample with empty flat bands and an electron-doped MLG tip. The $d^2I/dV^2$ map as a function of bias voltage and twist angle (around $\theta_c=0$) exhibits two types of sharp features in separate ranges of bias voltages. A linecut along the dotted vertical line at $\theta=0.58^{\circ}$ is plotted in Fig.~\ref{fig:theta_c_0_N_fermi_edge}b and shows that $d^2I/dV^2$ has two dips at large bias voltages $-eV\sim 80$\si{meV} and multiple peaks and dips of higher intensity at $-eV\sim 20$\si{meV}.

At large bias voltages, $-eV\sim\mu_T-\mu_S = 80$\si{meV}, the tip Dirac points are shifted in energy across the flat bands and induce Dirac-$\delta$ singularities in $d^2I/dV^2$. The blue lines in the color map trace out the dips of $d^2I/dV^2$ as a function of $\theta$. These lines %in the color map 
strikingly resemble the band structure of MATBG along the high symmetry line $\bm{\kappa}-\bm m - \bm \kappa' - \bm \gamma -\bm \kappa$ shown in Fig.~\ref{fig:theta_c_0_N_fermi_edge}c. This is consistent with our theory of Dirac-point spectroscopy.

At smaller bias voltages $-eV\sim -\mu_S = 20$\si{meV}, the Fermi level of the tip intersects the flat bands. However, for the parameters chosen in Fig.\ \ref{fig:theta_c_0_N_fermi_edge}a, 
the resulting features in the $d^2I/dV^2$ map are not in immediate correspondence with the sample band structure, as we explain below.
%{\color{brown}It is known \cite{xiao2024theory,inbar2023quantum} that at characteristic bias voltages $V^*$ where the tip Fermi circle is tangent to the sample bands, the singular tunneling phase space at the point of tangency generates square-root singularity in the tunneling current.} 

The bias dependence of the phase space for tunneling has a square-root singularity at a characteristic voltage $V^*(\theta)$, where the tip Fermi circle is tangent to the sample bands at a certain wave vector $\bm p^{*}(\theta)$ \cite{xiao2024theory,inbar2023quantum}. This nonanalyticity propagates into $I(V)$, producing a Fermi-edge singularity (see App.~\ref{app_sec:fs} for details)
\begin{align}
 %\frac{d^2I}{dV^2} \sim \frac{\Omega e^2}{h}|T_{\lambda\lambda'}(\bm p^*)|^2\sqrt{\frac{m^*}{\hbar^4 v_D^2 e|V-V^{*}|^{3}}} \Theta(\pm (V-V^{*})). \nonumber 
 \frac{d^2I}{dV^2} & \sim \mp \frac{\Omega e^2}{h}\frac{|T_{\lambda'\lambda}(\bm p^*)|^2}{\hbar^2v_D^2}
 \nonumber\\
 &\times
 \sqrt{\frac{v_D}{v_{\bm p^*}^S}\frac{|\mu_T|}{e|V-V^{*}(\theta)|^{3}}} \Theta(\pm [V-V^{*}(\theta)]). \label{eq:d2i_dv2_fs}
\end{align}
%
%\textcolor{red}{F: I left the equation unnumbered. But I am not sure that I understand the rationale behind it. I added a remark that it is derived in the limit of low doping in the tip. I also added a remark at the beginning of the paragraph stating that there is no general correspondence between fermi edge features and band structure. I find it easier to read if it is clear from the beginning what one is supposed to think about.}
This approximate expression is derived in the limit of low doping of the tip and generalizes related formulas in Refs.~\cite{inbar2023quantum,xiao2024theory} to flat-band samples with group velocity $v_{\bm p^*}^S$ at $\bm p^*$. One observes that the Fermi-edge contribution diverges more strongly than the Dirac-$\delta$ singularity induced by the tip's Dirac point and is quantitatively enhanced by the small velocity of the flat band. As a result, 
the strongest features in the $d^2I/dV^2$ map are due to the touching of the sample flat band by the tip's Fermi circle, as further illustrated by the line cut in Fig.~\ref{fig:theta_c_0_N_fermi_edge}b. 
In principle, this contribution to $d^2I/dV^2$ encodes band-structure information on the sample flat band through the dependence of the characteristic voltage $V^*$ on the angle $\theta$. However, the (necessarily) finite radius of the tip's Fermi circle complicates this dependence, as illustrated by Fig.~\ref{fig:theta_c_0_N_fermi_edge}d. The larger the tip's Fermi momentum the less direct is the  relation between the vector $\bm p^*$ and the angle $\theta$.
This reduces the spectral resolution as seen in Fig.~\ref{fig:theta_c_0_N_fermi_edge}a. The resolution can be improved by choosing a smaller Fermi momentum. But even then, the contribution of the tip's Fermi edge
provides at best the same band-structure information as scanning by the Dirac-point singularity. In addition to the sharpness of the features, the advantage of the latter is that it does not require fine-tuning of $\mu_T$ to the Dirac point. For this reason, we focus on 
Dirac-point spectroscopy, leaving an in-depth study of the Fermi-edge singularity beyond App.~\ref{app_sec:fs} for future work.

%The utility of the features in $d^2I/dV^2$ for spectroscopy of the sample bands relies on the dependence of the characteristic voltage $V^*$ on the angle $\theta$. A finite radius of the tip's Fermi circle complicates this dependence, as illustrated by Fig.~\ref{fig:theta_c_0_N_fermi_edge}d. The larger the tip's Fermi momentum the less direct is the  relation between the vector $\bm p^*$ and the angle $\theta$. This results in poorer spectral resolution, see Fig.~\ref{fig:theta_c_0_N_fermi_edge}a. Conversely, a smaller Fermi momentum improves resolution, but provides at best the same band-structure information as scanning by the Dirac-point singularity. In addition to the sharpness of the features, the advantage of the latter is that it does not require fine-tuning of $\mu_T$ to the Dirac point.
%However, the features have poorer resolution than those coming from the tip Dirac points, see Fig.~\ref{fig:theta_c_0_N_fermi_edge}a, and the extraction of the sample band structure is less immediate.
%The Fermi edge singularity is stronger than the Dirac-point singularity without temperature and quasiparticle broadening smearing. We will focus on the  Dirac-point singularity in the following discussions and leave the Fermi-edge singularity for future study.

\subsection[Dirac-point scanning at small twist angle]{Dirac-point scanning at small twist angles \texorpdfstring{($\theta_c=0$)}{0}}\label{sec:theta_0_N}
\begin{figure}
    \centering
    \includegraphics[width=1\linewidth]{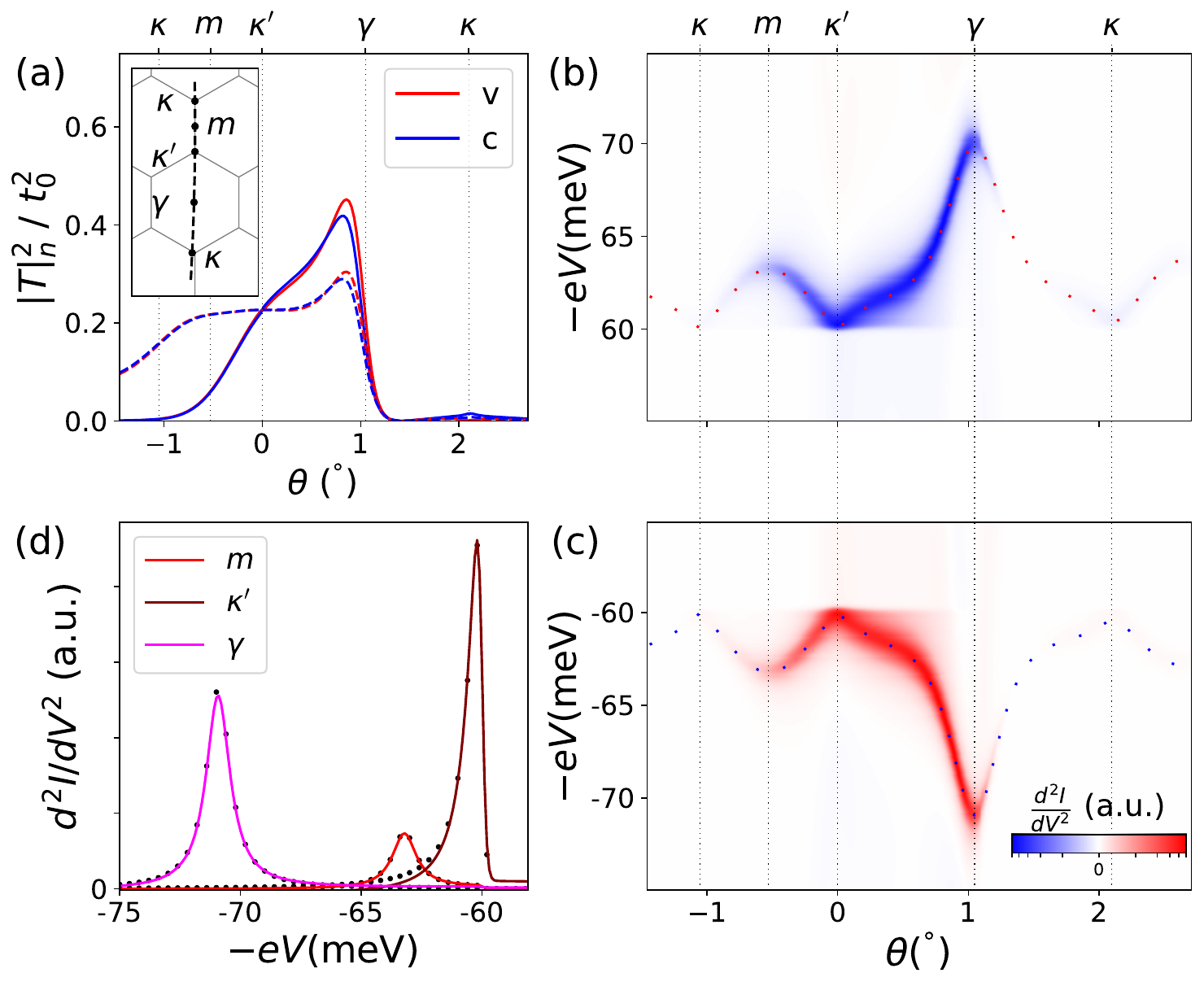}
    \caption{Dirac-point spectroscopy near $\theta_c=0$. (a) Solid lines represent the average tunneling matrix elements squared between the tip Dirac point and the MATBG Bloch states at wave vector $\bm K_1(\theta)$, see Eqs.~\eqref{eq:M_tbg}. Dashed lines plot the momentum distribution of the Bloch band $\lambda$ on the top layer, $\sum_{\alpha}|\psi_{t\alpha}^{\lambda}(\bm K_1)|^2$. The red and blue colors correspond to the flat valence ($\lambda=v$) and conduction ($\lambda=c$) bands, respectively. The vertical dotted lines mark the twist angles at which $\bm K_1(\theta)$ (almost) passes the high symmetry points in mBZs. The inset plots the trajectory of $\bm K_1(\theta)$. %with an arrow indicating the direction of increasing $\theta$. 
    (b) Simulated $d^2I/dV^2$ maps for charge-neutral MATBG at twist angle $\theta_{\text{TBG}}=1.05^{\circ}$ coupled with an electron-doped MLG with chemical potential $\mu_T=60$\si{meV}. The dots plot $\xi_{\bm K_1 c}^{S} + \mu_T$. (c) Same as (b) except for hole-doped MLG with $\mu_T=-60$\si{meV}. The dots plot $\xi_{\bm K_1 v}^{S} + \mu_T$. (d) Red, brown, and magenta curves are linecuts in (c) at $\bm m$, $\bm\kappa'$, and $\bm \gamma$, respectively. The dots calculated from Eq.~\eqref{eq:d2IdV2_dp_c3} with the Dirac $\delta-$function replaced by the spectral function $A_{\lambda}^{S}(\bm K_1,-eV-\mu_T)$ align with the simulated curves. %For these plots, $\gamma_S=0.6$\si{meV}.
    }
    \label{fig:theta_c_0_N}
\end{figure}
At small twist angles, the MLG tip scans the TBG band structure along $\bm K_{1}(\theta)$ [Eq.~\eqref{eq:kn}], see the black dashed line in Fig.~\ref{fig:k_trajectory}b and the inset of Fig.~\ref{fig:theta_c_0_N}a. As $\theta$ increases, $\bm K_1(\theta)$ sequentially traverses $\bm\kappa,\bm m, \bm\kappa'$, and $\bm \gamma$ before again passing the region near $\bm\kappa$ for $\theta\approx 2\theta_{\text{TBG}}$. 
%where the topmost $\bm\kappa$ corresponds to the bottom-layer Dirac point $\bm {K_b}$ and $\bm \kappa'$ corresponds to the top-layer Dirac point $\bm {K_t}$. THIS SEEMED REPETITIVE TO ME
%Let us begin with the matrix element effects in Eq.~\eqref{eq:d2IdV2_dp_c3}. 
Figure \ref{fig:theta_c_0_N}a depicts the square of the tunneling matrix element $|T|_n^2(\theta, \lambda)$ in Eq.~\eqref{eq:M_tbg} with $\chi=0$ (full lines) and the probability of the Bloch state $|\bm k\lambda S\rangle$ in the top layer at wave vector $\bm K_1(\theta)$, $\sum_{\alpha}|\psi_{t\alpha}^{\lambda}(\bm K_1)|^2$ (dashed lines), where $\bm k$ is the equivalent of $\bm K_1$ in the first mBZ. %the polarization of the Bloch state $| \bm k \lambda S\rangle$ to the plane-wave component at momentum $\bm K_n$ on the top layer, $\sum_{\alpha}|\psi_{t\alpha}^{\lambda'}(\bm K_n)|^2$
We highlight two features in the figure.%of the solid curves $|T|_1^2(\theta, \lambda)$:

% \begin{figure}
%     \centering
%     \includegraphics[width=1\linewidth]{matrixelements.pdf}
%     \caption{ (a) Solid lines stand for tunneling matrix elements squared from the tip Dirac point to the MATBG Bloch state $|\bm K_1\lambda S\rangle$ (see Eq.~\eqref{eq:M_tbg}) in the unit of $t_0^2$, while dotted lines are $\sum_{\alpha}|\psi_{t\alpha}^{\lambda}(\bm K_1)|^2$, the probability of the Bloch state on the top layer at momentum $\bm K_1$. The dependence of $\bm K_1$ on twist angle $\theta$ is given by Eq.~\eqref{eq:kn}. The blue and red colors correspond to the flat valence ($\lambda=v$) and conduction ($\lambda=c$) bands of TBG, respectively. (b) The same quantities except for $\theta\approx 38.2^{\circ}$ and in the unit of $t_{38.2^{\circ}}^2$. Solid curves swap colors at $\bm{K}_1=\bm {K_t}$ because of the double degeneracy of MATBG Bloch states at $\bm{\bm K_t}$. Comparing (a) and (b), we find that interference effects become more significant near the high symmetry line $\gamma-\kappa'-\kappa$ in Fig.~\ref{fig:k_trajectory}b. \textcolor{red}{to change the notations in the figure tonight.} }
%     \label{fig:matrixelements}
% \end{figure}

First, the results for the valence and conduction bands (red and blue lines, respectively) are nearly identical because of the $C_{2x}$ symmetry and the approximate particle-hole symmetry $P$ \cite{song2019all} of the BM Hamiltonian Eq.~\eqref{eq:H_bm}:
% They transform the plane-wave state in the following way,
% %
% \begin{align}
%     & C_{2x}|(p_x,p_y), \alpha, l\rangle = \sum_{\alpha' l'}\sigma_{\alpha\beta}^x \mu_{ll'}^x |(\bm p_x,-p_y), \alpha', l'\rangle, \\
%     & P|\bm p, \alpha, l\rangle = \sum_{l'} i\mu_{ll'}^y |\bm {K_t}+\bm{K_b}-\bm p, \alpha, l'\rangle,
% \end{align}
% %
% where $\mu^{i}$ are Pauli matrices in the layer pseudospin space. Therefore,
\begin{align}
    & C_{2x}: \psi_{t}^{\lambda}((p_x,p_y)) = \sigma^x\psi_{b}^{\lambda}((p_x,-p_y)), \label{eq:c_2x}\\
    & P: \psi_{t/b}^{c}(\bm p) = \pm \psi_{b/t}^{v}(\bm {K_t}+\bm{K_b}-\bm p).
\end{align}
When both symmetries are exact, their combination leads to $|T|_n^2(\theta, c) = |T|_n^2(\theta, v)$ for $\bm K_1(\theta)$ on the $\bm\kappa-\bm m- \bm\kappa'-\bm \gamma -\bm\kappa$ high symmetry line.

Second, the tunneling matrix elements are concentrated within a small range of twist angles, $|\theta| \lesssim \theta_{TBG}$. For $\theta>0$, the matrix elements follows the same trend as $\sum_{\alpha}|\psi_{t\alpha}^{\lambda}(\bm K_1)|^2$. When $\bm K_1$ is inside the Dirac cone centered at $\bm {K_t}$ of TBG, the phase difference between the two sublattice components of the top-layer wave function $\psi_{tA/B}^{\lambda}(\bm K_1)$ is close to $\pi/2$, so that the two sublattices essentially contribute independently to the square of the tunneling matrix element, see Eq.\ \eqref{eq:d2IdV2_dp_c3}. As a result, there is a close correspondence between the 
square of the tunneling matrix element and the probability of the Bloch state in the top layer for $\theta>0$.
As the probed wave vectors $\bm K_n(\theta)$ move further from the Dirac points $\bm {K_{t/b}}$, it is less likely for low-energy electrons to reach them via umklapp scattering of the moir\'e potential. As a result, both the wave function and the tunneling matrix elements of the flat bands drop rapidly at $\theta\gtrsim\theta_{TBG}$. In contrast, for $\theta<0$, there is a distinct difference between the 
square of the tunneling matrix element and the probability of the Bloch state in the top layer. We observe that 
$|T|_1^2(\theta,\lambda)$ decreases much faster than $\sum_{\alpha}|\psi_{t\alpha}^{\lambda}(\bm K_1)|^2$. This arises due to a strong interference suppression of the tunneling matrix elements.
In particular, tunneling is almost forbidden at $\bm {K_b}$ if $w_0=w_1$ in $\tilde{T}_n$. To see this, we treat $\tilde{T}_n$ as a perturbation in Eq.~\eqref{eq:H_bm} and obtain
\begin{equation}
    \psi_{t}^{\lambda}(\bm {K_b}) \approx \frac{1}{\hbar v_D k_M \sigma^y} \tilde{T}_1\psi_{b}^{\lambda}(\bm {K_b}) \propto
    \begin{pmatrix}
        1 \\
        -1
    \end{pmatrix},
\end{equation}
leading to vanishing tunneling matrix elements and $d^2I/dV^2$ at $\bm {K_b}$ according to Eq.~\eqref{eq:d2IdV2_dp_c3}. 

The strong suppression of $d^2I/dV^2$ when the MLG-Dirac point is twisted close to the bottom layer Dirac point of TBG was observed experimentally in large-angle TBG and  attributed to the low admixture of the top-layer component for Bloch wave functions near $\bm {K_b}$ \cite{inbar2023quantum}. Our results suggest that due to interference between the two sublattice components of the TBG Dirac-point wave functions, this phenomenon persists down to the magic angle despite the strong hybridization of the two layers.

This interference effect can be used to estimate the parameter $w_0/w_1$ of MATBG experimentally. In App.~\ref{sec:mee}, we derive for generic $w_0, w_1 (\ll \hbar v_D k_M)$ that
% ($\chi=0$ for $\theta\approx 0$),
%
\begin{equation}\label{eq:t_Kb}
    \sum_{\lambda=c,v}|T|_1^2(\theta_b,\lambda) \approx \frac{(w_1-w_0)^2}{\hbar^2 v_D^2 k_M^2}t_0^2\sum_{\lambda,\sigma}|\psi_{b\sigma}^{\lambda}(\bm {K_b})|^2,
\end{equation}
\begin{equation}\label{eq:t_Kt}
    \sum_{\lambda=c,v}|T|_1^2(\theta_t,\lambda) = t_0^2\sum_{\lambda,\sigma}|\psi_{t\sigma}^{\lambda}(\bm {K_t})|^2, 
\end{equation}
where $\theta_{b} = -\theta_{\text{TBG}}$ and $\theta_{t} =0$. By taking the ratio of the two equations, the unknown tunneling amplitude $t_0$ and the wave-function amplitudes cancel out as $|\psi_{t}^{\lambda}(\bm {K_t})|^2 = |\psi_{b}^{\lambda}(\bm {K_b})|^2$ due to the $C_{2x}$ symmetry Eq.~\eqref{eq:c_2x}. As the energy of the MLG Dirac point is tuned across the TBG Dirac point $\bm {K_l}$ by bias voltage, the size of the jump $\Delta(dI/dV)|_{\bm {K_l}}$ in $dI/dV$ is proportional to the tunneling matrix elements squared given by Eqs.~\eqref{eq:t_Kb} and~\eqref{eq:t_Kt}, and obeys the simple relation
\begin{equation}\label{eq:d2IdV2_ratio}
    \frac{\Delta\left(dI/dV\right)\big |_{\bm {K_b}}}{\Delta\left(dI/dV\right)\big |_{\bm {K_t}}} \approx \frac{(w_1-w_0)^2}{\hbar^2 v_D^2 k_M^2}.
\end{equation}
This allows for a determination of $w_0/w_1$ when combined with the Fermi velocity $v_F$ at the Dirac nodes of charge-neutral MATBG \cite{bernevig2021twisted}
\begin{equation}\label{eq:vf}
   \frac{v_{F}}{v_D}\approx \frac{1-3w_1^2/\hbar^2 v_D^2 k_M^2}{1+3(w_0^2+w_1^2)/\hbar^2v_D^2 k_M^2}, %\notag
\end{equation} 
which can be extracted from band structure measurements. For the commonly used value $w_0/w_1=0.8$ and $\hbar v_D k_M\approx \sqrt{3}w_1$ near the magic angle \cite{bistritzer2011moire}, the ratio in Eq.~\eqref{eq:d2IdV2_ratio} is $\sim 10^{-2}$. 

These results are derived using a continuum description of tip-sample tunneling. This is  applicable at small but finite angles $|\theta|$ and  $|\theta-\theta_{\text{TBG}}|$. Scanning the MATBG Dirac points at $\theta=0$ ($\bm {K_t}$) and $\theta_{\text{TBG}}$ ($\bm {K_b}$) violates this condition. In this case, the tunneling matrix elements vary with the lateral shift between the tip and sample \cite{bistritzer2010transport}. However, our results remain valid upon averaging $d^2I/dV^2$ over lateral shifts of the tip by distances of the order of the graphene lattice period.
%Strictly speaking, our expression for tunneling matrix elements Eq.~\eqref{eq:ht_squared} assume the incommensurability of the reciprocal lattice vectors $\bm Q$ in the sample and $\bm Q'$ in the tip. Scanning the Dirac points of the top (bottom) layer of MATBG results in commensuration of that layer and the MLG tip. %, while at $\theta= -\theta_{\text{TBG}}, \pm 38.2^{\circ}-\theta_{\text{TBG}}$, it is commensurate to the bottom layer of TBG. In either case, 
%In this case, the tunneling conductance depends on the relative in-plane translation vector $\bm d$ of the two commensurate lattices due to the interference of the three tunneling processes in each valley \cite{bistritzer2010transport}. Eqs.~\eqref{eq:ht_squared} and~\eqref{eq:fermigolden} average $\bm d$ over one graphene unit cell. 

We now illustrate the extraction of the sample band structure using Dirac-point spectroscopy based on numerical calculations of the current. %For each $\theta$, we use Eq.~\eqref{eq:i_basic} to compute tunneling current $I$ as a function of $V$ with fixed tip and sample chemical potentials, $\mu_{T}$ and $\mu_{S}$, respectively. 
We fix $\mu_T = \pm 60$\si{meV}, which is larger than the bandwidth of the flat bands. This separates the Dirac-point and Fermi-edge singularities in the tunneling spectra. The results are otherwise  insensitive to the choice of $\mu_T$. Figure \ref{fig:theta_c_0_N}b depicts the second derivative of the tunneling current between electron-doped MLG ($\mu_T = 60$\si{meV}) and charge-neutral MATBG. For $-eV>\mu_T$, $d^2I/dV^2$ exhibits dips as a function of $V$ whose positions track the conduction flat band of MATBG,
$ -eV_{n}= \xi_{\bm K_1(\theta),c}^{S} + \mu_T$ as represented by the dots. For $-eV < \mu_{T}$, $d^2I/dV^2\approx 0$ due to Pauli blocking when the MLG Dirac points and the MATBG single-particle states of the same energy are both occupied. 
%More precisely, as shown in App.~\ref{sec:dirac}, the $d^2I/dV^2$ singularity generated by touching of the MLG Dirac point and TBG flat bands contains the spectral function information of the flat bands. 
Figure \ref{fig:theta_c_0_N}c plots $d^2I/dV^2$ for hole-doped MLG ($\mu_{T}=-60$\si{meV}). In this case, the peaks in $d^2I/dV^2$ tracks the valence flat band (dotted line) below the MATBG Fermi level. $d^2I/dV^2$ flips sign as the states near the Dirac points in MLG are now empty and contribute to the tunneling current with opposite sign. The twist-angle dependence of the peak intensity in Figs. \ref{fig:theta_c_0_N}b,c is consistent with the matrix elements in Fig.~\ref{fig:theta_c_0_N}a.

We finally compare our numerical results to the approximate analytical result in Eq.~\eqref{eq:d2IdV2_dp_c3}. 
Figure \ref{fig:theta_c_0_N}d shows linecuts from Fig.~\ref{fig:theta_c_0_N}c at $\theta = -\theta_{\text{TBG}}/2$ (red), $0$ (brown), and $\theta_{\text{TBG}}$ (magenta), corresponding to the $\bm m$, $\bm\kappa'$, and $\bm \gamma$ points in the mBZs, respectively. They agree well with the dots calculated from Eq.~\eqref{eq:d2IdV2_dp_c3} after substituting the Dirac $\delta$-function with $A_{\lambda}^{S}(\bm K_1(\theta), -eV-\mu_T)$. The small discrepancies between the two results (e.g., on the tail of the brown line) should be attributed to the non-negligible momentum dependence of the tunneling matrix elements $T_{\lambda'\lambda}(\bm p)$ as hinted by the strong angle dependence of $|T|_n^2$ in Fig.~\ref{fig:theta_c_0_N}a. Overall, our numerical results indicate that Dirac-point spectroscopy allows one to extract the quasiparticle peaks of spectral functions from $d^2I/dV^2$ to a good approximation.

\subsection[Large commensurate angle]{Large commensurate angles (\texorpdfstring{$\theta_c=38.2^{\circ}$)}{38.2}}
\begin{figure}
    \centering
    \includegraphics[width=1\linewidth]{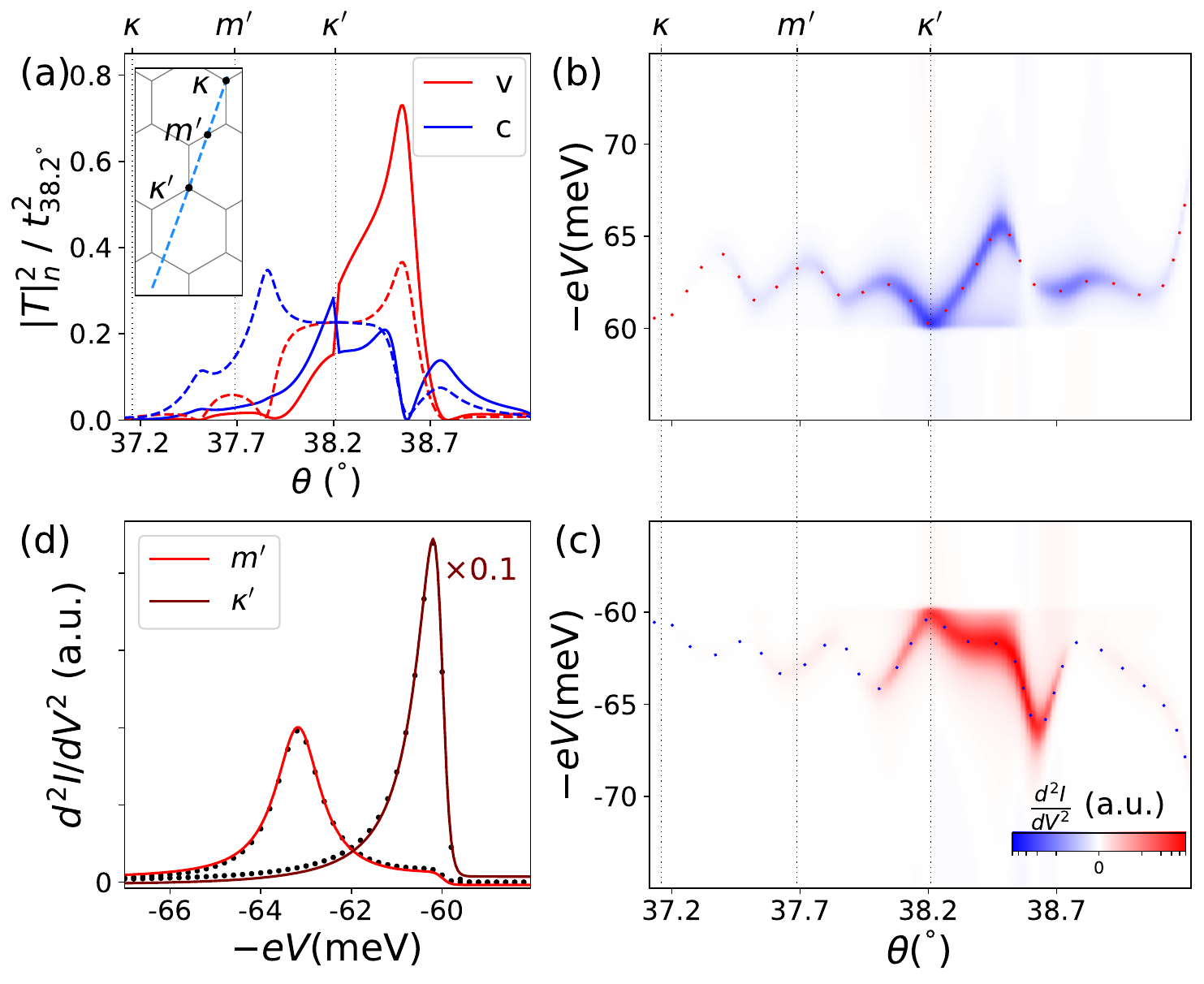}
    \caption{Same quantities as in Fig.~\ref{fig:theta_c_0_N}, but for $\theta_c \approx 38.2^{\circ}$. The solid curves have a discontinuity at $\theta_c$ because the relative phase between $\psi_{tA}^{c/v}(\bm K_1)$ and $\psi_{tB}^{c/v}(\bm K_1)$ changes by $\pi$ as $\bm {K}_1$ passes the top-layer Dirac point $\bm {K_t}$ at $\theta=\theta_c$. The inset of (a) shows the $\bm K_1(\theta)$ trajectory in mBZs, which crosses the three high symmetry points $\bm\kappa, \bm m'$, and $\bm\kappa'$(\textit{i.e.}, $\bm {K_t}$) with increasing $\theta$. (b) and (c) are simulated $d^2I/dV^2$ maps for charge-neutral MATBG with the tip chemical potential $\mu_T=\pm 60$\si{meV}. (d) Linecuts from (c) at $\bm m'$ and $\bm\kappa'$.}
    \label{fig:theta_c_38.2_N}
\end{figure}
%The interference effect is more significant at small twist angles when the tip Dirac point traverse a high-symmetry line of MATBG as shown by the blue line in Fig.~\ref{fig:k_trajectory}b

In this section, we analyze the matrix elements and numerical simulation of the tunneling current around the commensurate angle $\theta_c \approx 38.2^{\circ}$. The inset of Fig.~\ref{fig:theta_c_38.2_N}a shows the path $\bm K_1(\theta_c+\delta\theta)$ scanned by the tip Dirac point for $|\delta\theta|\sim 1^{\circ}$. The path of $\bm K_1$ passes $\bm\kappa$ at $\delta\theta=-\theta_{\text{TBG}}$, $\bm\kappa'$ at $\delta\theta = 0$, and their midpoint $\bm m'$ at $\delta\theta = -\theta_{\text{TBG}}/2$. %Here, $\bm \kappa'$ corresponds to the top-layer Dirac point $\bm {K_t}$ of TBG.

Figure \ref{fig:theta_c_38.2_N}a depicts $|T|_1^2(\theta,\lambda)$ and $\sum_{\alpha}|\psi_{t\alpha}^{\lambda}(\bm K_1)|^2$ in solid and dotted lines, respectively. Similar to the small-angle scanning, these two quantities are well correlated at twist angles slightly larger than $\theta_c$, while the interference suppression of tunneling matrix elements become evident for angles smaller than $\theta_c$.
%as $\bm K_1$ moves close to $\bm \kappa'$. 
The suppressed matrix elements at a few twist angles are due to almost vanishing wave function amplitudes $\psi_{t}^{\lambda}(\bm K_1)$. In addition, the tunneling matrix elements exhibit a discontinuity across $\theta_c$. Near the top-layer Dirac point $\bm {K_t}$, the top-layer wave functions $\psi_{t}^{\lambda}(\bm K_1)$ are approximately the eigenstates of the effective Dirac Hamiltonian $\hbar v_F (\bm K_1-\bm{K_t})\cdot\bm\sigma$. Thus, $\psi_{tA}^{c/v}(\bm K_1)/\psi_{tB}^{c/v}(\bm K_1) \approx e^{i(\theta_c \pm \pi\text{sgn}(\theta-\theta_c))/2}$ along the trajectory of the tip Dirac point in the inset, cf. Eq.~\eqref{eq:k_1}. Thus,
%
% \begin{equation}
%     |T|_1^2/t_{38.2^{\circ}}^2 = |\psi_{t B}^{c}(\bm K_1)|^2|e^{i\frac{\theta_c - \pi}{2}-i\frac{2\pi}{3}} + 1 |^2  \approx 0,
% \end{equation}
%
\begin{equation}
    \frac{|T|_1^2(\theta_c^-, c)}{|T|_1^2(\theta_c^+, c)} \approx \frac{\cos^2({\frac{2\chi + \theta_c - \pi}{4}})}{\cos^2({\frac{2\chi + \theta_c + \pi}{4}})},
\end{equation}
%, where the twist angle dependence of tunneling matrix elements are resulted from rapid change of the relative phase between two sublattice components $\psi_{t A/B}^{\lambda}(\bm K_1)$ in Eq.~\eqref{eq:M_tbg} near the Dirac point $\bm \kappa'$.
%
with $\chi = 0$ in our numerical calculations.
%\textcolor{brown}{Optional: This interference suppression of tunneling matrix elements on one side of the twist angle $\theta_c=38.2^{\circ}$ pertains to the Dirac cone in the sample and is applicable to a MLG sample as well, which is distinct from the interference effect near $\bm{K_b}$ analyzed in Sec.~\ref{sec:theta_0_N}.} Additionally, compared with the small-twist-angle scanning, $|T|_1^2(\theta,\lambda)$ now shows a stronger asymmetry between the flat conduction and valence bands. 

For the simulated $d^2I/dV^2$ maps in Figs.~\ref{fig:theta_c_38.2_N}b and c, we find again a good match of the dip or peak positions and the MATBG band structures represented by dots. Figure \ref{fig:theta_c_38.2_N}d shows linecuts at $\bm m'$ and $\bm\kappa'$ points, which align with the black dots derived from Eq.~\eqref{eq:d2IdV2_dp_c3} after replacing the Dirac $\delta-$function by the broadened spectral function.
 
Our analysis generalizes straightforwardly to other special twist angles $\theta_c$.
Using the invariance under $C_{6z}$ rotations of the MLG tip, we obtain the tunneling current near $\theta_c \approx -21.8^{\circ}$ by the substitution $\theta\rightarrow \theta-\pi/3$. In addition, the approximate $C_{2x}P$ symmetry of the present model maps the conduction and valence bands at wave vector $\bm K_1(\theta_c+\delta\theta)$ to the valence and conduction bands at $\bm K_{1}(-\theta_c+\delta\theta)$, respectively, for small $\delta\theta$. For this reason, we do not repeat the analysis for $\theta\approx -38.2^{\circ}$ or $\theta\approx 21.8^{\circ}$. If the $C_{2x}P$ symmetry is strongly broken in MATBG, however, scanning near these commensurate angles generally provides additional information on the electronic structure, beyond what is obtained from scans near $0^{\circ}$ and $38.2^{\circ}$.

\begin{figure}
    \centering
    \includegraphics[width=1\linewidth]{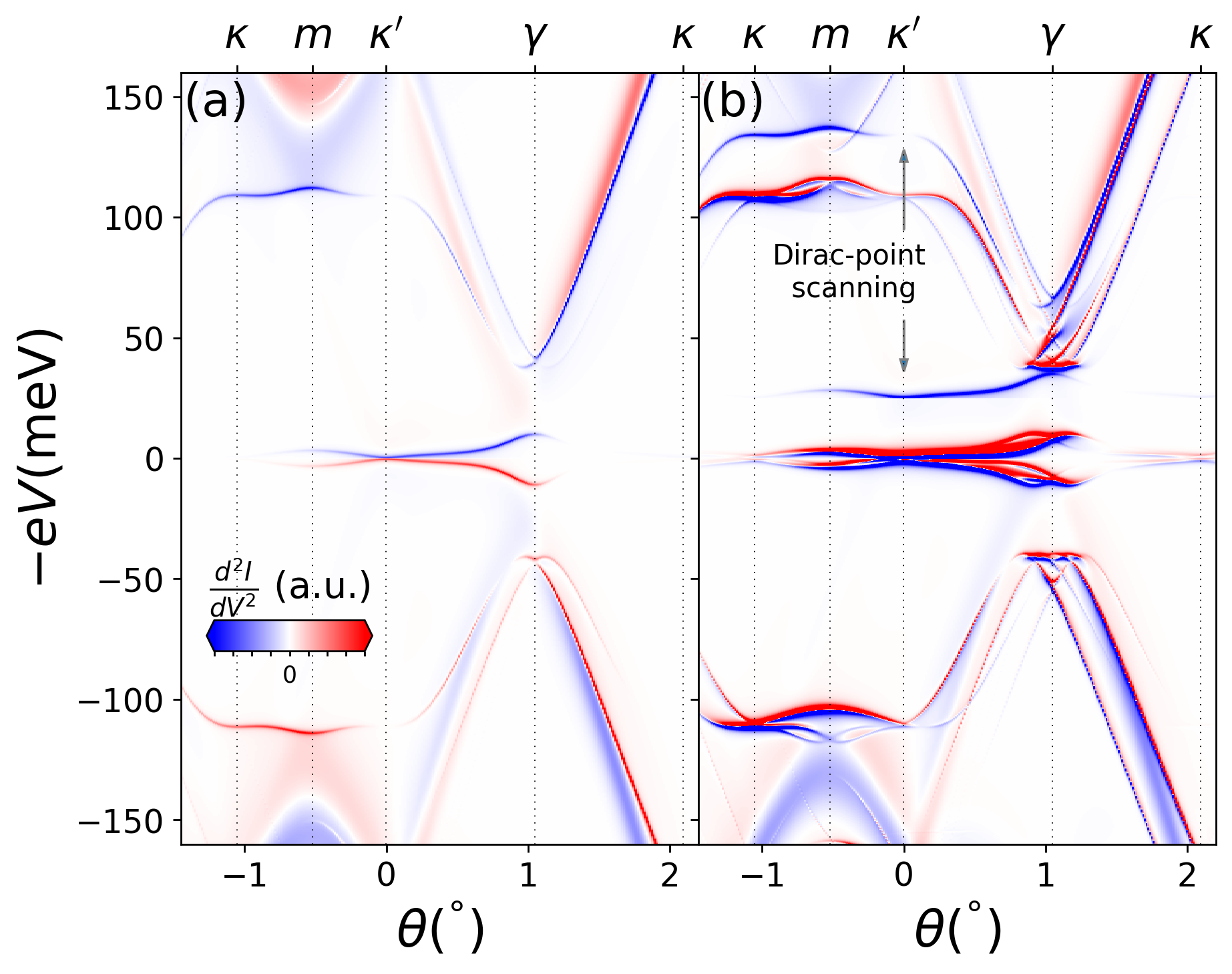}
    \caption{Simulated $d^2I/dV^2$ maps for scanning  $1.05^{\circ}$-TBG at charge neutrality ($\mu_S=0$) with a MLG tip, including the remote bands of TBG. (a) MLG tip tuned to charge neutrality ($\mu_T=0$). The main features in $d^2I/dV^2$ are produced by Dirac-point scanning. (The Fermi-edge contribution vanishes due to the zero density of states at the chemical potential of the tip.) (b) MLG chemical potential tuned into the conduction band ($\mu_T=25$\si{\milli\electronvolt}). The tip's Fermi level remains inside the gap between the flat and remote bands of TBG when the tip's Dirac points are scanning the sample's flat bands. Features appearing due to Dirac-point scanning (sharp blue lines without red satellites) are highlighted by arrows. Blue and red double lines 
    result from the Fermi-edge singularities. Note that due to the nonzero $\mu_T$, lines due to Dirac-point scanning in panel (b) are shifted up by $25$\si{\milli\electronvolt} relative to their counterparts in panel (a) and relative to the corresponding Fermi-edge features in panel (b).} %The Fermi-edge singularity at small bias encodes the TBG band structure information, but has complicate features near $\bm\gamma$ point due to the finite radius of the tip's Fermi circle.
    \label{fig:theta_0_mu_t_0}
\end{figure}

We conclude Sec.~\ref{sec:normal} with a discussion of QTM maps over a broader energy range, which includes the remote bands. We also contrast the case of $\mu_T$ and $\mu_S$ parked at the respective charge neutrality points (Fig.~\ref{fig:theta_0_mu_t_0}a) with the case of the tip chemical potential detuned from the neutrality point (Fig.~\ref{fig:theta_0_mu_t_0}b). In the former case, one observes sharp traces produced by both the conduction and valence flat bands. In addition, the remote bands are clearly resolved. At a finite chemical potential, $\mu_T=25$meV, the overall intensity of $d^2I/dV^2$ increases. Scanning with the Dirac point produces a trace of the conduction flat and remote bands only. (As discussed in Sec.\ \ref{sec:general}, scanning of the valence flat and remote bands would require tuning to a negative $\mu_T$.) Moreover, we observe that the features produced by the Fermi edge crossing the flat bands are blurred. (We note that the features are more closely related to the sample bands structure than in Fig.\ \ref{fig:theta_c_0_N_fermi_edge}a due to the smaller choice for $\mu_T$ consistent with the discussion in Sec.\ \ref{sec:DiracFermi}.)
The traces associated with the Fermi edge crossing the remote bands are also less sharp than those produced by the Dirac points of the tip, especially around the $\bm\gamma$ points.
%Note that the Dirac points of MATBG have much smaller Fermi velocity than the MLG. As pointed out in Sec.~\eqref{sec:main}, they cannot induce additional singularities when touching the MLG bands. This explains why our simulation of current second derivative at low temperature does not reveal a noticeable MLG Dirac cone that centered at the MATBG Dirac point.   

%For large deviations from commensurate angles, tunneling current diminishes because Bloch wave functions in MATBG flat bands concentrate on one or two shells of moir\'e reciprocal lattice vectors $\bm {g_m}$'s and $\psi_{t \beta}^{S\lambda' }(\bm{p})\approx 0$ if $\bm p$ is sufficiently away from the Dirac points in either layer of MATBG, $|\bm{k}-\bm {K_{t/b}}|\gtrsim |\bm g|$.

\section[Effects of strain]{Effects of $C_{3z}$ symmetry breaking}\label{sec:strain}
\begin{figure*}
    \centering
    \includegraphics[width=1.8\columnwidth]{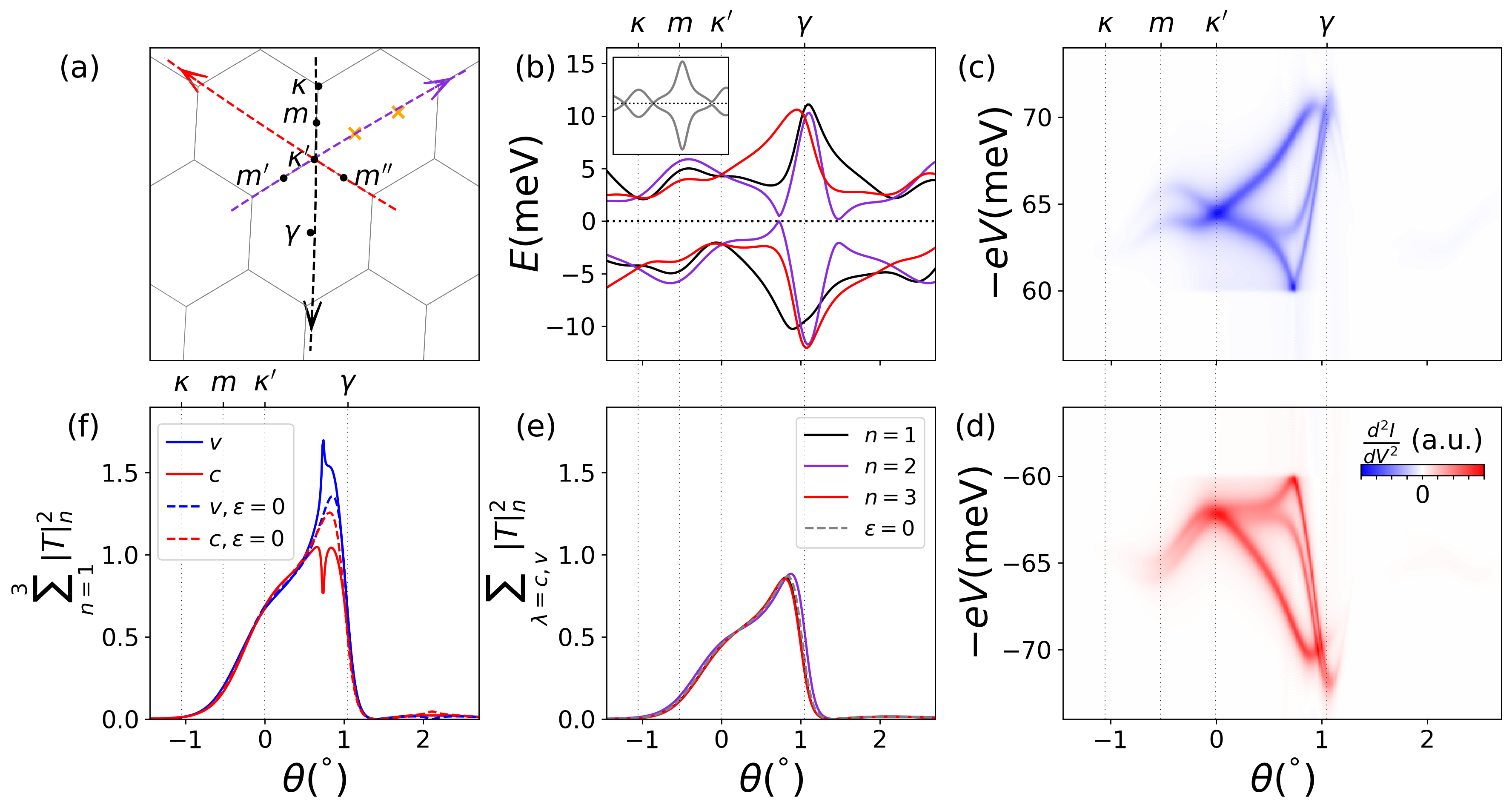}
    \caption{Effect of breaking $C_{3z}$ symmetry. (a) Deformed mBZs of MATBG with $\theta_{\text{TBG}}=1.05^{\circ}$ and uniaxial heterostrain $\epsilon=0.1\%$, $\varphi = 15^{\circ}$. The two crosses mark $K$ valley Dirac points of strained MATBG. The $K$ valley Dirac point of the MLG tip couples with the three wave vectors $\bar{\bm K}_{1,2,3}(\theta)$ in the mBZs (via umklapp scattering). The black, red, and violet dashed lines represent an interval of $K_{1,2,3}(\theta)$, respectively, across the same range of $\theta$ with arrows indicating the direction of increasing $\theta$. Note that $\bar{\bm K}_{1,2,3}$ exhibit small deviations from $\bm\kappa,\bm\kappa'$ proportional to $\epsilon$. For heterostrain, their deviations from $\bm m,\bm m',\bm m''$ at $\theta = -\theta_{\text{TBG}}/2$, respectively, are even smaller and of second order in $\theta_{\text{TBG}}$ and $\epsilon$. (b) $K$ valley single-particle energy dispersion of MATBG along the three intervals of trajectories in (a). Energy is measured relative to the chemical potential $\mu_S$ at charge neutrality. $\theta=0$ when the MLG tip is angle aligned with the top layer of MATBG. The vertical lines mark the twist angles at which $\bar{\bm K}_1(\theta)$ reaches closest to the four labelled wave vectors $\bm\kappa,\bm m, \bm \kappa',$ and$\bm\gamma$. The inset plots energy dispersions of strain-free MATBG, where the dispersions along the three trajectories become identical due to $C_{3z}$ symmetry. (c)-(d) Simulated $d^2I/dV^2$ maps for strained MATBG at charge neutrality coupled to (c) electron-doped MLG with $\mu_T=60$\si{meV} and (d) hole-doped MLG with $\mu_T=-60$\si{meV}. $C_{3z}$ symmetry breaking generally triples the number of singularities, which occur at distinct bias voltages and have different intensities. (e) Solid lines in three colors plot the total tunneling matrix elements squared of MATBG flat bands along the three trajectories $\bar{\bm K}_{n}(\theta)$ in (a). The gray dashed line shows the same quantity in strain-free MATBG and is identical for $n=1,2,3$. (f) Total tunneling matrix elements squared along the three trajectories $\bar{\bm K}_{n}(\theta)$. The blue and red solid (dashed) lines correspond to the valence (v) and conduction (c) flat bands of strained (strain-free) MATBG.}
    \label{fig:strain}
\end{figure*}
%represent trajectories $\bm K_{1,2,3}(\theta)$ of the tip $K$ valley Dirac points in mBZs, 

As dicussed above, the tip scans the sample band structure along three lines $\bm K_n(\theta)$ within each valley. As these three lines are related by $C_{3z}$ symmetry, this provides a sensitive probe of $C_{3z}$-symmetry breaking in the sample, provided that $C_{3z}$ symmetry is preserved in the tip. 

%In this section, we show that $C_{3z}$ symmetry breaking can triple the tunneling current singularity in the Dirac-point spectroscopy which provides band structure information of the sample along three distinct lines in reciprocal space.

Strain-induced distortions of the moir\'e superlattice which break the $C_{3z}$ symmetry are commonly seen in STM experiments on TBG \cite{kerelsky2019maximized,choi2019electronic,xie2019spectroscopic}. Consider two graphene layers $l=t/b$ that are uniformly deformed from a honeycomb lattice $\bm R_{i}$ to new lattice coordinates $\bar{\bm R}_{l,i} = \mathcal{M}_{l}\bm R_{i} + \bm d_l$, where $\bm d_l$ is a layer-dependent in-plane translation  and $\mathcal{M}_l$ can be decomposed into rotation and strain,
\begin{equation}\label{eq:deformation}
    \mathcal{M}_{l} = 
    \left(1 + \mathcal{E}_l \right) O(\theta_l).
\end{equation}
We can fix the twist angles to $\theta_t = -\theta_b = \theta_{\text{TBG}}/2$. The strain tensor $\mathcal{E}_l$ is real symmetric. For simplicity, we only consider uniaxial heterostrains so that $\mathcal{E}_t = -\mathcal{E}_b = \mathcal{E}/2$ and $\mathcal{E}$ can be described by two parameters, the strain magnitude $\epsilon$ and the strain direction $\varphi$, 
\begin{equation}
\mathcal{E} \equiv 
\begin{pmatrix}
    \epsilon_{xx} & \epsilon_{xy}\\
    \epsilon_{xy} & \epsilon_{yy}
\end{pmatrix}
= O(\varphi)^{T}
    \begin{pmatrix}
        \epsilon & 0\\
        0 & -\nu_{\scriptscriptstyle{P}}\epsilon
    \end{pmatrix} O(\varphi),
\end{equation}
where $\nu_{\scriptscriptstyle{P}}\approx 0.16$ is the Poisson ratio of monolayer graphene. 

Next, we analyze the strain-induced corrections to the single-particle Hamiltonian. To leading order in $\epsilon$, the single-layer Dirac Hamiltonian Eq.~\eqref{eq:h} becomes \cite{bi2019designing}
\begin{equation}
    h_{\bm p,l}' = \hbar v_D \left(\mathcal{M}_l^{T} \bm p - \tau\bm K -  \tau\bm A_l \right) \cdot (\tau\sigma^x, \sigma^y), 
    %\approx \hbar v_D \mathcal{M}^{T} (\bm p -\bm K' + \bm A)\cdot\bm\sigma, 
\end{equation}
where $\tau=\pm$ denotes the valley index of the wave vector $\bm p$ and $\bm A_t = -\bm A_b = \frac{\sqrt{3}\beta}{4a_0}(\epsilon_{xx} - \epsilon_{yy}, -2\epsilon_{xy})$ (with $\beta \approx 3.14$). The gauge field $\bm A_l$ originates from the dependence of the hopping amplitudes on the distance between neighboring sites. For small twist angles and strains, the Dirac points of layer $l$ are shifted from $\pm\bm K = (\pm 4\pi/3a_0, 0)$ to $\pm(\bar{\bm {K_l}} + \bm A_l)$, where $\pm\bar{\bm {K_l}} = \pm(\mathcal{M}_l^{-1})^{T} \bm K$ are the corners of the deformed Brillouin zone. The gauge field $\bm{A}_l$ shifts the Dirac points of strained graphene away from the Brillouin zone corners. 

The continuum model for the interlayer tunneling Hamiltonian becomes
\begin{equation}\label{eq:H_T_strain}
    \int d^2r \sum_{n=1}^3 \Psi_{\tau t}^{\dagger}(\bm r) \tilde{T}_{\tau n} e^{-i\tau\bar{\bm g}_n\cdot\bm r} \Psi_{\tau b}(\bm r) + h.c.,
\end{equation}
where $\bar{\bm g}_1=0$, $\bar{\bm g}_{2,3} = (\mathcal{M}_{t}^{-1} - \mathcal{M}_{b}^{-1})^{T}\bm G_{\pm}$, and $\bm G_{\pm}=(-\frac{2\pi}{a_0}, \pm\frac{2\pi}{\sqrt{3} a_0})^{T}$ are two primitive reciprocal lattice vectors of the original lattice $\bm R_i$. Because we have neglected the momentum dependence of the interlayer tunneling matrix $\tilde{T}_{\tau n}$, strain does not modify $\tilde{T}_{\tau n}$. In summary, the BM Hamiltonian of the $K$ valley electrons in TBG subject to uniform uniaxial heterostrain reads \cite{bi2019designing}
\begin{align}\label{eq:H_bm_strain}
&\hat{H}_{ij}'(\bm p) =
\begin{pmatrix}
    h_{\bm p + \tau\bar{\bm g}_i, t}'\delta_{i,j} & \sum\limits_{n=1}^{3}\tilde{T}_{\tau n} \delta_{\bar{\bm g}_i+\bar{\bm g}_n, \bar{\bm g}_j}\\
    \sum\limits_{n=1}^{3}\tilde{T}_{\tau n}^{\dagger} \delta_{\bar{\bm g}_i, \bar{\bm g}_j+ \bar{\bm g}_n} & h_{\bm p + \tau\bar{\bm g}_i, b}'\delta_{i,j}
\end{pmatrix}.
\end{align}
This Hamiltonian breaks $C_{3z}$ symmetry, but preserves time-reversal symmetry, $\hat{H}_{ij}^{\prime}(\bm p) = \hat{H}_{ij}^{\prime}(-\bm p)^{*}$. 

The deformation of the sample reciprocal lattice changes the three wave vectors in the $K$ valley of the sample which couple (via umklapp scattering) most strongly to the tip Dirac points. The change is from $\bm K_{n} = \bm p_n - \bm G_n$ given by Eqs.~\eqref{eq:pn} and~\eqref{eq:kn} to $\bar{\bm K}_{n}\equiv \bm p_n - (1+\mathcal{E}_t)^{-1}\bm G_{n} \approx \bm {K}_n(\theta) + \mathcal{E}_t \bm G_n$ for $n=1,2,3$. Here, the $\bm G_n$ are the reciprocal lattice vectors of the top layer of undeformed TBG as defined in Sec.~\ref{sec:commensuration}.
% For weak strain $\epsilon$,
% \begin{equation}\label{eq:kn_strain}
%     \bar{\bm K}_n(\theta)  = \bm {K}_n(\theta) + \mathcal{E}_t \bm G_n.
% \end{equation}
%
Figure \ref{fig:strain}a depicts $\bar{\bm K}_{1,2,3}$ as the black, red, and violet dashed lines, respectively. The arrows point along the positive $\theta$ direction and each forms angles of $2\pi/3$ with the other two. If the sample is subject to heterostrains only, the three lines will pass the points $\bm m, \bm m'$, and $\bm m''$ in the mBZ.

Figure \ref{fig:strain}b depicts band dispersions of MATBG with a strain field $\epsilon=0.1\%$ and $\varphi = 15^{\circ}$ along the three trajectories $\bar{\bm K}_{n}(\theta)$. In the absence of $C_{3z}$ symmetry, the quasiparticle energies $\xi_{\bar{\bm K}_{1,2,3}(\theta) \lambda}^{S}$ are generally all different. In contrast, without strain, $\epsilon=0$, $\mathcal{M}_{t/b} = O(\pm\theta_{\text{TBG}}/2)$, and $\bar{\bm g}=\bm g$. In this limit, Eq.~\eqref{eq:H_bm_strain} reduces to Eq.~\eqref{eq:H_bm}, and the energy dispersions along all three $\bm K_{n}(\theta)$ become identical, see the inset of Fig.~\ref{fig:strain}b. The Dirac points of MATBG marked by the crosses in Fig.~\ref{fig:strain}a shift away from the mBZ corners and do not lie precisely on the three trajectories. %exactly.%, \textcolor{red}{although we find numerically that for a wide range of parameters $\epsilon$ and $\phi$ they are very close to one of the trajectories.}
%Comparing band structures of MATBG with and without strains, it is readily noticed that the flat band structures are highly susceptible to heterostrains \cite{huder2018electronic, parker2021strain}.

Time-reversal symmetry ensures identical contributions to the tunneling current from the two valleys and therefore simplifies the general expression Eq.~\eqref{eq:d2IdV2_dp} for the second derivative of the current to
\begin{align}\label{eq:d2IdV2_dp_trs}
    &\frac{d^2I}{dV^2} = \frac{4 \Omega e t_{\theta_c}^2}{\hbar^3 v_D^2}\left(\frac{d\phi}{dV}\right)^2 (f_{-eV-\mu_T} - f_{-\mu_{T}}) \sum_{\lambda}\sum_{n=1}^{3}\notag\\
    &\quad \left|e^{i\frac{2\pi (\chi + n-1)}{3}} \psi_{tA}^{\lambda}+\psi_{tB}^{\lambda} \right|_{\bar{\bm{K}}_{n}}^2 \delta(eV+\mu_T+\xi_{\bar{\bm{K}}_{n}\lambda}^S). 
\end{align}
Since neither the band structure $\xi_{\bar{\bm K}_{n}(\theta) \lambda}^{S}$ nor the tunneling matrix elements are $C_{3z}$ invariant, the  singularity in $d^2I/dV^2$ splits into three with different intensities. 

Figure \ref{fig:strain}c,d depicts the simulated $d^2I/dV^2$ maps as a function of bias voltage $V$ and twist angle $\theta$ between the top layer of MATBG and the MLG tip. As $\theta$ varies, the dips (peaks) in $d^2I/dV^2$ trace out the dispersion of the conduction (valence) band of strained MATBG
%MATBG conduction (valence) band dispersion
along these three trajectories when the MLG tip is electron (hole)-doped. The dispersion agrees well with the band structure in Fig.~\ref{fig:strain}b. In particular, for $0<\theta<1^{\circ}$, our results confirm the tripling of the current singularities when $C_{3z}$ symmetry is broken. (Outside of this twist angle range, at least one of the singularities becomes too weak to be visible.) 
Despite the dramatic changes in the energy dispersions in weakly strained MATBG, the twist angle dependence of the peak intensities in $d^2I/dV^2$ shares similarities with Figs.~\ref{fig:theta_c_0_N}b and c for unstrained MATBG. %The intensities  $\theta<0$ and $\theta>1^{\circ}$, and the tunneling matrix elements near $\bar{\bm K}_{1}(-\theta_{\text{TBG}}) \approx \bar{\bm {K_b}}$ is much less than at $\bar{\bm K}_1(0)\approx \bar{\bm {K_t}}$.
This can be rationalized by an approximate sum rule of the tunneling matrix elements of MATBG flat bands for sufficiently weak strain,
\begin{equation}\label{eq:sumrule}
     \sum_{\lambda= c,v} |T|_n^2(\theta, \lambda) \approx t_{\theta_c}^2\sum_{\rho= c,v} \left|\tilde{\psi}_{tA}^{\rho}+ \tilde{\psi}_{tB}^{\rho} \right|_{\bm K_1(\theta)}^2. 
\end{equation}
Here, $\tilde{\psi}$ are reciprocal-space wave functions of strain-free MATBG. The right hand side is the sum of the tunneling matrix elements squared over the MATBG flat bands without strain and is plotted as a dashed line in Fig.~\ref{fig:strain}e. The left hand side, depicted as solid lines, is evaluated in the presence of weak strain, $\epsilon=0.1\%$ and $\varphi=15^{\circ}$. We observe that the ``sum rule" holds very well along all three trajectories $\bar{\bm K}_{n}$. To derive Eq.~\eqref{eq:sumrule}, we assume the strain to be sufficiently weak such that $\bar{\bm K}_n \approx \bm K_n$ and the strain field induces little mixing between the flat and remote bands of MATBG. Nevertheless, even a weak strain field can strongly mix the two flat bands
\begin{equation}\label{eq:psi_unitary}
    {\psi}^{\lambda}(\bm K_n) = \sum_{\rho = c,v} U_{\rho}^{\lambda}(\bm K_n) \tilde{\psi}^{\rho}(\bm K_n),
\end{equation}
where $U_{\rho}^{\lambda}$ is a $2\times 2$ unitary matrix and $\tilde{\psi}^{\rho}$ denotes the flat band Bloch wave functions in the absence of strain. One can derive Eq.~\eqref{eq:sumrule} by plugging Eq.~\eqref{eq:psi_unitary} and the orthonormality $\sum_{\lambda}U_{\rho}^{\lambda}(U_{\rho'}^{\lambda})^{*}= \delta_{\rho,\rho'}$ into Eq.~\eqref{eq:M_tbg} and using the $C_{3z}-$invariance of the tunneling matrix elements squared in strain-free MATBG. 
%The approximate sum rule ensures that the tunneling matrix elements of strained MATBG exhibit the same trend as those in unstrained MATBG as the twist angle varies. %{\color{brown}{\bf skip to avoid repetitiveness and for the sake of brevity?} In particular, the interference suppression of tunneling between the MLG Dirac points and the flat bands near $\bm {K_b}$, which we explained in Sec.~\ref{sec:normal} in the absence of strain, is robust against perturbations to the Bistritzer-MacDonald model in Eq.~\eqref{eq:H_bm} much weaker than the energy separation $\sim \hbar v_D k_M\sim 100$\si{meV} of the flat and remote bands at $\bm {K_b}$.}

Ideally, the  approximate sum rule of the tunneling matrix elements can be useful to extract band-structure information from $d^2I/dV^2$ maps when the $C_{3z}$ symmetry is broken. For instance, Fig.~\ref{fig:strain}c shows that for $0<\theta<1^{\circ}$ the strongest singularity is the one at the highest bias voltage, while Fig.~\ref{fig:strain}d shows that the middle singularity has the smallest intensity. Therefore, these two singularities correspond to the dispersion of the conduction and valence flat bands at the same wave vector, consistent with the two red bands in Fig.~\ref{fig:strain}b.

Figure \ref{fig:strain}f depicts $\sum_{n=1}^{3}|T|_n^2(\theta, \lambda)$. We find numerically that this sum is also insensitive to weak strain for a range of twist angles $\theta$. However, for $0.5^{\circ}<\theta<1^{\circ}$, we observe a large difference between samples with strain ($\epsilon=0.1\%$ and $\varphi=15^{\circ}$, solid lines) and strain-free MATBG (dashed lines). In this regime, $\bar{\bm K}_{3}(\theta)$ is close to the MATBG Dirac points, see the violet dashed line in Fig.~\ref{fig:strain}a, near which the relative phases between two sublattice components of the top-layer wave functions $\arg (\psi_{tB}^{\lambda}/\psi_{tA}^{\lambda})$ becomes sensitive to $\bar{\bm K}_3(\theta)$ and leads to sharp variations in the tunneling matrix elements squared Eq.~\eqref{eq:M_tbg}.

Finally, we note that although this section focuses on the effects of $C_{3z}$ symmetry breaking induced by uniaxial heterostrains, there could be other sources of $C_{3z}$ symmetry breaking, including extrinsic ones such as homostrains or intrinsic ones such as nematic order. Additionally, even for $C_{3z}$-symmetric samples, the MLG tip could also be subject to strain fields \cite{tang2024chip}. Either of these factors could in principle triple the number of current singularities provided that the QTM has sufficient energy resolution. 

\section{Conclusions and discussion}\label{sec:discussion}
In this work, we have developed a theory for Dirac-point tunneling spectroscopy of the momentum-resolved energy dispersion and single-particle spectral functions of the flat bands of moir\'e graphene with a quantum twisting microscope (QTM). To this end, we have studied elastic momentum-conserving tunneling through a planar junction between a monolayer graphene (MLG) tip and a moir\'e graphene sample. The QTM provides independent control of the doping levels and energy-band offsets of the tip and sample layers via bias and gate voltages, while the tip Dirac momenta can be shifted relative to the Brillouin zones of the sample by twisting. The second derivative $d^2I/dV^2$ of the tunneling current exhibits singularities at characteristic bias voltages $V_n$, at which the Dirac-point energy aligns with the bands of the sample at the same momentum. Measuring $V_n$ as a function of twist angle $\theta$ maps out the energy dispersion of both occupied and unoccupied states in the sample along the trajectories of the Dirac points in reci\-procal space. These trajectories  traverse all high symmetry points of the mBZ. This method should be contrasted with ARPES, which measures band structures over a finite area of the Brillouin zone but only below the Fermi level at low temperature. 

In addition to the spectrum, Dirac-point spectroscopy provides information regarding the electron wave functions of the sample, especially for flat bands where the twist-angle dependence of the intensity of the features in $d^2I/dV^2$ is solely determined by
%Dirac-point singularity 
%depends on 
the sample wave functions and not by the band dispersion. For samples made of magic angle twisted bilayer graphene (MATBG), we propose to extract the ratio of the intrasublattice and intersublattice tunneling parameters, $w_0/w_1$, a key parameter in the Bistritzer-MacDonald model, from the intensity of $d^2I/dV^2$ singularities at the two Dirac points $\bm{K_{b}}$ and $\bm {K_{t}}$ of MATBG. 

In experiment, the resolution
%momentum resolution 
of the QTM is restricted by the dimensions of the tunnel junction as well as by disorder and the quasiparticle lifetimes in sample and tip. 
The quasiparticle lifetime is fundamentally limited by electron-electron scattering.
%while the energy resolution of Dirac-point spectroscopy at low temperature ($k_BT\ll \mu_T$ \textcolor{brown}{N: to discuss}) could be potentially limited by at least the following factors.
%(i) The broadening of quasiparticle spectral peaks of the tip Dirac point due to the disorder scattering and electron-electron interactions. 
In doped MLG, the electron-electron scattering rate of quasiparticles at a Dirac point due to the Coulomb interaction increases
approximately linearly with the Fermi energy $\mu_T$ \cite{polini2008plasmons,hwang2008quasiparticle}. 
%\cite{kotov2012electron,bostwick2010observation,brar2010observation}
%(Within the second order perturbation theory in Coulomb interactions, the scattering rate of quasiparticles at the Dirac points given by the Fermi's golden rule is proportional to the Fermi energy). 
Using a MLG tip at a small doping level reduces this interaction-induced broadening, although tuning the Fermi level precisely to the MLG Dirac point is challenging due to charge inhomogeneity \cite{rhodes2019disorder}. More generally, the inhomogeneity may help mask the effects of a residual gap in the MLG tip spectrum induced by the hexagonal-boron nitride (hBN) substrate. %Thus, we do not expect additional limitations of the resolution due to such a gap. 
The sharpness of the resonances in $d^2I/dV^2$ vs.\ $\theta$ also relies on the weakness of the tip-sample tunneling. (The tunneling strength between two stacked graphene layers twisted by $\theta_c=\pm 38.2^{\circ}$ is about $t_{38.2^{\circ}}\sim 1$\si{meV} \cite{scheer2022magic,shallcross2008quantum,jiang2022tunable,talktington2023terahertz}.)

We note that a periodic electrostatic potential $V(\bm r)=\sum_{\bm g} V(\bm g)e^{i\bm g\cdot\bm r}$ imposed by the sample moir\'e superlattice on the tip does not degrade the energy or momentum resolution of Dirac-point spectroscopy. $V(\bm r)$ cannot gap the Dirac cone because the tip Hamiltonian
\begin{equation}
    \hbar v_D \bar{\bm k}\cdot \sigma + V(\bm r)\sigma_0, \notag
\end{equation}
has a Kramer's degeneracy at $\bar{\bm k}=0$ protected by the antiunitary symmetry $\mathcal{R}=\sigma^y\mathcal{K}$, where $\mathcal{K}$ is complex conjugation and $\mathcal{R}^2 =-1$.
%The Dirac crossing is also protected by $C_{2z}\mathcal{T} =\sigma^x\mathcal{K}$ symmetry \cite{po2018origin,zou2018band}.
Since $V(\bm r)$ has the same periodicity as the sample, the tip wave functions with and without modulation of $V(\bm r)$ couple to the same quasimomenta in the sample via the momentum-conserving tunneling. Thus, the umklapp scattering due to $V(\bm r)$ only results in corrections to the tunneling matrix elements on the order of $|V_{\bm g}|/\hbar v_D|\bm g|$ and does not  broaden the Dirac-point singularities. 
%
%According to Fermi liquid theory, this approximation works best at the Fermi level where the zero-temperature spectral function $A_{\lambda,\lambda'}^{t}(k_{F}, \epsilon)=Z\delta(\omega-\xi_{k_{F},\lambda}^{t})\delta_{\lambda,\lambda'}$ in the . Therefore, we expect that at sufficiently low temperature, the singularity induced by the tangency of the MLG Fermi surface and the remain correct.

While our work primarily focuses on Dirac-point spectroscopy, investigating protocols to extract momentum-resolved energy dispersion from the Fermi-edge singularities would also be highly valuable. First, as discussed in Sec.~\ref{sec:DiracFermi}, the Fermi-edge singularity could have a stronger intensity than the Dirac-point singularity. Second, although the finite extent of the Fermi lines of the tip complicates the extraction of band structure information, it also offers the advantage of probing momenta away from the momentum-space trajectories %that are distant from the trajectories 
of tip Dirac points.
%, which can be difficult to access with Dirac-point spectroscopy
Moreover, Fermi-edge singularities are %singularity is 
ubiquitous in momentum-conserving tunneling between two-dimensional electron systems, making them less restrictive with regard to the choice of tip material. 
In this work, we studied tunneling spectra of the flat bands of TBG within the single-particle approximation as  described by the Bistritzer-MacDonald model. Band deformations arising from Hartree-Fock effects are believed to depend strongly on the filling of flat bands and play an important role in the physics of TBG such as its fermiology \cite{guinea2018electrostatic,zhu2024gw,ezzi2024self}. Scanning the flat bands as a function of filling is therefore of much interest. Our theory also paves the way for the exploration of momentum-resolved tunneling spectroscopy of phases with spontaneously broken symmetries  \cite{sharpe2019emergent,serlin2020intrinsic,zhou2021half,wolf2024magnetism}. Probing collective excitations \cite{xie2024long,lewandowski2019intrinsically,kumar2021lattice,bernevig2021twistedv,khalaf2020soft,peri2024probing,pichler2024probing} in flat-band systems by exploiting inelastic tunneling in QTM constitutes another interesting direction.

\begin{acknowledgments}
We thank Paco Guinea, Shahal Ilani, and Jiewen Xiao for insightful discussions and comments. Calculations were conducted at the Yale Center for Research Computing. N.W.\ acknowledges support through the Yale Prize Postdoctoral Fellowship in Condensed Matter Theory. Research at Yale was supported by NSF Grant No.\ DMR-2410182 and by the Office of Naval Research (ONR) under Award No.\ N00014-22-1-2764. Research at Freie Universit\"at Berlin and Yale was supported by Deutsche Forschungsgemeinschaft through CRC 183 (project C02 and a Mercator Fellowship).   Research at Freie Universit\"{a}t Berlin was further supported by Deutsche Forschungsgemeinschaft through a joint ANR-DFG project (TWISTGRAPH). This work was performed in part at the Aspen Center for Physics, which is supported by a grant from the Simons Foundation (1161654, Troyer) and by National Science Foundation grant PHY-2210452.
\end{acknowledgments}

%\clearpage
%\newpage
\appendix
\onecolumngrid

\section{Tunneling-current singularity induced by Dirac points}\label{app_sec:dos}
%
%Near Dirac points $\bm p_n$, $\delta\epsilon_{\bm p}^{\pm, \lambda} \approx \bm v_{n}\cdot (\bm p -\bm p_n) \mp v_D|\bm p -\bm p_n| + \delta\epsilon_n$ with the sample group velocity at $\bm p_n$, $\bm v_{n}=\partial_{\bm p}\xi_{\bm p\lambda}^{S}|_{\bm p_n}$. 

\subsection{Singularity in the limit of zero level broadening and $T=0$}\label{sec:zerotemperature}

Without loss of generality, we assume that the tip band has a Dirac point at wave vector $\bm p = 0$. Near the Dirac point, the two-band Hamiltonian of the tip can be expressed as $H^{T}(\bm k')= \hbar v_D\bm k' \cdot\bm \sigma$. We use $\lambda'= \pm 1$ to label its conduction and valence bands, which have energy dispersions $\xi_{\bm k' \lambda'}^T = \lambda' \hbar v_D |\bm k'| - \mu_T$ relative to the tip's Fermi level, and two-component Bloch wave functions $\psi^{\lambda'}(\bm k') = (\lambda', e^{i\theta_{\bm k'}})^{\text{T}}/\sqrt{2}$. To simplify notations, we consider a \textit{single} band $\lambda$ in the sample with energy dispersion $\xi_{\bm p \lambda}^S$ relative to the sample's Fermi level.

The tunneling current derived from Fermi's golden rule reads
\begin{align}\label{app_eq:i}
    I = \frac{2\pi e \Omega}{\hbar}\sum_{\bm p}\sum_{\lambda'}(f(\xi_{\bm p \lambda}^{S}) - f(\xi_{\bm p \lambda'}^T))  |T_{\lambda'\lambda}(\bm p)|^2 \delta (eV- \xi_{\bm p\lambda'}^{T} + \xi_{\bm p\lambda}^{S})
\end{align}
where the wave vector $\bm p$ is summed over the 2D reciprocal space. Because we are only interested in the singular contribution to the tunneling current from the Dirac point at $\bm p=0$, we truncate the wave vectors to a sufficiently small region $\mathcal{R}$ centered around $\bm p=0$ where the energy dispersion of the sample band $\lambda$ can be linearized as $\xi_{\bm p\lambda}^S\approx \xi_{0\lambda}^S + \bm v_0 \cdot \bm p$ and the two Fermi-Dirac distributions are approximately constant. Eq.~\eqref{app_eq:i} then reduces to 
\begin{align}
    I \approx \frac{2\pi e \Omega}{\hbar}\sum_{\lambda'}  (f(\xi_{0\lambda'}^{T}-eV) - f(\xi_{0\lambda'}^{T}))\sum_{\bm p\in \mathcal{R}}|T_{\lambda'\lambda}(\bm p)|^2\delta (eV - \xi_{\bm p\lambda'}^{T} + \xi_{\bm p\lambda}^{S})  +.... \label{app_eq:i_dp}
\end{align}
Here, ``$\ldots$" includes contributions to the tunneling current from regions outside of $\mathcal{R}$, which will be omitted below. %$T_{\lambda\lambda'}(\bm p)$ depends on the interlayer tunneling Hamiltonian which will be specified in Sec.~\ref{sec:H_tun_derivation}. Nevertheless, 
Based on the wave function $\psi^{\lambda'}(\bm k')$ of the Bloch state $|\bm k' \lambda' T\rangle$ and the assumption that the wave functions of the sample are smooth at $\bm p=0$, the tunneling matrix elements can be generally written as
\begin{equation}\label{app_eq:T_general}
    T_{\lambda'\lambda}(\bm p) = \frac{1}{\sqrt{2}}\left( a(\bm p) \lambda'+ b(\bm p) e^{-i\theta_{\bm p}}\right) \approx \frac{1}{\sqrt{2}}\left( a \lambda'+  b e^{-i\theta_{\bm p}} \right),
\end{equation}
where $a(\bm p)$ and $b(\bm p)$ are two smooth functions and $a=a(0)$, $b = b(0)$.

Plugging Eq.~\eqref{app_eq:T_general} into Eq.~\eqref{app_eq:i_dp} and defining $\epsilon \equiv -eV -\mu_T - \xi_{0\lambda}^S$, we arrive at
\begin{equation}
\label{app_eq:i2}
    I = \frac{2\pi e \Omega}{\hbar}  (f(-eV-\mu_T) - f(-\mu_T))\left(\frac{|a|^2 + |b|^2}{2} \mathcal{I}_1(\epsilon) + \frac{a b^{*} }{2}e^{i\theta_0}\mathcal{I}_2(\epsilon) + \frac{a^{*} b}{2}e^{-i\theta_0}\mathcal{I}_2(\epsilon)^{*} \right). 
\end{equation}
with $\bm v_0 = v_0 (\cos\theta_0, \sin\theta_0)$ and
\begin{align}
    \mathcal{I}_1(\epsilon) \equiv &\sum_{ \lambda'=\pm 1} \int\frac{d^2p}{(2\pi)^2} \delta(\epsilon - \hbar \bm v_0\cdot\bm p+\lambda' \hbar v_D |\bm p|), \label{app_eq:I1}\\
    \mathcal{I}_2(\epsilon) \equiv &\sum_{ \lambda'=\pm 1} \lambda' \int\frac{d^2p}{(2\pi)^2} e^{i(\theta_{\bm p}-\theta_0)} \delta(\epsilon - \hbar \bm v_0\cdot\bm p+\lambda' \hbar v_D |\bm p|). \label{app_eq:I2}
\end{align}
Note that $\mathcal{I}_1(\epsilon)$ is the joint density of states of the combined spectrum $\delta\epsilon_{\bm p}^{\lambda'\lambda} = \xi_{\bm p \lambda}^S + \mu_S - \xi_{\bm p\lambda'}^{T} - \mu_T$. The contours of integration as defined by the Dirac $\delta$-functions in Eqs.~\eqref{app_eq:I1} and~\eqref{app_eq:I2} can be ellipses ($v_0<v_D$), parabolas ($v_0=v_D$), or hyperbolas $(v_0>v_D)$. $\mathcal{I}_{1,2}(\epsilon)$ behave qualitatively different in the three regimes of $\alpha=v_0/v_D$:

(i) $v_0 < v_D$: For convenience, we redefine the positive $x-$direction as being parallel to $\bm{v}_0$. The elliptic intersections are $(1-\alpha^2)( p_x+\frac{\alpha}{1-\alpha^2}p_\epsilon )^2 + p_y^2 = \frac{1}{1-\alpha^2}p_{\epsilon}^2$ with $p_{\epsilon}=\epsilon/\hbar v_D$. Equations~\eqref{app_eq:I1} and~\eqref{app_eq:I2} simplify to
\begin{align}
    \mathcal{I}_1(\epsilon) = & \frac{|\epsilon|}{2\pi \hbar^2 v_D^2 \left(1-\frac{v_0^2}{v_D^2} \right)^{\frac{3}{2}}}, \label{app_eq:I1_v0<vd}\\
    \mathcal{I}_2(\epsilon) = & \frac{v_0}{v_D} \frac{|\epsilon|}{2\pi \hbar^2 v_D^2\left (1-\frac{v_0^2}{v_D^2} \right)^{\frac{3}{2}}}. \label{app_eq:I2_v0<vd}
\end{align}
%
%{\color{brown}(?) Exclude the intermediate-step integration in A(10), (A11), formulate the result for I(V) in Eq. (A4)}

(ii) $v_0=v_D$: We truncate the parabolic intersection $2p_\epsilon p_x = p_\epsilon^2 - p_y^2$ to $-\Lambda_-/v_D < p_x < \Lambda_+/v_D$. Then
\begin{align}
    \mathcal{I}_1(\epsilon) =  \mathcal{I}_2(\epsilon) = \frac{\sqrt{2}}{6\pi^2 \hbar^2 v_D^2}\frac{\Lambda_{\text{sgn}(-\epsilon)}^{\frac{3}{2}}}{\sqrt{|\epsilon|}}. \label{app_eq:I1_v0=vd}
\end{align}
It is interesting to point out that for $v_0= v_D$, the entire $p_x-$axis is the $\epsilon=0$ equal-energy line of $\delta\epsilon_{\bm p}^{\lambda'\lambda}$.  $\delta\epsilon_{\bm p}^{\lambda'\lambda}$ increases with $p_y$ quadratically on both sides of the $p_x-$axis. The inverse square root singularity of $\nu$ therefore has the same origin as the density of states singularity of a 1D parabolic band.
%{\color{brown}(?) Exclude the intermediate-step integration in A(12), (A13), formulate the result for I(V) in Eq. (A4)}

(iii) $v_0 > v_D$: The equal energy contours consist of two branches of hyperbolas. We split the integral into the two pieces which correspond to the left and right branches of the hyperbola with $x-$intercepts $p_{\pm} = \frac{\alpha}{\alpha^2-1}p_\epsilon \pm \frac{1}{\alpha^2-1}|p_\epsilon|$, respectively. We find that both branches of hyperbola can give singular contributions $\propto \epsilon \ln |\epsilon|$ to $\mathcal{I}_{1,2}$ but the signs of these two singular contributions are opposite. As a result, the two integrals,
\begin{equation}\label{app_eq:I1_hyperbola}
      \mathcal{I}_1(\epsilon) = \frac{v_0}{v_D}\frac{\Lambda_{-}+\Lambda_{+}}{2\pi^2 \hbar^2 v_D^2\left(\frac{v_0^2}{v_D^2}-1 \right)^{\frac{1}{2}}} + \frac{\epsilon}{2\pi^2 \hbar^2 v_D^2\left(\frac{v_0^2}{v_D^2}-1 \right)^{\frac{3}{2}}} \ln \frac{\Lambda_{+}}{\Lambda_{-}},
\end{equation}
\begin{equation}\label{app_eq:I2_hyperbola}
      \mathcal{I}_2(\epsilon) = \frac{\Lambda_{-}+\Lambda_{+}}{2\pi^2 \hbar^2 v_D^2\left(\frac{v_0^2}{v_D^2}-1 \right)^{\frac{1}{2}}} + \frac{v_0}{v_D}\frac{\epsilon}{2\pi^2 \hbar^2 v_D^2\left(\frac{v_0^2}{v_D^2}-1 \right)^{\frac{3}{2}}} \ln \frac{\Lambda_{+}}{\Lambda_{-}},
\end{equation}
are analytic functions of $\epsilon$.
%{\color{brown}(?) Exclude Eqs. (A14) - (A16), formulate the result for I(V) in Eq. (A4)}
%

We recall now that the energy $\epsilon \equiv -eV -\mu_T - \xi_{0\lambda}^S$ in Eqs.~(\ref{app_eq:i2})-(\ref{app_eq:I2_hyperbola}) is related to the band offset $\phi = -eV - \mu_T + \mu_S$ between the tip and sample. The Dirac point intersects with the probed band $\lambda$ at $\epsilon=0$. This corresponds to the band offset $\phi$ tuned to $\phi_0 \equiv \xi_{0\lambda}^S + \mu_S$. The combination of Eq.~(\ref{app_eq:i2}) with Eqs.~(\ref{app_eq:I1_hyperbola}) and (\ref{app_eq:I2_hyperbola}) shows that the dependence of $I$ on $\phi-\phi_0$ is analytical if $v_0>v_D$. In contrast, at $v_0<v_D$, \textit{i.e.}, for a ``sharp'' Dirac cone of the tip (Fig.~\ref{fig:intersections}a), the current $I\propto |\phi-\phi_0|$ mirrors the non-analytic behavior of the density of states $\nu(\phi)=\mathcal{I}_1(\epsilon)$, see Eqs.~(\ref{app_eq:i2})-(\ref{app_eq:I2_v0<vd}) and Eq.~\eqref{eq:dos} of the main text.

% \\{\color{brown}(?) out the rest} To summarize, when the band offset $\phi = -eV - \mu_T + \mu_S$ between the tip and sample is tuned to $\phi = \phi_0 \equiv \xi_0^T+ \mu_T - \xi_{0\lambda}^S - \mu_S$ such that $\epsilon =0$ and the Dirac point intersects with the probed band $\lambda$, tunneling current has a $|\phi-\phi_0|$ singularity if and only if the Dirac velocity $v_D$ is not smaller than the group velocity $v_0$ of the probed band $\lambda$ at the intersecting point, $v_0\leq v_D$. Tunneling current has a concise expression
% %
% \begin{gather}\label{app_eq:i_result}
%     I = \frac{2\pi e \Omega}{\hbar}(f(\xi_{0\lambda}^{S})-f(\xi_{0}^{T}))|T|^2 \nu(\phi) + ..., \\
%     |T|^2 = \frac{|a|^2 + |b|^2}{2} + \frac{v_0}{v_D} \left(\frac{a b^{*}}{2} e^{i\theta_0} + c.c. \right). \label{app_eq:T2}
% \end{gather}
% %
% where the joint density of states $\nu(\phi)$ is given by Eqs.~\eqref{app_eq:I1_v0<vd} and~\eqref{app_eq:I1_v0=vd} and $...$ denote contributions to tunneling current outside the reciprocal space region $\mathcal{R}$. Examples of $|T|^2$ will be given in Sec.~\ref{sec:me_bloch} for tunneling between a MLG tip and a graphene-based sample near commensurate twist angles.

\subsection{Effects of quasiparticle broadening and temperature on the Dirac-point singularity}\label{sec:broadening}
In this section, we explain how nonzero quasiparticle relaxation rates and/or temperature smear the tunneling current singularities induced by Dirac points. Our main conclusion is that the singularity $d^2I/dV^2\propto\delta(\phi-\phi_0)$ is smeared by the broadening of quasiparticle spectra but not by temperature. Instead, finite temperature $T$ entering the Fermi distribution functions in Eq.~(\ref{app_eq:i}) reduces the contrast of the $d^2I/dV^2$ vs. $V$ pattern.

% The main conclusions are summarized below:
% \begin{enumerate}
%     \item $v_0<v_D$: The Dirac-$\delta$ singularity at $\epsilon=0$ can be smeared by finite quasiparticle relaxation rates $\gamma_d/\hbar$ and $\gamma_0/\hbar$ of the Dirac point and the probed Bloch state, respectively, but are insensitive to finite temperature $T$ if the broadening energy scale is much less than either $k_BT$ or the Fermi energies $\mu_{0}$ and $\mu_d$.
%     \item $v_0>v_D$: Only when the Dirac cone is undoped, $d^2I/dV^2$ exhibits $1/\epsilon$ singularity. This singularity can be smeared by temperature and quasiparticle broadening, $\text{max}|d^2I/dV^2| \sim  1/\text{max}\{\gamma_0, \frac{v_0}{v_D}\gamma_d, \frac{v_0}{v_D}k_BT\}$. A slower Dirac velocity leads to stronger smearing by the temperature and finite relaxation rate of the Dirac cone.
% \end{enumerate}
%

The expression for the tunneling current Eq.~\eqref{eq:i_basic} can be rewritten as follows
\begin{equation}\label{app_eq:i_dirac}
    I = \frac{2 \pi e}{\hbar}\int_{-\infty}^{\infty}d\omega \left(f(\omega -eV-\mu_T)-f(\omega - \mu_T)\right) \sum_{\bm p,\lambda'} |T_{\lambda' \lambda}(\bm p)|^2 A_{\lambda}^{0}(\bm p,\omega-eV-\mu_T)A_{\lambda'}^{d}(\bm p, \omega-\mu_T).
\end{equation}
For convenience, $\omega$ is the energy measured relative to the Dirac point. (Note that this is different from the definition used in the main text.) For Lorentzian spectral functions $A_{\lambda}^{S}(\bm p, \omega) = \frac{1}{\pi}\frac{\gamma_S}{(\omega - \xi_{\bm p\lambda}^S)^2 + \gamma_S^2} $ and $A_{\lambda'}^{T}(\bm p, \omega) = \frac{1}{\pi}\frac{\gamma_T}{(\omega - \xi_{\bm p\lambda'}^T)^2 + \gamma_T^2} $, the momentum summation simplifies to
\begin{subequations}\label{app_eq:P}
\begin{align}
    &\frac{1}{\Omega}\sum_{\bm p,\lambda'} |T_{\lambda' \lambda}(\bm p)|^2 A_{\lambda}^{S}(\bm p,\omega-eV-\mu_T)A_{\lambda'}^{T}(\bm p, \omega-\mu_T) \\
    =&\sum_{\lambda'}\iint \frac{d\bm p}{(2\pi)^2} \frac{|a|^2+|b|^2 + \lambda'(a b^{*} e^{i\theta_0}+c.c.) \cos\theta_{\bm p}}{2\pi^2} \frac{\gamma_S}{(\omega  + \epsilon- \hbar v_0 p_x)^2 + \gamma_S^2}\frac{\gamma_T}{(\omega -\lambda'\hbar v_D |\bm p|)^2 + \gamma_T^2} \notag\\
    %
    %=&  \frac{1}{\pi^2}\text{Re}\iint_{-\infty}^{\infty} \frac{dp dp_x}{(2\pi)^2}  \frac{(|a|^2+|b|^2) p + (a b^* e^{i\theta_0}+c.c.) p_x}{\sqrt{(p+i0^+)^2 - p_x^2}} \frac{\gamma_0}{(\omega + \epsilon-\hbar v_0 p_x)^2 + \gamma_0^2} \frac{\gamma_d}{(\omega - \hbar v_D p)^2 + \gamma_d^2}\notag\\
    %
    =& \frac{1}{4\pi^2\hbar^2v_D}\text{Re}\int_{-\infty}^{\infty} \frac{dp_x}{2\pi}\frac{(|a|^2+|b|^2) \frac{\omega + i\gamma_T}{\hbar v_D} + (a b^* e^{i\theta_0}+c.c.) p_x }{\sqrt{\left(\frac{\omega + i\gamma_T}{\hbar v_D}\right)^2 - p_x^2}} \frac{\gamma_S}{(\omega + \epsilon-\hbar v_0 p_x)^2 + \gamma_S^2}. \label{app_eq:P2}
\end{align}
\end{subequations}
Here $\epsilon = -eV - \mu_T - \xi_{0\lambda}$, and parameters $a$ and $b$ are defined in Eq.~(\ref{app_eq:T_general}). We chose the branch cut of $\sqrt{z}$ to be $[0, +\infty)$.
Plugging Eqs.~\eqref{app_eq:P} into Eq.~\eqref{app_eq:i_dirac} and defining $\xi= \hbar v_0 p_x $ yield
\begin{align}
    &I =  \frac{2\pi e\Omega}{\hbar} \left(\frac{|a|^2 + |b|^2}{2} \tilde{\mathcal{I}}_1(\epsilon) +  \frac{ab^{*}}{2}e^{i\theta_0}\tilde{\mathcal{I}}_2(\epsilon) +  \frac{a^{*}b}{2}e^{-i\theta_0}\tilde{\mathcal{I}}_2(\epsilon)^{*} \right), \label{app_eq:i_I1I2}\\
    &\tilde{\mathcal{I}}_1(\epsilon) =  \frac{1}{2\pi^2 \hbar^2 v_0 v_D}\text{Re}\iint_{-\infty}^{\infty}d\omega d\xi \left(f(\omega -eV-\mu_T)-f(\omega - \mu_T)\right) \frac{\alpha(\omega + i\gamma_T)}{\sqrt{\alpha^2\left(\omega + i\gamma_T\right)^2 - \xi^2}}\frac{\gamma_S}{(\omega +\epsilon-\xi)^2 + \gamma_S^2}, \label{app_eq:tildeI_1}\\
    &\tilde{\mathcal{I}}_2(\epsilon) =  \frac{1}{2\pi^2 \hbar^2 v_0 v_D}\text{Re}\iint_{-\infty}^{\infty}d\omega d\xi \left(f(\omega -eV-\mu_T)-f(\omega - \mu_T)\right) \frac{\xi}{\sqrt{\alpha^2\left(\omega + i\gamma_T\right)^2 - \xi^2}}\frac{\gamma_S}{(\omega +\epsilon-\xi)^2 + \gamma_S^2}. \label{app_eq:tildeI_2}
\end{align}
%
%\subsection{Smearing of Dirac-point singularity}
%
For $v_0< v_D$ (\textit{i.e.}, $\alpha\equiv v_0/v_D < 1$), we consider the limit where the energy of the Dirac point nearly aligns with the band of the sample at the same wave vector,
\begin{equation}\label{app_eq:criterion}
   \frac{|\epsilon|}{|1-\alpha|} \ll \Lambda < \text{max}\{|\mu_T|, k_BT\}, \text{max}\{|eV+\mu_T|, k_BT\}  .
\end{equation}
Notice that the Lorentzian spectral peak in Eqs.~\eqref{app_eq:tildeI_1} and~\eqref{app_eq:tildeI_2} is concentrated in the region $\xi\sim \omega + \epsilon$. In this region, the real part of the integrand becomes small unless $\alpha^2\omega^2\gtrsim (\omega+\epsilon)^2$ (\textit{i.e.}, $|\omega| \lesssim |\epsilon/(1-\alpha)|$). Thus, the condition \eqref{app_eq:criterion} implies that the Fermi functions are approximately $\omega-$independent in the energy window that gives the dominant contribution to the tunneling current. We can pull the Fermi functions outside the integral and then integrate $\omega$ first along the contour that closes the upper half plane of $\omega$. After some algebra, we arrive at
\noindent
\begin{align}
    \tilde{\mathcal{I}}_1(\epsilon) &\approx  \frac{1}{2\pi^2 \hbar^2 v_D^2}\left(f(- eV -\mu_T)-f(-\mu_T)\right)\text{Re}\int_{-\Lambda}^{\Lambda}d\xi \frac{\xi -\epsilon + i\gamma}{\sqrt{\frac{\alpha^2}{1-\alpha^2}\left(\epsilon - i\gamma\right)^2 - (1-\alpha^2)\left[\xi + \frac{\alpha^2(\epsilon - i\gamma)}{1-\alpha^2}\right]^2}} \notag\\
    & = \frac{1}{2\pi^2 \hbar^2 v_D^2 (1-\alpha^2)^{\frac{3}{2}}} \left(f( - eV -\mu_T)-f(-\mu_T)\right)  \text{Re}\left\{ i\left(\epsilon - i\gamma\right)\ln\left(\frac{1-\alpha^2}{\alpha}\frac{2i\Lambda}{\epsilon - i\gamma}\right)^2 \right\},
\end{align}
where $\gamma\equiv\gamma_S+\gamma_T$ and the branch cut of $\ln z$ is $z\in (-\infty,0]$.
It is straightforward to show that $\tilde{\mathcal{I}}_2(\epsilon) =\alpha\tilde{\mathcal{I}}_1(\epsilon)$. Thus, with the approximation $d^2\epsilon/dV^2= d^2(\mu_S-\mu_T)/dV^2 \approx 0$ (e.g., when $\mu_T$ and $\mu_S$ are independent of bias), we find
\begin{align}
    \frac{d^2I}{dV^2} = \frac{2\pi e\Omega}{\hbar}\left(\frac{d\epsilon}{dV}\right)^2\left(f( - eV -\mu_T)-f(-\mu_T)\right)|T|^2 \frac{1}{\pi^2 \hbar v_D^2 (1-\frac{v_0^2}{v_D^2})^{\frac{3}{2}}}\frac{\gamma_S + \gamma_T}{\epsilon^2 + (\gamma_S + \gamma_T)^2}.
    \label{eq:broadening_effect}
\end{align}
Here $|T|$ can be interpreted as the  absolute value of the tunneling matrix elements at the Dirac point. Its expression will be specified in App.~\ref{sec:me_bloch} but is unimportant for the discussion here. The main implication of the above equation is that the broadening of the spectral peaks of the Dirac cone and the probed band can both lead to broadening of the $d^2I/dV^2$ singularity. This is reflected in the last factor in Eq.~(\ref{eq:broadening_effect}). Finite temperature can change the intensity of $d^2I/dV^2$ via the Fermi functions but does not directly broaden the singularity.  
% \begin{figure}
%     \centering
%     \includegraphics[width=0.9\linewidth]{d^2I_dV2_two_dirac_cones.pdf}
%     \caption{Enter Caption}
%     \label{fig:enter-label}
% \end{figure}

\section{Fermi-edge singularity}\label{app_sec:fs}

In this section, we analyze the tunneling-current singularity induced by the discontinuity in the  electron occupation  across the Fermi level of the tip or sample. We first show that the zero-temperature differential tunneling conductance exhibits inverse square root divergences at characteristic bias voltages $V^* (\neq 0)$, where the tip's Fermi line and the sample band structure are tangent at a wave vector $\bm p^*$,
\begin{equation}\label{app_eq:fs_tangency}
 \xi_{\bm p^*}^{T} = \xi_{\bm p^*}^{S} + eV^* = 0, \quad \bm v_{\bm p^*}^{T} \parallel  \bm v_{\bm p^*}^{S}. 
\end{equation}
Here, we hide the band indices to shorten the notation. Let us denote $v_{\bm p}^{T} = |\bm v_{\bm p}^{T}|$ and $v_{\bm p}^{S} = \bm v_{\bm p}^{S}\cdot\bm v_{\bm p}^{T}/|\bm v_{\bm p}^{T}|$, and expand the band structures of the tip and sample around $\bm p^*$,
\begin{equation}\label{app_eq:xi_linearized}
    \xi_{\bm p^{*}+\bm q}^{T/S} \approx \xi_{\bm p^*}^{T/S} + \hbar v_{\bm p^*}^{T/S}q_{\perp} + \frac{\hbar^2 q_{\scriptscriptstyle{\parallel}}^2}{2m_{\bm p^*}^{T/S}},
\end{equation}
where $q_{\perp}$ ($q_{\scriptscriptstyle{\parallel}}$) is the component of the wave vector $\bm q$ along (transverse to) the direction of $\bm v_{\bm p^{*}}^{T}$. For $V\sim V^*$, 
\begin{align}
    \frac{dI}{dV} & = \frac{2\pi e \Omega}{\hbar}\sum_{\bm q} |T(\bm p^*+\bm q)|^2 (f(\xi_{\bm p^*+\bm q}^{S}) - f(\xi_{\bm p^*+\bm q}^T)) \frac{d}{dV}\delta (eV + \xi_{\bm p^*+\bm q}^{S}- \xi_{\bm p^*+\bm q}^{T}) \notag\\
    & \approx \frac{2\pi e^2 \Omega}{\hbar^2}\int \frac{dq_{\perp}d q_{\scriptscriptstyle{\parallel}}}{(2\pi)^2} |T(\bm p^*+\bm q)|^2 (f(\xi_{\bm p^*+\bm q}^{S}) - f(\xi_{\bm p^*+\bm q}^T)) \frac{1}{v_{\bm p^*+\bm q}^S-v_{\bm p^*+\bm q}^T}\frac{\partial}{\partial q_{\perp}}\delta (eV + \xi_{\bm p^*+\bm q}^{S}- \xi_{\bm p^*+\bm q}^{T}) \notag \\
    &\approx  \frac{2\pi e^2 \Omega}{\hbar}|T(\bm p^*)|^2 \frac{v_{\bm p^*}^T}{v_{\bm p^*}^T - v_{\bm p^*}^S}\int \frac{dq_{\perp}dq_{\scriptscriptstyle\parallel}}{(2\pi)^2} \delta(\xi_{\bm p^*+\bm q}^T)\delta (eV + \xi_{\bm p^*+\bm q}^{S}). \label{app_eq:dI_dV_fs}
\end{align}
To arrive at the last line, we neglect the $\bm q-$dependences of the tunneling matrix elements and the group velocity, and use integration by parts. Note that $\partial_{q_{\perp}}f(\xi_{\bm p^* + \bm q}^{S})\approx 0$ for $V^*\neq 0$. The two Dirac $\delta$-functions in Eq.~\eqref{app_eq:dI_dV_fs} restrict the wave vector $\bm p^*+\bm q$ to the intersections of the tip Fermi lines and the sample band. Plugging Eq.~\eqref{app_eq:xi_linearized} into Eq.~\eqref{app_eq:dI_dV_fs}, we arrive at
\begin{align}
    \frac{dI}{dV} & =  \frac{e^2 \Omega}{\hbar^2}|T(\bm p^*)|^2 \frac{1}{v_{\bm p^*}^T - v_{\bm p^*}^S} \int\frac{d q_{\scriptscriptstyle\parallel}}{2\pi}\delta \left(-e(V-V^*) + \Big(\frac{v_{\bm p^*}^S}{2m_{\bm p^*}^{T}v_{\bm p^*}^T}-\frac{1}{2m_{\bm p^*}^S}\Big)\hbar^2q_{\scriptscriptstyle\parallel}^2\right) \notag\\
    & = \frac{e^2 \Omega}{\pi \hbar^3}|T(\bm p^*)|^2  \frac{1}{v_{\bm p^*}^T - v_{\bm p^*}^S}\frac{1}{\sqrt{2e\Big(\frac{v_{\bm p^*}^S}{m_{\bm p^*}^{T}v_{\bm p^*}^T}-\frac{1}{m_{\bm p^*}^S}\Big)(V-V^*)}}\Theta\left(\Big(\frac{v_{\bm p^*}^S}{m_{\bm p^*}^{T}v_{\bm p^*}^T}-\frac{1}{m_{\bm p^*}^S}\Big)(V-V^*)\right). \label{app_eq:dI1_dV}
\end{align}

Similarly, $dI/dV$ exhibits an inverse square root singularity at a characteristic bias voltage $V^{**}$ where the sample Fermi line and tip band are tangent at a wave vector $\bm p^{**}$,
\begin{equation}\label{app_eq:fs_tangency2}
 \xi_{\bm p^{**}}^{S} = \xi_{\bm p^{**}}^{T} - eV^{**} = 0,  \quad \bm v_{\bm p^{**}}^{T} \parallel  \bm v_{\bm p^{**}}^{S}. 
\end{equation}
We denote $v_{\bm p^{**}}^{S} = |\bm v_{\bm p^{**}}^{S}|$ and $v_{\bm p^{**}}^{T} = \bm v_{\bm p^{**}}^{T}\cdot\bm v_{\bm p^{**}}^{S}/|\bm v_{\bm p^{**}}^{S}|$. For $V\sim V^{**}$, we can derive an equation analogous to Eq.~\eqref{app_eq:dI_dV_fs},
\begin{align}
    \frac{dI}{dV} &\approx  \frac{2\pi e^2 \Omega}{\hbar}|T(\bm p^{**})|^2 \frac{v_{\bm p^{**}}^S}{v_{\bm p^{**}}^S - v_{\bm p^{**}}^T}\int \frac{dq_{\perp}dq_{\scriptscriptstyle\parallel}}{(2\pi)^2} \delta(\xi_{\bm p^{**}+\bm q}^T-eV)\delta (\xi_{\bm p^{**}+\bm q}^{S}) \notag\\
    & =  \frac{e^2 \Omega}{\hbar^2}|T(\bm p^{**})|^2\frac{v_{\bm p^{**}}^S}{|v_{\bm p^{**}}^T|}\frac{1}{v_{\bm p^{**}}^S - v_{\bm p^{**}}^T}\int\frac{d q_{\scriptscriptstyle\parallel}}{2\pi}\delta \left(-e\frac{v_{\bm p^{**}}^S}{v_{\bm p^{**}}^T}(V-V^{**}) + \Big(\frac{v_{\bm p^{**}}^S}{2m_{\bm p^{**}}^{T}v_{\bm p^{**}}^T}-\frac{1}{2m_{\bm p^{**}}^S}\Big)q_{\scriptscriptstyle\parallel}^2\right) \notag\\
    & = \frac{e^2 \Omega}{\pi \hbar^3}|T(\bm p^{**})|^2\frac{1}{v_{\bm p^{**}}^S - v_{\bm p^{**}}^T}\frac{1}{\sqrt{2e\Big(\frac{1}{m_{\bm p^{**}}^{T}}-\frac{v_{\bm p^{**}}^T}{m_{\bm p^{**}}^Sv_{\bm p^{**}}^S}\Big)(V-V^{**})}}\Theta\left(\Big(\frac{1}{m_{\bm p^{**}}^{T}}-\frac{v_{\bm p^{**}}^T}{m_{\bm p^{**}}^Sv_{\bm p^{**}}^S}\Big)(V-V^{**})\right). \label{app_eq:dI2_dV}
\end{align}
This result is consistent with Eq.~\eqref{app_eq:dI1_dV} if we make the substitutions $T\leftrightarrow S, V \rightarrow -V, V^{**} \rightarrow -V^{*}$.

To connect this general formalism with the main text, we consider scanning of a flat band by a MLG tip at a relatively low doping. In this case, $v_{\bm p^*}^{T}=v_D\gg |v_{\bm p^*}^S|$ and $m_{\bm p^*}^T = \lambda' \hbar |\bm p^*|/v_D = \mu_T/v_D^2$, where $\lambda'=1(-1)$ if the Fermi level is in the conduction (valence) band of MLG. Under the assumption $|\mu_T|\ll (v_D/|v_{\bm p^*}^S|)W$ (W is the bandwidth of the flat band), % (which is valid at least in the limit of small $|\mu_T|$), 
Eq.~\eqref{app_eq:dI1_dV} becomes
%
%\begin{equation}\label{app_eq:dI_dV_dirac_old}
%    \frac{dI}{dV}=\frac{\Omega e^2}{h} \frac{|T(\bm p^*)|^2 }{\hbar^2 v_D^2}\sqrt{2\frac{v_D}{|\bm v_{\bm p^*}^S|}\frac{|\mu_T|}{e|V-V^*|}}\Theta\left(\frac{v_{\bm p^*}^S}{\mu_T}(V-V^*)\right).
%\end{equation}
%
\begin{equation}\label{app_eq:dI_dV_dirac}
    \frac{dI}{dV}=\frac{\Omega e^2}{h} \frac{|T(\bm p^*)|^2}{\hbar^2 v_D^2}
\sqrt{2\frac{v_D|\mu_T|}{|\bm v_{\bm p^*}^S|}}{\rm Re}\left\{\frac{1}{\pm e(V-V^*)-i\gamma_S}\right\}^{1/2}.
%\Theta\left(\frac{v_{\bm p^*}^S}{\mu_T}(V-V^*)\right).
\end{equation}
The $\pm$ sign here is defined by the sign of $\mu_T/v_{\bm p^*}^S$. We also included the effect of spectral broadening $\gamma_S$ in the sample and evaluated $dI/dV$ in the limit $|\mu_T|\gg (v_D/|v_{\bm p^*}^{S}|)\gamma_S$. This limit is compatible with the small-$\mu_T$ assumption, as long as $W\gg\gamma_S$. Equation~(\ref{app_eq:dI_dV_dirac}) at $\gamma_S=0$ yields Eq.~\eqref{eq:d2i_dv2_fs} in the main text.

The broadening of the spectrum $\xi_{\bm p}^S$ introduces a momentum uncertainty $\Delta p\sim \gamma_S/|v_{\bm p^*}^{S}|$. Once the uncertainty exceeds the Fermi momentum in the tip, the expansion of the dispersion relation $\xi_{\bm p}^T$, see Eq.~\eqref{app_eq:xi_linearized}, becomes inapplicable. This provided the limit of applicability of Eq.~(\ref{app_eq:dI_dV_dirac}) quoted above, 
$|\mu_T|\gg (v_D/|v_{\bm p^*}^{S}|)\gamma_S$. In the opposite case, $|\mu_T|\lesssim (v_D/|v_{\bm p^*}^{S}|)\gamma_S$, we can simplify the general result as
%Let us comment on the effects of quasiparticle broadening. For simplicity, we consider the case where quasiparticles in the tip have infinite lifetime and those in the sample have a Lorentzian spectral peak with broadening parameter $\gamma_S$. Interestingly, quasiparticle broadening in the sample smears the singularities induced by the tip's and the sample's Fermi edges to different extents: 
%

%(i) Singularities induced by the tip's Fermi edge: The divergence of Eqs.~\eqref{app_eq:dI1_dV} and~\eqref{app_eq:dI_dV_dirac} is cutoff by the broadening at $|V-V^{*}| \sim \gamma_S$. Because the energy broadening introduces a momentum uncertainty $\Delta p\sim \gamma_S/|v_{\bm p^*}^{S}|$, the inverse square root behavior for $V\sim V^{*}$ relies on an implicit assumption that the linearized dispersion Eq.~\eqref{app_eq:xi_linearized} remains valid for $|\bm p-\bm p^{*}|\sim \gamma_S/|v_{\bm p^*}^{S}|$, which becomes unjustified for a small Fermi momentum of the tip. For instance, when a MLG tip at a small chemical potential $|\mu_T|\lesssim v_D\gamma_S/|v_{\bm p_n}^S|$ ($\bm p_n$ labels the tip's Dirac point) scans the flat bands in the sample, 
%
\begin{align}
    \frac{dI}{dV} &\approx -\frac{2e^2\Omega}{\hbar} \int_{-\infty}^{\infty} d\xi \sum_{\bm p}  |T(\bm p)|^2 (f(\xi-eV) - f(\xi)) \delta(\xi-\xi_{\bm p}^T) \frac{\partial}{\partial \xi} \frac{\gamma_S}{(eV+\xi_{\bm p}^S - \xi)^2 + \gamma_S^2} \notag\\
    &\approx -\frac{2 e^2\Omega}{\hbar} |T|^2 \int_{-\infty}^{\infty} d\xi\ \nu_{T}(\xi+\mu_T)  (f(\xi-eV) - f(\xi)) \frac{\partial}{\partial \xi} \frac{\gamma_S}{(eV+\xi_{\bm p_n}^S - \xi)^2 + \gamma_S^2} \notag\\
    &\approx \frac{2 e^2\Omega}{\hbar} |T|^2 \nu_{T}(\mu_T)  \frac{\gamma_S}{(eV+\xi_{\bm p_n}^S)^2 + \gamma_S^2} +... . \label{app_eq:dIdV_small_fs}
\end{align}
Here $\nu_T(\epsilon) = |\epsilon|/2\pi\hbar^2 v_D^2$ is the density of states of the MLG tip, and $|T|^2\approx |T|_n^2$ (see Eq.~\eqref{eq:t_averaged}) as $\bm p^{*}$ approaches the Dirac point $\bm p_n$ of the tip and the Bloch wave functions of the flat band become approximately the same at these two points. To arrive at the last line, we used integration by parts and the relation $\partial_\xi f(\xi)=-\delta(\xi)$ valid at $T=0$. We also neglected $\partial_{\xi} \nu_T(\xi+\mu_T)$ which includes the singularity associated with the Dirac point.

Equation~\eqref{app_eq:dIdV_small_fs} conveys two messages. First, the same information about the energy dispersion and tunneling matrix elements of the sample's flat bands can be obtained from the Fermi-edge singularity in $dI/dV$ and from the Dirac-point singularity in $d^2I/dV^2$, cf.\ Eq.\eqref{eq:d2IdV2_dp}. The former requires $0<|\mu_T|\lesssim \gamma_S v_D/|v_{\bm p^*}^{S}|$, while the latter requires scanning with both signs of $\mu_T$. Second, we can compare the intensity of these two singularities by computing the ratio of the maximal $dI/dV$ contributed by the Fermi-edge singularity and the jump of $dI/dV$ associated with the Dirac-point singularity, cf.\ Eqs.~\eqref{eq:i_approximate} and~\eqref{eq:dos}. We find that this ratio equals $\mu_T/2\pi\gamma_S$ if $\mu_{T,S}$ are independent of $V$.

Next, we briefly address the singularities induced by the sample's Fermi edge scanning the Dirac cones of the tip. Using the linearized dispersion Eq.~\eqref{app_eq:xi_linearized} and the Lorentzian spectral function in the sample, we derive
\begin{equation}\label{app_eq:dIdV_fs_S}
    \frac{d I}{dV} \propto \text{Re}\left\{\Big[\pm e(V-V^{**}) - i \frac{v_{\bm p^{**}}^T}{v_{\bm p^{**}}^S}\gamma_S\Big]^{-\frac{1}{2}}\right\},\qquad (V\sim V^{**}).
\end{equation}
We provide an intuitive explanation for the additional factor $v_{\bm p^{**}}^T/v_{\bm p^{**}}^S$ in the broadening parameter in Eq.~\eqref{app_eq:dIdV_fs_S}. The energy uncertainty $\sim \gamma_S$ of the electrons at the Fermi level of the sample is associated with a momentum uncertainty $\Delta p \sim \gamma_S/ |v_{\bm p^{**}}^S|$. Because of momentum conservation, this momentum uncertainty translates to an energy uncertainty $\Delta\epsilon \sim |v_{\bm p^{**}}^T|\Delta p \sim \gamma_S |v_{\bm p^{**}}^T/v_{\bm p^{**}}^S|$ in the tip (which is being probed in this case). A finite temperature can also introduce an energy uncertainty $\sim k_BT$ and this ``lever-arm effect" implies that the temperature smearing of $dI/dV$ singularity induced by the Fermi edge in the flat bands of the sample is enhanced by a factor of $|v_{\bm p^{**}}^T/v_{\bm p^{**}}^S|\gg 1$.

The above analysis of singularities associated with the Fermi edge assumes that the Fermi energies in the tip and sample are away from the respective Dirac points. The type of singularities change if one of the Fermi levels coincide with the Dirac point.
When the Fermi level of the tip is tuned to the Dirac point $\bm p^{*}$ and $v_{\bm p^{*}}^{T} > v_{\bm p^{*}}^{S} $, the Fermi-edge and Dirac-point singularities merge into one feature in $dI/dV\propto \Theta(\pm (V-V^{*}))$ at zero temperature, which is qualitatively the same as the Dirac-point singularity. When the Fermi level is positioned at a Dirac point $\bm p^{**}$ of the sample's flat bands and the band intersection approaches a hyperbola near $\bm p^{**}$, only one branch of the hyperbola contributes to the tunneling current at zero temperature and $dI/dV\sim \ln|V-V^{**}|$. The hyperbolic type of intersection is enforced by the condition $v_{\bm p^{**}}^S < v_{\bm p^{**}}^T$. However, unlike the case depicted in Fig.~\ref{fig:intersections} and considered in (iii) part of App.~\ref{sec:zerotemperature}, only {\sl one} branch of the hyperbola contributes to the current.
%where smooth Fermi functions are implicitly assumed at the Dirac point. 
This singularity can be significantly broadened by a finite quasiparticle relaxation rate and finite temperature, as discussed in the preceding paragraph. This is consistent with our numerical results in Figs.~\ref{fig:theta_c_0_N}b-c and Figs.~\ref{fig:theta_c_38.2_N}b-c, which  show that the Dirac point of the flat bands does not scan the tip. 

\section{Interlayer tunneling Hamiltonian between twisted graphene bilayers}\label{sec:H_tun_derivation}

\subsection{Symmetry analysis of the continuum Hamiltonian}\label{app_sec:symmetry}

% We have so far assumed rigidly twisted graphene lattices with only $p_z$ orbitals. The lattice relaxation and corrugation, however, becomes non-negligible when two graphene lattices are in contact and are twisted by a small but finite angle, which then modify the tunneling Hamiltonian. Nevertheless, the general expressions for $H_{\text{tun}}$ can be constrained by symmetries of TBG at commensurate angles. Consider two graphene lattices rotated about the out-of-plane axis by $\pm\theta/2$, respectively, with zero displacement $\bm d=0$. The system enjoys $C_{3z}, C_{2x}$, $C_{2z}\mathcal{T}$, and moir\'e translation symmetries.

Let us consider two graphene layers $l$ and $l'$ rotated clockwise by $\pm \theta/2$ from the AA-aligned configuration, where $\theta$ is close to a commensurate angle $\theta_c$. An additional lateral shift $\bm d$ between the two layers 
%in this section 
will be analyzed in Appendices~\ref{app_sec:tightbinding} and~\ref{app_sec:d}.

The two layers have $K$ valley Dirac points at $\bm {K_l}= O(-\frac{\theta}{2}) (\frac{4\pi}{3a_0},0)^{T}$ and $\bm {K_{l'}}= O(\frac{\theta}{2}) (\frac{4\pi}{3a_0},0)^{T}$, with $a_0$ denoting the graphene lattice constant. The three shortest Dirac wave vectors in the reciprocal lattice of layer $l (l')$ which overlap Dirac points in the other layer at $\theta=\theta_c$ are $\bm {K_l} + \bm G_{1,2,3} (\bm {K_{l'}} + \bm G_{1,2,3}')$, where $\bm G_{n}(\bm G_{n}')$ are reciprocal lattice vectors of layer $l(l')$. 
One can define continuum Dirac fields $\Psi_{\tau l/l'}^{\dagger}=(\Psi_{\tau l/l'A}^{\dagger}, \Psi_{\tau l/l'B}^{\dagger})$ in layer $l/l'$ and express the continuum Hamiltonian $H_0^{\tau}$ of the individual graphene layers $l$ and $l'$ in the $\bm k\cdot \bm p$ approximation relative to a common reference wave vector $\tau(\bm {K_l}+\bm G_1)$. In the $K$ valley ($\tau=+$),
\begin{align}\label{app_eq:H0}
    H_0^{\tau=+} = \int d^2r \Psi_{+l}^{\dagger}(\bm r) (-i\bm \sigma\cdot\bm \nabla) \Psi_{+l}(\bm r) + \Psi_{+l'}^{\dagger}(\bm r) (-i\bm \sigma\cdot\bm \nabla - \delta \bm K) \Psi_{+l'}(\bm r).
\end{align}
Note that the Dirac point $\bm {K_{l'}} + \bm G_1'$ of the layer $l'$ differs from the reference point $\bm {K_l}+\bm G_1$ by $\delta\bm K = \bm {K_{l'}} + \bm G_1' - \bm {K_l} -\bm G_1$ with $|\delta\bm K|\ll |\bm K|$. Therefore, the continuum fields in the two layers transform differently under $C_{3z}$ rotations,
\begin{subequations}\label{app_eq:c3}
    \begin{equation}
        \Psi_{+l}(\bm r) \rightarrow  e^{-i\frac{2\pi n}{3}\sigma^z}\Psi_{+l} \left(O(\frac{2\pi n}{3})\bm r \right),
    \end{equation}
    \begin{equation}
        \Psi_{+l'}(\bm r) \rightarrow e^{-i\frac{2\pi n}{3}\sigma^z} e^{i\delta\bm K\cdot \bm r-i\delta\bm K\cdot O(\frac{2\pi n}{3})\bm r} \Psi_{+l'} \left(O (\frac{2\pi n}{3})\bm r \right).
    \end{equation}
\end{subequations}
As long as the deviation from the commensurate configuration is small, different valleys in the two layers remain distant in reciprocal space \cite{scheer2022magic}. Therefore, momentum conservation suppresses intervalley tunneling and ensures valley conservation. Equations~\eqref{app_eq:c3} require that the local interlayer tunneling consist of at least three Fourier components per valley to respect $C_{3z}$ symmetry,
\begin{equation}\label{app_eq:H_tun_continuum}
    H_{\text{tun}}^{\tau=+} = \int d^2r \sum_{n=1}^3 \Psi_{+l'}^{T\dagger}(\bm r) \hat{T}_{n} e^{-i\delta\bm G_n\cdot\bm r} \Psi_{+l}^{S}(\bm r) + h.c..
\end{equation}
with the wave vectors of the tunneling amplitude $\bm G_1=0$ and $\delta \bm G_{n+1} \equiv O(\frac{2\pi n}{3})^{-1}\delta\bm K - \delta\bm K $.
%
% \begin{equation}
%     \delta \bm G_{n+1} \equiv O(\frac{2\pi n}{3})^{-1}\delta\bm K - \delta\bm K = \bm G_{n+1}'-\bm G_1' - (\bm G_{n+1} - \bm G_1), (n=1,2).
% \end{equation}
%
These relations of $\bm G_n$ yield Eq.~\eqref{eq:DeltaG} in the main text. 
The three tunneling matrices $\hat{T}_{n}$'s are related by
\begin{equation}\label{eq:c3_z}
    C_{3z}:\ e^{i\frac{2\pi}{3}\sigma^z}\hat{T}_{n}e^{-i\frac{2\pi}{3}\sigma^z}=\hat{T}_{n+1}.
\end{equation}
The $\hat{T}_{1}$ is further constrained by
\begin{align}
    &C_{2x}:\ \sigma^x\hat{T}_{1}^{\dagger}\sigma^x=\hat{T}_{1},\label{eq:c2_x}\\
    &C_{2z}\mathcal{T}:\ \sigma^{x}\hat{T}_{1}^{*}\sigma^{x}=\hat{T}_{1}.
\end{align}
The hermitian conjugation in Eq.~\eqref{eq:c2_x} arises because the $\pi$ rotation flips two layers. The most general expression for $\hat{T}_1$ is $ \hat{T}_1 = w_0 e^{i\chi\sigma^z}+w_1\sigma^x$. Together with Eq.~\eqref{eq:c3_z}, we obtain
\begin{equation}
    \hat{T}_{n} = w_0 e^{i\chi\sigma^z}+w_1\left[\cos\frac{2\pi (n-1)}{3}\sigma^{x}+\sin\frac{2\pi (n-1)}{3}\sigma^{y}\right],\ n=1,2,3.
\end{equation}
At $\theta_c=0^{\circ}$, two AA-aligned layers have an additional mirror symmetry with respect to the $xy$-plane, which implies that $\hat{T}_{1}=\sigma^x\hat{T}_{1}\sigma^x$ and therefore $\chi = 0$.  
The $-K$ valley Hamiltonian can be obtained by time-reversal symmetry. By defining $\hat{T}_{-n}= \hat{T}_{n}^{*}$, Eq.~\eqref{app_eq:H_tun_continuum} is generalized to
\begin{equation}
    H_{\text{tun}}^{\tau} = \int d^2r \sum_{n=1}^3 \Psi_{\tau l'}^{T\dagger}(\bm r) \hat{T}_{\tau n} e^{-i\tau\delta\bm G_n\cdot\bm r} \Psi_{\tau l}^{S}(\bm r) + h.c..
\end{equation}

\subsection{Tight-binding description of tunneling Hamiltonian between two rigidly twisted graphene layers}\label{app_sec:tightbinding}
\begin{figure}
    \centering
    \includegraphics[width=1\linewidth]{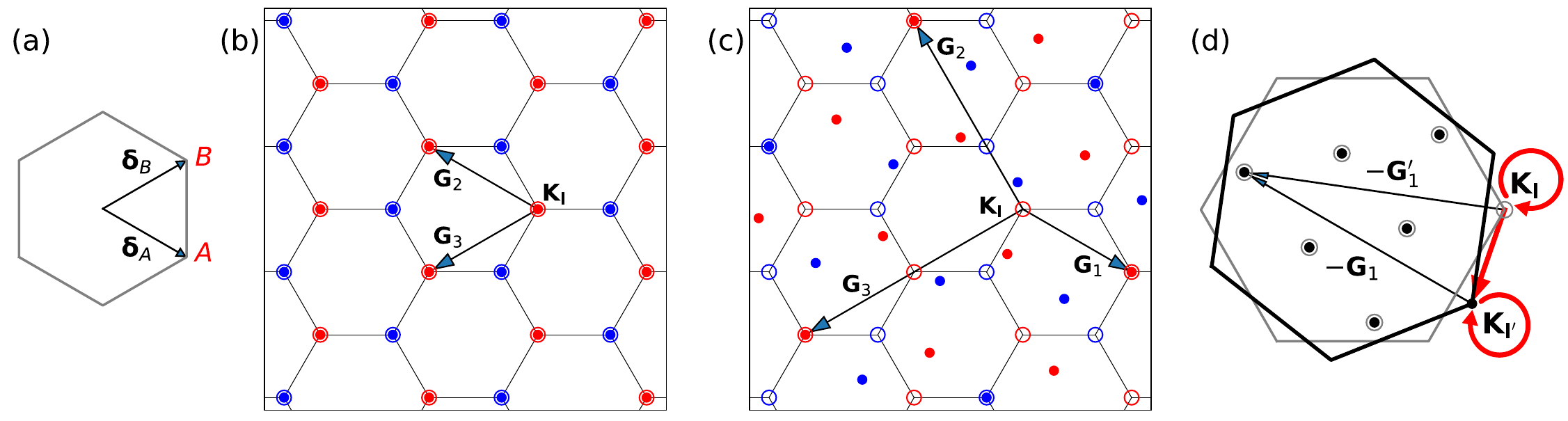}
    \caption{(a) A unit cell of graphene with two sublattices, A and B. (b) The reciprocal lattices of the layer $l$ (dots) and $l'$ (circles) without twist. $\pm K$ valleys are colored in red and blue, respectively. For the three shortest overlapping Dirac wave vectors in the $K$ valley, $\bm {K_l} + \bm{G}_{1,2,3}$, $\bm G_1=0$ and $\bm G_{2,3}$ are indicated by arrows. (c) The layer $l'$ is rotated clockwise by $\theta_c\approx38.2^{\circ}$. (d) The dots and circles mark the seven wave vectors in the first Brillouin zone of the graphene layer $l$ (gray) and $l'$ (black), respectively, which correspond to the same crystal momentum of the moir\'e superlattice at the commensurate angle $\theta_c\approx 38.2^{\circ}$. The three red arrows constitute a three-step tunneling process, with the intermediate one enabled by umklapp scattering due to $\bm {K_{l'}}-\bm G_{1} = \bm {K_l} - \bm G_{1}'$. This three-step process primarily contributes to the tunneling matrix $\hat{T}_1$, cf. Eq.~\eqref{app_eq:tn_3rd}.}%The dashed arrows stand for $\bm G_{i_n}$, cf. Eq.~\eqref{app_eq:G_in}. }
    \label{app_fig:lattice}
\end{figure}

We derive a continuum model for the tunneling Hamiltonian $H_{\text{tun}}$ near commensuration from a tight-binding description of interlayer tunneling, following Ref.~\cite{bistritzer2011moire,scheer2022magic}. We label the positions of the carbon atoms in a graphene layer $l$ as $\bm R+ \bm\delta_{\alpha}$. $\bm R$ represents the centers of the honeycomb unit cells and $\bm\delta_{A/B}=(\frac{a_0}{2}, \mp\frac{a_0}{2\sqrt{3}})^{T}$ are the positions of A and B sublattices, see Fig.~\ref{app_fig:lattice}a. The graphene layer $l'$ is twisted clockwise by an angle $\theta$ relative to the layer $l$, starting with the AA stacking configuration, and laterally shifted by $\bm d$, $\bm{R}'=O(\theta)\bm{R}+\bm{d}$ and $\bm\delta_{\alpha}'=O(\theta)\bm{\delta}_{\alpha}$. Primed vectors are rotated. $|\bm R,\alpha,l\rangle$ and $|\bm R',\alpha',l'\rangle$ denote $2p_z$ orbitals on the lattice sites $\bm R+\bm\delta_{\alpha}$ in the layer $l$ and $\bm{R}'+\bm\delta_{\alpha'}'$ in the layer $l'$, respectively. 
In the two-center approximation, the tunneling Hamiltonian reads 
\begin{align}\label{app_eq:HT_r}
H_{\text{T}}(\bm d) & =\sum_{\bm R, \alpha} \sum_{\bm{R}',\alpha} t\left(\bm R'+\bm\delta_{\alpha'}'-\bm R-\bm\delta_\alpha\right) | \bm R', {\alpha}', l'\rangle\langle \bm R, {\alpha}, l| +h . c. .
\end{align}
Substituting $|\bm R, \alpha, l\rangle =\sum_{\bm p} e^{-i \bm{p} \cdot\left(\bm{R} +\bm\delta_{\alpha}\right)} |\bm p, \alpha, l\rangle/\sqrt{N}$ ($N$ is the number of unit cells) into Eq.~\eqref{app_eq:HT_r} yields
\begin{align}
\langle \bm p', \alpha', l'|H_{\text{T}}(\bm d)|\bm p, \alpha, l\rangle =\sum_{\bm{G}, \bm{G}'} \hat{t}\left(\bm {p}+\bm{G}\right) e^{i \bm{G}'\cdot (\bm\delta_{\alpha'}^{\prime}+\bm{d}) - i \bm{G}\cdot \bm \delta_\alpha} \delta_{\bm{p}+\bm{G}, \bm{p}^{\prime}+\bm{G}'}, \label{eq:tunnelingmatrix_full}
\end{align}
where $\bm{G}$ and $\bm{G}'$ represent the reciprocal lattice vectors of two layers, and $\hat{t}(\bm {q}) = \iint d^2r\  e^{-i\bm q\cdot \bm r} t(\bm r)/A$ ($A$ is the unit cell area). The Kronecker-$\delta$ enforces momentum conservation,
% \begin{equation}
%     \bm{q}_n=\left[O\left(-\frac{\delta\theta}{2}\right)-O\left(\frac{\delta\theta}{2}\right)\right]\bm{K}_n(\theta_c).
% \end{equation}
allowing us to express the matrix elements of the effective tunneling Hamiltonian between low-energy states in the $K$ valley as follows,
\begin{equation}\label{app_eq:tunnelingmatrix_eff}
    \langle \bm{k}', \alpha', l'|H_{\text{tun}}(\bm d)|\bm{k},\alpha, l\rangle\approx \sum_{n}(\hat{T}_{n})_{\alpha'\alpha}\delta_{\bm k+\bm {G}_n,\bm k'+\bm{G}_n'},
\end{equation}
%
%For $\theta\approx 0$, $\chi= 0$ and $|\bm{K} +\bm{G}_n'| = |\bm K|$. For $\theta\approx \pm 38.2^{\circ}$, $ \chi = \pm \frac{2\pi}{3}$ and  $|\bm{K} +\bm{G}_n'| = \sqrt{7}|\bm K|$.
where $|\bm k-\bm{K_{l}}|,|\bm k'-\bm{K_{l'}}| \ll |\bm K|$,
%
%, meaning that we start from an AA aligned bilayer graphene and rotate the top layer about the center of a hexagon. 
% Eq.~\eqref{app_eq:overlapping_dp} implies that
% %
% \begin{equation}
%     \bm G_{n}'\cdot\bm \delta_{\alpha'}' = \left[O(\theta-\theta_c)\left(\bm K_l + \bm G_{n}\right) - \bm K_{l'} \right]\cdot O(\theta)\bm \delta_{\alpha'} %= \bm G_{i_n} \cdot \bm \delta_{\alpha'}, \label{app_eq:Gtau}
% \end{equation}
% %
and $\bm {K_{l'}} = O(\theta)\bm {K_l}$ represented a Dirac wave vector in the rotated layer $l'$. Because the interlayer hopping amplitude $t(\bm{r})$ is a smooth function of the in-plane distance, $\hat{t}(\bm {q})$ decreases rapidly with increasing $|\bm q|$ on the scale of the Brillouin zone \cite{bistritzer2011moire}. It is sufficient to retain three terms ($n=1,2,3$) in Eq.~\eqref{app_eq:tunnelingmatrix_eff} such that $\bm {{K}_{l'}} + \bm{G}_{n}'$ are the three shortest Dirac wave vectors that satisfy 
\begin{equation}\label{app_eq:overlapping_dp}
    \bm {{K}_{l'}} + \bm{G}_n' = O\left(\theta - \theta_c \right)(\bm {K_l} + \bm{G}_n),
\end{equation}
%Below, we derive $\hat{T}_n$ near three commensurate angles.
%, and the unrotated reciprocal lattice vectors $\bm G_{i_n}$ are given by
%
% \begin{equation}\label{app_eq:G_in}
%     \bm G_{i_n} \equiv O(-\theta)\bm G_{n}' = O(-\theta_c)\left(\bm K_l + \bm G_{n}\right) - \bm K_l.
% \end{equation}
%
as illustrated in Fig.~\ref{app_fig:lattice}b and c. Below we derive $\hat{T}_n$ near three commensurate angles.
\begin{enumerate}[label=(\roman*)]
    \item $\theta_c=0^{\circ}$: As $|\bm {K_l} + \bm{G}_n|=|\bm K|$ (Fig.~\ref{app_fig:lattice}b), $\hat{t}(\bm k+\bm G)$ is negligible if $\bm G\neq \bm G_{1,2,3}$; Hence, Eq.~\eqref{app_eq:tunnelingmatrix_eff} indicates that
    \begin{align}\label{app_eq:tn}
    (\hat{T}_{n})_{\alpha'\alpha} &\approx \hat{t}(\bm{K_l}+\bm G_n)e^{i \bm G_n'\cdot \bm \delta_{\alpha'}'-i \bm G_n\cdot \bm\delta_\alpha}e^{i\bm{G}_n\cdot\bm d}.
\end{align}
    Using $\bm G_{n}\cdot\bm\delta_{\alpha} = \bm G_{n}'\cdot\bm\delta_{\alpha}'$ and denoting the identity in the sublattice space as $\sigma^{0}$, we obtain
\begin{equation}
    \hat{T}_{n} = \hat{t}(\bm {K_l})e^{i\bm G_n\cdot\bm d}\left[\sigma^{0}+\cos\frac{2\pi (n-1)}{3}\sigma^{x}+\sin\frac{2\pi (n-1)}{3}\sigma^{y}\right]. \label{app_eq:tn_0}
\end{equation}

\item $\theta_c\approx 38.2^{\circ}$: The two graphene layers form a moir\'e superlattice consisting of $7$ graphene unit cells. Umklapp scattering couples the Dirac point of one layer to $7$ wave vectors in the first Brillouin zone of the other graphene layer, see Fig.~\ref{app_fig:lattice}d. The effective tunneling Hamiltonian between the low-energy states near the Dirac points should take into account higher-order processes in $H_{\text{T}}$,
\begin{align}
    \langle \bm{k}', \alpha', l'|H_{\text{tun}}(\bm d)|\bm{k},\alpha, l\rangle &= \langle \bm{k}', \alpha', l'|H_{\text{T}}(\bm d) + H_{\text{T}}(\bm d)\frac{1}{E-\hat{H}_{l}} H_{\text{T}}(\bm d) \frac{1}{E-\hat{H}_{l'}} H_{\text{T}}(\bm d)|\bm{k},\alpha, l\rangle +... \notag\\
    &\approx \sum_{n=1}^{3}\left(\hat{T}_{n}^{(1)}+\hat{T}_{n}^{(3)}\right)_{\alpha'\alpha}\delta_{\bm k+\bm {G}_n,\bm k'+\bm{G}_n'}.
\end{align}
where $H_{l}$ and $H_{l'}$ denote the Hamiltonian of two decoupled graphene layer with a relative twist angle $\theta_c$ and $E$ is the energy of their Dirac points. The first-order term $\hat{T}_n^{(1)}$ is given by Eq.~\eqref{app_eq:tn} with $\bm G_{n=1,2,3}$ depicted in Fig.~\ref{app_fig:lattice}a and $|\bm{K_l} + \bm G_{n}| =\sqrt{7}|\bm K|$. After some algebra, we obtain
%
% \begin{subequations}
%  \begin{alignat}{3}
% &\bm{G}_{1} = |\bm{K}|\left(\frac{3}{2},-\frac{\sqrt{3}}{2}\right)^{T},\ \ \ &&\bm{G}_{2} = |\bm{K}|\left(-\frac{3}{2},\frac{3\sqrt{3}}{2}\right)^{T},\ \ \ &&\bm{G}_{3} = |\bm{K}|\left(-3,-\sqrt{3}\right)^{T};  \label{app_eq:Gn_38}\\  
% &\bm{G}_{i_1} = |\bm{K}|\left(\frac{3}{2},\frac{\sqrt{3}}{2}\right)^{T},&&\bm{G}_{i_2} = |\bm{K}|\left(-3,\sqrt{3}\right)^{T}, &&\bm{G}_{i_3} =|\bm{K}|\left(-\frac{3}{2},-\frac{3\sqrt{3}}{2}\right)^{T}. \label{app_eq:Gin_38}
% \end{alignat}
% \end{subequations}
% %
% Combining Eqs.~\eqref{app_eq:tn},~\eqref{app_eq:Gtau}, and~\eqref{app_eq:Gin_38} yields
\begin{equation}
    \hat{T}_{n}^{(1)} = \hat{t}(\bm{K_l} + \bm G_{1})e^{i\bm G_n\cdot\bm d}\left[e^{-i\frac{2\pi}{3}\sigma^z}+\cos\frac{2\pi (n-1)}{3}\sigma^{x}+\sin\frac{2\pi (n-1)}{3}\sigma^{y}\right]. \label{app_eq:tn_38}
\end{equation}
The $\hat{T}_{1}^{(3)}$ is illustrated by the three-step process in Fig.~\ref{app_fig:lattice}d,
\begin{align}\label{app_eq:tn_3rd}
    \hat{T}_1^{(3)} &\approx \hat{t}(\bm {K_l})(\sigma^0+\sigma^x)\frac{1}{E-\hat{H}_{l}(\bm {K_{l'}})} \hat{t}\left(\bm {K_{l'}}-\bm G_1\right)(e^{-i\frac{2\pi}{3}\sigma^z}+\sigma^x)\ \frac{1}{E-\hat{H}_{l'}(\bm {K_{l}})}\hat{t}(\bm {K_l})(\sigma^0+\sigma^x) e^{i\bm G_1\cdot\bm d}\notag\\
    & = t_{38.2^{\circ}}e^{i\bm G_1\cdot \bm d}(\sigma^0+\sigma^x).
\end{align}
We note that the sublattice structure of this matrix elements is fully defined by the states in the low-energy sub-space and independent of the details of the states in higher-energy bands, as it is clear from Eq.~(\ref{app_eq:tn_3rd}).
There are two complementary three-step processes which contribute to $\hat{T}_{2,3}^{(3)}$, where $t_\parallel$ is the intralayer tunneling amplitude in the monolayer graphene. The results can be summarized as $\hat{T}_{n}^{(3)}=t_{38.2^{\circ}}e^{i\bm G_1\cdot \bm d}(\sigma^0+\sigma^x e^{2\pi i(n-1)/3\sigma^z})$. Despite being the third order in tunneling, the contribution ${\hat T}_n^{(3)}$ may dominate over the first-order one, ${\hat T}_n^{(1)}$. The respective condition for that reads $\hat{t}(2|\bm K|/\sqrt{7})\hat{t}(|\bm K|)^2/t_{\scriptscriptstyle\parallel}^2\gg \hat{t}(\sqrt{7}|\bm K|)$. It can be satisfied if $\hat t (q)$ falls off rapidly with $q$. For simplicity, here we assumed that $\hat{t}(\bm q)$ depends only on $|\bm q|$.
%If $\hat{t}(\bm q)$ depends only on $|\bm q|$, $t_{38.2^{\circ}}\sim \hat{t}(2|\bm K|/\sqrt{7})\hat{t}(|\bm K|)^2/t_{\scriptscriptstyle\parallel}^2$ and $t_{\scriptscriptstyle\parallel}$ is nearest neighbor hoping amplitude of monolayer graphene. The relative strength of $\hat{T}_n^{(1)}$ and $\hat{T}_n^{(3)}$ depends on the roughness of the atomic-scale wave functions, which determines how fast $\hat{t}(\bm p)$ decreases with respect to $|\bm p|$, as well as the energy dispersion of monolayer graphene. 
For the model used in Ref.~\cite{moon2013optical,scheer2022magic}, $\hat{t}(|\bm K|) \approx 0.11$\si{\electronvolt}, $\hat{t}(2|\bm K|/\sqrt{7}) \approx 0.24$\si{\electronvolt}, $t_{\scriptscriptstyle\parallel}=2.7$\si{\electronvolt}, and $t_{38.2^{\circ}}\sim 1$\si{\milli\electronvolt} $\gg \hat{t}(\sqrt{7}|\bm K|) \approx 60$\si{\micro\electronvolt}. Based on this model, we used
\begin{equation}\label{app_eq:tn_38.2}
    \hat{T}_{n} = \hat{T}_{n}^{(1)} + \hat{T}_{n}^{(3)} \approx t_{38.2^{\circ}}e^{i\bm G_n\cdot\bm d}\left[\sigma^{0}+\cos\frac{2\pi (n-1)}{3}\sigma^{x}+\sin\frac{2\pi (n-1)}{3}\sigma^{y}\right]
\end{equation}
to produce Fig.~\ref{fig:theta_c_38.2_N}.
%We note that when additional graphene layers besides $l$ and $l'$ are present, virtual tunneling into extra graphene layers results in corrections to $H_{\text{tun}}$ of higher order in the tunneling strength, which can be neglected.
\\
\item $\theta_c \approx -38.2^{\circ}$: the lattice configuration of the twisted bilayers are related to the one at $38.2^{\circ}$ by a mirror transformation about the $xz$-plane $\mathcal{M}$. The transformation also switches the sublattices and its representation in the sublattice Hilbert space is $\sigma^x$. The tunneling matrix $\hat{T}_n$ can be obtained from Eq.~\eqref{app_eq:tn_38.2} after applying the transformation $\hat{T}_1 \rightarrow\sigma^x\hat{T}_1\sigma^x, \hat{T}_2 \rightarrow\sigma^x\hat{T}_3\sigma^x, \hat{T}_3 \rightarrow\sigma^x\hat{T}_2\sigma^x$. As a result, $\hat{T}_n$ is also given by Eq.~\eqref{app_eq:tn_38.2}.
%$\hat{T}_1 =\sigma^x\hat{T}_1'\sigma^x, \hat{T}_2 =\sigma^x\hat{T}_3'\sigma^x, \hat{T}_3 =\sigma^x\hat{T}_2'\sigma^x$, where $\hat{T}_n'$ is the tunneling matrix at $\theta\approx 38.2^{\circ}$, see Eq.~\eqref{app_eq:tn_38.2}. Thus, we find that $\hat{T}_n$ is also given by Eq.~\eqref{app_eq:tn_38.2}.
\end{enumerate}
%
%For two graphene layers without dielectric separation, Ref.~\cite{scheer2022magic} estimates $\hat{t}(|\bm K|) \approx 113$\si{meV} and $\hat{t}(\sqrt{7}|\bm K|) \approx 1$\si{meV}.

%
\subsection[d-independence of tunneling current at incommensurate angles]{$d$-independence of the tunneling current at incommensurate angles}\label{app_sec:d}
Plugging the tunneling matrix elements Eq.~\eqref{app_eq:tunnelingmatrix_eff} into Eq.~\eqref{eq:ht_squared} which assumes no interference between different umklapp processes and is valid for incommensurate angles, we find that
\begin{align}
    |\left\langle \bm k' \lambda' T\left|H_{\text{tun}}(\bm d)\right| \bm k \lambda S\right\rangle |^2 & = \sum_{n=1}^3 |\left\langle \bm k' \lambda' T\left|H_{\text{tun}}(0)e^{i\bm{G}_{n}\cdot\bm d}\right| \bm k \lambda S\right\rangle |^2 \delta_{\bm{k}+\bm G_n,\bm{k}'+\bm G_n'}\notag\\
    & = \sum_{n=1}^3|\left\langle \bm k' \lambda' T\left|H_{\text{tun}}(0)\right| \bm k \lambda S\right\rangle|^2 \delta_{\bm{k}+\bm {G}_n,\bm{k}'+\bm {G}_n'}. \label{app_eq:ht2}
\end{align}
Note that for twisted bilayer graphene, Ref.~\cite{bistritzer2011moire} has pointed out that the $\bm d-$dependence in the tunneling Hamiltonian can be gauged out by multiplying each plane wave state $\ket{\bm k,\alpha, l}$ with a proper phase factor, and therefore the electronic spectrum does not depend on the lateral shift between the two layers. However, the same argument is inapplicable to multilayer systems with multiple moir\'e patterns, whose band structures can indeed depend on the relative shift between moir\'e patterns \cite{devakul2023magic}. Nevertheless, we show that in the weak-tunneling regime where Fermi's golden rule applies, the tunneling conductance remains independent of the lateral shift of the tip at incommensurate twist angles. 
At commensurate twist angles, however, interference between different terms in Eq.~\eqref{eq:ht} renders the absolute values of the tunneling matrix elements %squared 
dependent on $\bm d$ and lead to variations by a factor of order one relative to their average value, which is given by Eq.~\eqref{app_eq:ht2} \cite{bistritzer2010transport}. To simplify our analysis, we neglect this subtlety of commensurate angles and use Eq.~\eqref{app_eq:ht2} for all angles.

\section{Tunneling matrix elements between MLG Dirac points and the sample's Bloch bands}
\label{sec:me_bloch}
In this section, we derive %the absolute square of tunneling matrix elements, 
$|T|_{\tau n}^2$ for tunneling between the Dirac points $\bm p_{\tau n}$ of MLG and the Bloch band $\lambda$ of a graphene-based sample whose group velocity is smaller than the Dirac velocity $v_D$. Here, $\tau=\pm 1$ is the valley index and $n=1,2,3$. Equations~\eqref{app_eq:i2},~\eqref{app_eq:I1_v0<vd}, and~\eqref{app_eq:I2_v0<vd} lead to the following expressions for the singular part of $d^2I/dV^2$
%the tunneling current second derivatives,
%
\begin{gather}\label{app_eq:i_result}
    \frac{d^2I}{d\phi^2} = \frac{2 e \Omega}{\hbar^3v_D^2}(f(-eV-\mu_T)-f(-\mu_T)) \Big(1-\frac{v_{\tau n}^2}{v_D^2}\Big)^{-\frac{3}{2}} |T|_{\tau n}^2 \delta(\phi-\phi_{\tau n}), \\
    |T|_{\tau n}^2 = \frac{|a|^2 + |b|^2}{2} + \frac{v_{\tau n}}{v_D} \left(\frac{a b^{*}}{2} e^{i\theta_{\tau n}} + c.c. \right). \label{app_eq:T2}
\end{gather}
Here, $\phi=-eV-\mu_T+\mu_S$, $\bm v_{\tau n}$ is the group velocity in the energy band $\lambda$ at $\bm p= \bm p_{\tau n}$, and $v_{\tau n} = |\bm v_{\tau n}| $. $\theta_{\tau n}$ denotes the angle between $\bm v_{\tau n}$ and the six Dirac wave vectors at the corners of the first Brillouin zone of MLG, $O(\frac{2\pi(n-1)}{3}) \tau\bm K_{\theta}$.
By matching Eq.~\eqref{app_eq:T_general} with Eq.~\eqref{eq:T_tbg} we find
\begin{equation}
    a = e^{i\frac{2\pi\tau\chi}{3}} b =  t_{\theta_c}\left (e^{i\frac{2\pi\tau(\chi+n-1)}{3}} \psi_{tA}^{\lambda}(\bm K_{\tau n}) + \psi_{tB}^{\lambda}(\bm K_{\tau n}) \right), \label{app_eq:a_b_mlg_tbg}
\end{equation}
with $\bm K_{\tau n} = \bm p_{\tau n} - \tau \bm G_{n}$. Plugging Eq.~\eqref{app_eq:a_b_mlg_tbg} into Eq.~\eqref{app_eq:T2}, we arrive at
\begin{equation}\label{app_eq:T2_mlg_tbg}
    |T|_{\tau n}^2 = t_{\theta_c}^2\left[1+\frac{v_{\tau n}}{v_D}\cos(\theta_{\tau n} + \frac{2\pi\tau\chi}{3}) \right] \left |e^{i\frac{2\pi\tau(\chi+n-1)}{3}}\psi_{tA}^{\lambda}(\bm K_{\tau n}) + \psi_{tB}^{\lambda}(\bm K_{\tau n}) \right|^2,
\end{equation}
In $C_{3z}-$invariant systems, $\theta_{\tau n}$ and $v_{\tau n}$ are $n-$independent. In comparison to Eq.~\eqref{eq:M_tbg} derived for flat bands, the above equation shows that a finite group velocity of the sample bands rescales the tunneling matrix elements by a velocity-dependent factor.

\section{Numerical Details}\label{sec:numerics}
In our numerical calculations, we assumed that the tip quasiparticles have infinite lifetime, $A_{\lambda}^{T}(\bm{k},\omega) = \delta(\omega - \xi_{\bm k, \lambda}^{T})$, whereas the spectral function of the sample $A^S(\bm k, \omega)$ is Lorentzian with a broadening parameter $\gamma_S$, see Eq.~\eqref{eq:As}. The tunneling current Eq.~\eqref{eq:i_basic} can be simplified,
\begin{align}
    I & = \frac{2\pi e}{\hbar}\sum_{\bm p}\sum_{\lambda\lambda'}\left(f(\xi_{\bm p\lambda'}^{T}-eV) - f(\xi_{\bm p\lambda'}^{T})\right)A_{\lambda}^{S}(\bm p,\xi_{\bm p \lambda'}^{T}-eV)|T_{\lambda^{\prime} \lambda}(\bm p)|^2 \notag\\
    & = \frac{2\pi e}{\hbar}\sum_{n=1}^{3}\sum_{\bm k\in \text{mBZ}} \sum_{\bm g} \sum_{\lambda\lambda'}\left(f(\xi_{\bm p\lambda'}^{T}-eV) - f(\xi_{\bm p\lambda'}^{T})\right)A_{\lambda}^{S}(\bm p,\xi_{\bm p \lambda'}^{T}-eV)|T_{\lambda^{\prime} \lambda}(\bm p)|^2 \Big |_{\bm p = \bm k+\bm g+\bm {G}_n},
    \label{eq:i_num}
    % & = \frac{2\pi N_fe}{\hbar}\sum_{n=1}^{3}\sum_{\bm k\lambda\lambda'}\left(f(\xi_{\bar{\bm k}\lambda}) - f(\xi_{\bar{\bm k}\lambda}-eV)\right)A_{\lambda'}^{S}(\bar{\bm k}+\bm{K}_n(\theta),\xi_{\bar{\bm k}\lambda}-eV)|T_{\lambda \lambda^{\prime}}(\bar{\bm k} + \bm K_{n}(\theta)+\bm {G}_n')|^2.
    % & = \frac{2\pi N_fe}{\hbar}\sum_{n=1}^{3}\sum_{\bm k,\bm{g_m},\lambda} \sum_{\lambda'}\left(f(\bar{\xi}_{\bar{\bm p}, \lambda}^{T}) - f(\bar{\xi}_{\bar{\bm p}, \lambda}^{T}-eV)\right)A_{\lambda'}^{S}(\bm k,^{\xi_{\bar{\bm p}+\bm K_n, \lambda}T}-eV)|T_{\lambda \lambda^{\prime}}(\bm p+\bm G_n')|^2 \Big |_{\bm p = \bm k+\bm g_m}.
\end{align}
% \begin{equation}
%     T_{\lambda^{\prime} \lambda}(\bar{\bm k}+\bm K_{n}(\theta)+\bm {G}_n) = \left(\frac{1}{\sqrt{2}},\frac{\lambda^{\prime} e^{-i(\theta_{\bar{\bm k}}+\theta_t)}}{\sqrt{2}} \right)\hat{T}_n \psi_{t}^{\lambda}(\bar{\bm k}+\bm K_n(\theta)).
% \end{equation}
by using the periodicity of the Bloch band dispersion, $\xi_{\bm p\lambda}^{S}=\xi_{\bm k\lambda}^{S}, A_{\lambda}^S(\bm p,\omega) = A_{\lambda}^S(\bm k, \omega)$. 

We define a grid of wave vectors $\bm k = \frac{i}{N_1} \bm g_2  + \frac{j}{N_2}\bm g_{3}\ (i,j,N_{1,2}\in \mathbb{Z}, 0\leq i < N_1,  0\leq j < N_2)$. We then diagonalize the single-particle Hamiltonian Eq.~\eqref{eq:H_bm} for each $\bm k$ to obtain band energies $\xi_{\bm k\lambda}^{S}+\mu_{S}$ and eigenstates $\psi_{l\beta}^{\lambda}(\bm k + \bm g)$. 
The tunneling matrix elements $|T|_{\lambda'\lambda}^2(\bm p)$ can be computed via Eq.~\eqref{eq:T_tbg} for a given twist angle $\theta$. Finally, these results and the MLG dispersion
\begin{equation}
    \xi_{\bm k+\bm g+\bm {G}_n \lambda'}^{T}  = \xi_{\bm k+\bm g+\bm {G}_n - \bm G_n'\lambda'}^T = \lambda'\hbar v_D|\bm k + \bm g - \bm K_n(\theta)|-\mu_T,
\end{equation}
with $\bm K_n(\theta)$ given by Eq.~\eqref{eq:kn}, are all plugged into Eq.~\eqref{eq:i_num} to calculate current. The $\bm g$ summation is truncated according to the criterion $\hbar v_D|\bm k + \bm g - \bm K_n(\theta)| < 0.4$\si{eV}. The last two steps are repeated for a list of twist angles $\theta$.

Note that Eq.~\eqref{eq:i_num} can suffer large numerical noise due to the sharp Dirac cone dispersion and low density of states at small energy of MLG, especially at low temperature and with small quasiparticle broadening $\gamma_S$. The situation worsens when calculating the current second derivative. For this reason, we choose a dense $\bm k-$grid with $N_{1}=N_{2}=3600$ to ensure the convergence of our results. 
% Due to the time-reversal and spin SU(2) symmetries, four spin-valley flavors contribute equally and $N_f=4$. Likewise, because we consider $C_{3z}$ invariant systems in this section, $n=1,2,3$ three terms yield identical contributions to tunneling current.

\begin{figure}
    \centering
    \includegraphics[width=0.9\linewidth]{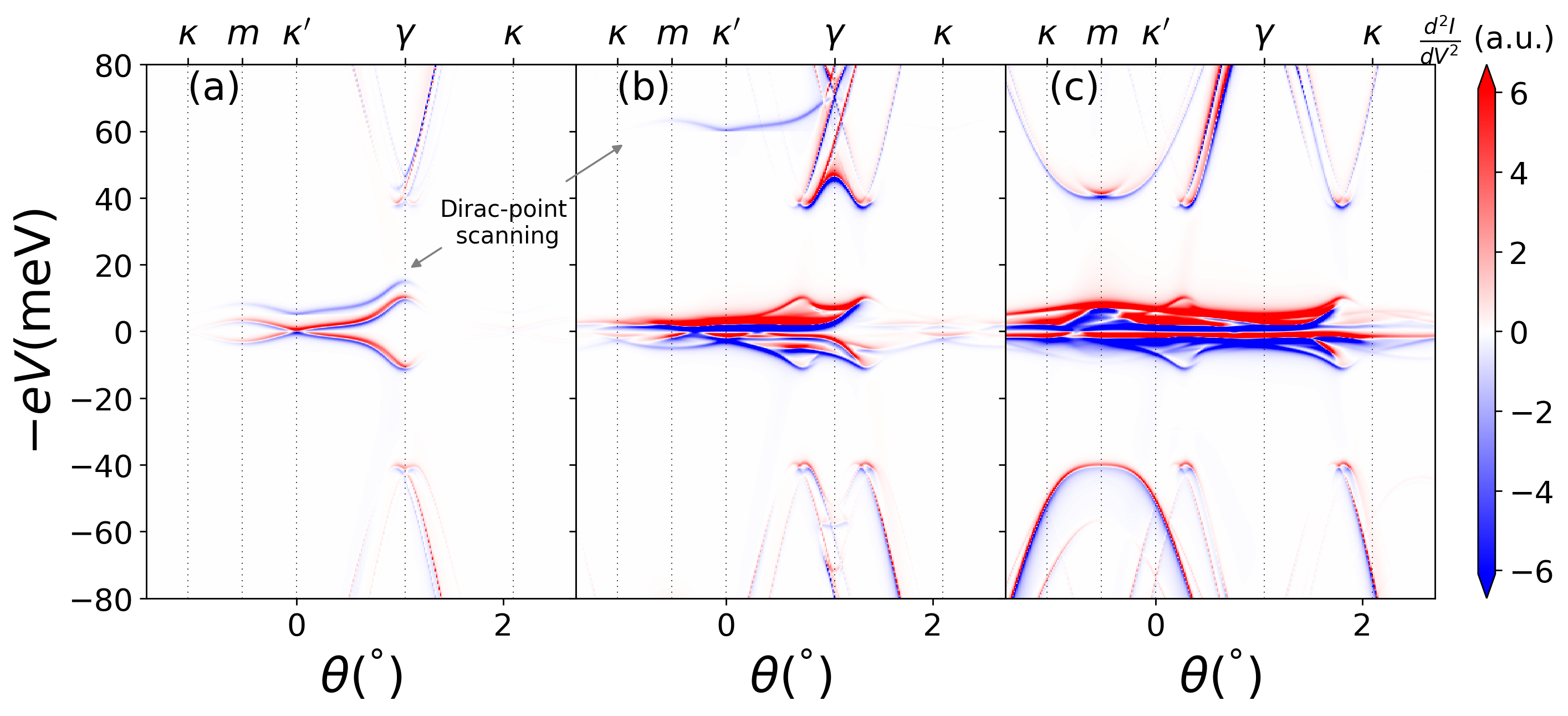}
    \caption{Dependence of $d^2I/dV^2$ maps on the chemical potential $\mu_T$ of the MLG tip, with the sample being charge neutral $1.05^{\circ}$-TBG ($\mu_S=0$). (a) $\mu_T=5$\si{meV}. The red and blue lines at small bias outline the energy dispersion of the TBG flat bands. These singularities arise from the Fermi edge of the tip crossing the spectrum of TBG flat bands. The Dirac-point singularity marked by the arrow is shifted up by $\mu_T$ relative to the Fermi-edge singularities. (b) $\mu_T=60$\si{meV}. The finite radius of the Fermi circle blurs the Fermi-edge singularities. Strong singularities can also occur as the Fermi edge touches the remote bands. Those features overlap with the Dirac-point singularity near the $\gamma$ point at large bias. (c) $\mu_T=150$\si{\milli\electronvolt}. The Dirac-point singularity falls outside the plotted range of bias voltages. All plots are generated with $T=1$\si{\kelvin} and $\gamma_S=0.6$\si{meV}.}
    \label{app_fig:mu_t}
    \includegraphics[width=0.85\linewidth]{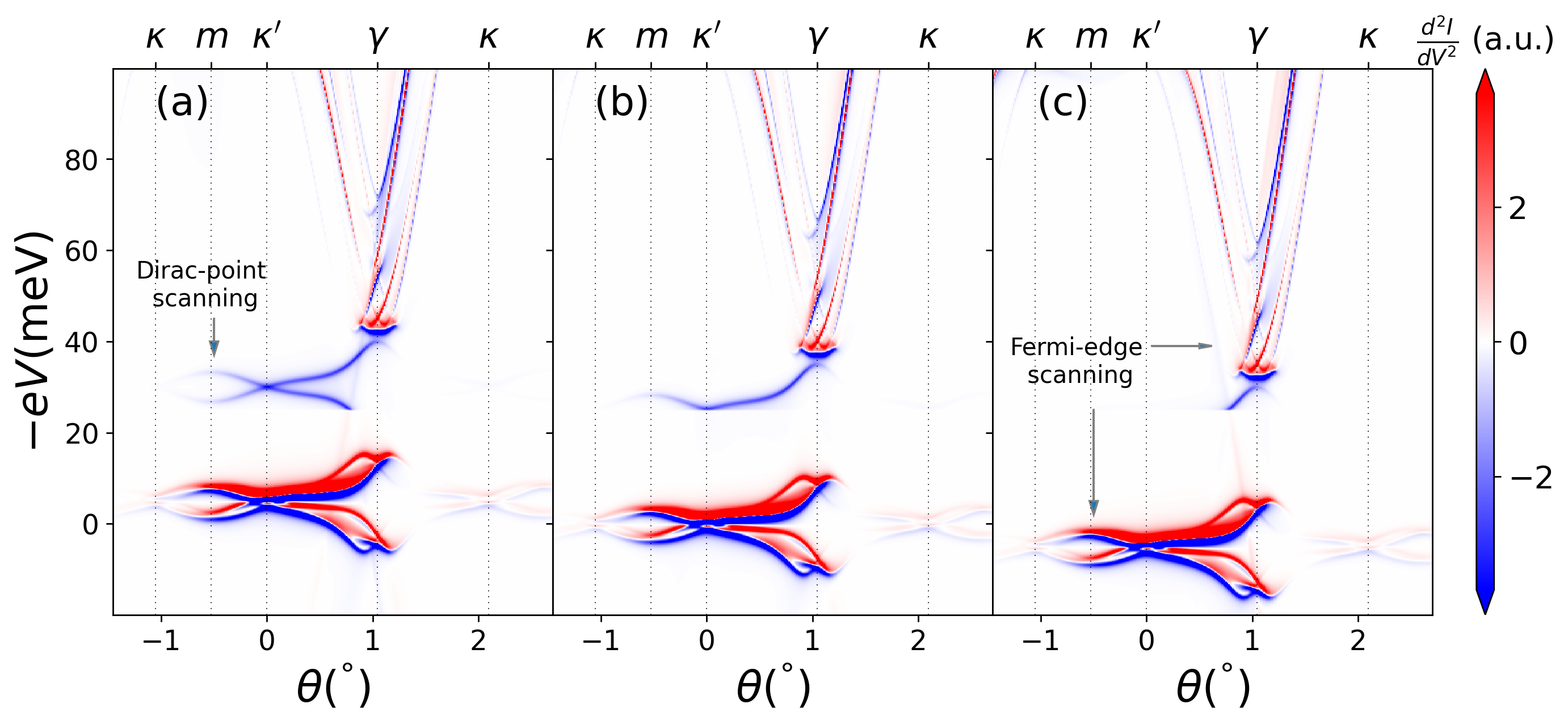}
    \caption{Simulated $d^2I/dV^2$ map for tunnel junctions between $1.05^{\circ}$-TBG and an electron-doped MLG tip with chemical potential $\mu_T=25$\si{\milli\electronvolt}. The chemical potentials of TBG are (a) $\mu_S=-5$\si{meV}, (b) $0$\si{meV}, and (c) $5$\si{meV}. The singularities at small bias are due to the Fermi edge of the tip crossing the spectrum of the flat bands of TBG. The arrow in panel (a) points to a trace produced by the Dirac-point singularity. At large bias, the tip's Dirac point and Fermi edge both scan the remote bands, leading to crowded features near $\gamma$.  In (a) and (c), a faint, steep line intersects the Dirac-point singularities at $V=\mu_T$. This line is induced by the Fermi edge of the sample crossing the tip's Dirac cone. %and traces the tip's Dirac dispersion.
    The same values of $T$ and $\gamma_S$ as in Fig.~\ref{app_fig:mu_t} were used.}
    \label{app_fig:remote}
\end{figure}

Figure~\ref{app_fig:mu_t} presents simulations of $d^2I/dV^2$ for tunnel junctions between $1.05^{\circ}$-TBG and electron-doped MLG tip. It supplements Fig.~\ref{fig:theta_0_mu_t_0} in the main text to illustrate the influence of the tip's chemical potentials $\mu_T$ on $d^2I/dV^2$. Arrows point to Dirac-point singularities which trace the unoccupied part of TBG flat bands, while the red and blue lines at $V\sim 0$ are Fermi-edge singularities. The Dirac-point singularities generally have weaker intensity, but provide a better resolution of the band structure regardless of $\mu_T$. The Fermi-edge singularities can also clearly image the energy dispersion of TBG when $|\mu_T|$ is small (Fig.~\ref{app_fig:mu_t}a), although lowering $\mu_T$ reduces the intensity of Fermi-edge singularities due to a decreased density of states at the Fermi level.

Figure~\ref{app_fig:remote} depicts $d^2I/dV^2$ maps for single-particle TBG at different chemical potentials $\mu_S$. Due to finite density of states at the sample's Fermi level in (a) and (c), a very weak feature traces the Dirac dispersion of the tip as the sample's Fermi edge cross the tip's Dirac cone. It is consistent with Eq.~\eqref{app_eq:dIdV_fs_S} that this type of singularity can be strongly suppressed by quasiparticle broadening in the sample.

\section{Interference suppression of tunneling matrix elements at the bottom-layer Dirac point of TBG}\label{sec:mee}

In this section, we use perturbation theory to analyze the TBG wave functions near the bottom-layer Dirac point $\bm {K_b}=O(-\theta_{\text{TBG}}/2)\bm K$. Our analysis provides analytic insights into the suppression of tunneling matrix elements due to interference effects when scanning the TBG band structures near $\bm {K_b}$ in Fig.~\ref{fig:theta_c_0_N}a. Let us define $\bar{\bm k} = \bm k - \bm{K_b}$. For $|\bar{\bm k}|\ll k_M= |\bm {K_t} - \bm{K_b}|$, we truncate the Bistritzer-MacDonald Hamiltonian to the four plane-wave states of the lowest kinetic energy and obtain the following $8\times 8$ matrix equation,
\begin{equation}\label{eq:perturbation}
    \begin{pmatrix}
        h_{\bm k}(-\frac{\theta_{\text{TBG}}}{2}) &\tilde{T}_{1} &\tilde{T}_{2} &\tilde{T}_{3} \\
       \tilde{T}_{1}^{\dagger} & \hat{h}_{\bm k+\bm g_1}(\frac{\theta_{\text{TBG}}}{2}) &0 &0\\
       \tilde{T}_{2}^{\dagger}  & 0 & \hat{h}_{\bm k+\bm g_2}(\frac{\theta_{\text{TBG}}}{2}) & 0\\
       \tilde{T}_{3}^{\dagger}  & 0 & 0 & \hat{h}_{\bm k+\bm g_3}(\frac{\theta_{\text{TBG}}}{2}) \\
    \end{pmatrix}
    \begin{pmatrix}
        \psi_{b}^{\lambda}(\bm k)\\
        \psi_{t}^{\lambda}(\bm k+\bm g_1)\\
        \psi_{t}^{\lambda}(\bm k+\bm g_2)\\
        \psi_{t}^{\lambda}(\bm k+\bm g_3)
    \end{pmatrix} = \epsilon_{\bm k, \lambda}^{S}    
    \begin{pmatrix}
        \psi_{b}^{\lambda}(\bm k)\\
        \psi_{t}^{\lambda}(\bm k+\bm g_1)\\
        \psi_{t}^{\lambda}(\bm k+\bm g_2)\\
        \psi_{t}^{\lambda}(\bm k+\bm g_3)
    \end{pmatrix}.
\end{equation}
At a small twist angle $\theta_{\text{TBG}}\ll 1$, we can neglect the twist angle dependence of $\hat{h}_{\bm k}$. This allows us to
%and 
derive a low-energy effective Dirac Hamiltonian $\hbar v_F \bar{\bm k}\cdot \sigma$ with Fermi velocity given by Eq.~\eqref{eq:vf}
% \begin{equation}
%     v_F=v_D \left(1-\frac{3w_1^2}{\hbar^2 v_D^2 k_M^2}\right)/\left[1+\frac{3(w_0^2+w_1^2)}{\hbar^2v_D^2 k_M^2} \right].
% \end{equation}
%
and reduce the Schr\"{o}dinger equation, Eq.~(\ref{eq:perturbation}), to $\hbar v_F \bar{\bm k}\cdot \sigma \psi_{b}^{\lambda}(\bm k)=\epsilon_{\bm k,\lambda}^{S}\psi_{b}^{\lambda}(\bm k)$. The two eigenvectors $\psi_{b}^{\lambda=c/v}(\bm k)$ of the effective Dirac Hamiltonian obey the completeness relation, 
\begin{equation}\label{app_eq:completeness}
    \sum_{\lambda}\psi_{b\sigma_1}^{\lambda}(\bm k)\psi_{b\sigma_2}^{\lambda}(\bm k)^{*}=\frac{1}{2} \delta_{\sigma_1,\sigma_2}\sum_{\lambda\sigma}|\psi_{\sigma}^{\lambda}(\bm k)|^2.
\end{equation}
We mention in passing that this relation is exact at $\bm k=\bm {K_b}$ due to the $C_{3z}$ and $C_{2z}\mathcal{T}$ symmetries.

To evaluate the tunneling matrix elements between the tip and the MATBG sample Eq.~\eqref{eq:M_tbg}, one needs to know the wave function on the top layer of MATBG at $\bm k= \bm K_1$. For $\bm K_1$ close to $\bm {K_b}$, it can be derived from the second equation in Eqs.~\eqref{eq:perturbation}, %Plugging this equation into the tunneling matrix element Eq.~\eqref{eq:matrixelement_band},
\begin{align}
    \psi_{t}^{\lambda}(\bm K_1) &\approx \frac{\hbar v_D\bm q\cdot\bm\sigma - \epsilon_{\bm K_1, \lambda}^{S}\sigma^0}{\hbar^2v_D^2\bm q^2-(\epsilon_{\bm K_1, \lambda}^{S})^2}\tilde{T}_1\psi_{b}^{\lambda}(\bm K_1), %\notag\\
    % &= \frac{1}{\hbar^2v_D^2\bm q^2-(\epsilon_{\bm K_1, \lambda}^{S})^2}
    % \begin{pmatrix}
    %     -\epsilon_{\bm K_1\lambda}^{S} w_0 + \hbar v_D (q_x-iq_y)w_1 & \hbar v_D (q_x-iq_y)w_0 -\epsilon_{\bm K_1\lambda}^{S} w_1\\
    %     \hbar v_D (q_x + iq_y) w_0 - \epsilon_{\bm K_1\lambda}^{S}w_1 & - \epsilon_{\bm K_1\lambda}^{S}w_0 + \hbar v_D (q_x + iq_y) w_1  \\
    % \end{pmatrix}\psi_{b}^{\lambda}(\bm K_1)
\end{align}
Here we defined $\bm q= \bm K_1- \bm {K_t}$. Energy $\epsilon_{\bm K_1,\lambda}^{S}$ in the 
%energy 
denominator can be dropped as it becomes very small near the magic angle, $\epsilon_{\bm K_1, \lambda}^{S}\ll \hbar v_{D}q$. Using $\tilde{T}_1=w_0\sigma^0 + w_1\sigma^1$, we arrive at
\begin{equation}
    \psi_{tA}^{\lambda}(\bm K_1) + \psi_{tB}^{\lambda}(\bm K_1) \approx \frac{1}{\hbar^2v_D^2\bm q^2}\left((\hbar v_D q_x - \epsilon_{\bm K_1\lambda}^{S})(w_1 + w_0) - i\hbar  v_D q_y (w_1-w_0), c.c.  \right)\cdot \psi_{b}^{\lambda}(\bm K_1)
\end{equation}
By combing this equation with Eq.~\eqref{app_eq:completeness}, we derive
\begin{equation}\label{eq:T2_perturbation}
    \sum_{\lambda}|T|_1^2 (\theta, \lambda) \approx t_{0}^2\frac{(\hbar v_D q_x - \epsilon_{\bm K_1(\theta)\lambda}^{S})^2(w_1 + w_0)^2 + \hbar^2  v_D^2 q_y^2 (w_1-w_0)^2}{\hbar^4v_D^4q^4}\sum_{\lambda\sigma}|\psi_{b\sigma}^{\lambda}(\bm K_1(\theta))|^2.
\end{equation}
%
%decreases as $1/p_1^2$ near $\kappa$, consistent with the numerical results depicted as the dashed line in Fig.~\ref{fig:matrixelements}a.
%
At twist angle $\theta=\theta_b\equiv -\theta_{TBG}$, $\bm K_1(\theta_b) = \bm {K_b}$, $\bm q= k_M \hat{y}$, and $\epsilon_{\bm K_1\lambda}^{S}=0$; Eq.~\eqref{eq:T2_perturbation} reduces to
\begin{equation}\label{eq:T2_Kb}
    \sum_{\lambda} |T|_{1}^2 ({\theta}_b, \lambda) \approx t_{0}^2\frac{(w_1-w_0)^2}{\hbar^2 v_D^2 k_M^2}\sum_{\lambda\sigma}|\psi_{\sigma}^{\lambda}(\bm {K_b})|^2.
\end{equation}
This is Eq.~\eqref{eq:d2IdV2_ratio} in the main text. It suggests that the tunneling matrix elements and thus the intensity of the $d^2I/dV^2$ singularity at $\bm K_1=\bm {K_b}$ contain information on the interlayer tunneling parameters of MATBG. We discussed in the main text how to exploit this equation to extract $w_0/w_1$ in MATBG from differential tunneling conductance measurements. 

%Here, we would like to point out potential challenges in that proposal. Because of the inevitable strains in both the sample and the tip, their Dirac points cannot be perfect aligned, inducing a small but finite $p_x$. In addition, as particle-hole symmetry is broken in MATBG, $\epsilon_{\bm {K_b}\lambda}\neq 0$ in general. Including these effects makes the first term in Eq.~\eqref{eq:T2_perturbation} finite and increases the tunneling matrix elements, which would cause an error if we used Eq.~\eqref{eq:T2_Kb} to estimate $w_0/w_1$, especially for $w_0/w_1 \sim 1$ (\textit{i.e.}, $w_1+w_0 \gg w_1-w_0$).

% \begin{figure}
%     \centering
%     \includegraphics[width=0.5\linewidth]{w0_w1.pdf}
%     \caption{$r=\sum_{\lambda} \overline{|T|^2}(\bm {K_b}, \lambda)/\sum_{\lambda} \overline{|T|^2}(\bm {K_t}, \lambda)$ is the ratio of the tunneling matrix elements with respect to when the Dirac point $\bm {K}_\theta$ of the tip align with that of the top layer of TBG $\bm {K_t}$ ($\theta=0$). The matrix elements are computed based on the particle-hole symmetric BM model. The solid line correspond to the twist angle $\theta=\theta_{\text{TBG}}$ at which the Dirac points of the tip and the lower layer of TBG coincide. The red and blue dashed lines are obtained by assuming a horizontal displacement of tip Dirac point by $\pm 0.02\% \bm K$ and  $\pm 0.04\% \bm K$, respectively. These two curves show the sensitivity of the tunneling matrix elements to the misalignment of the two Dirac points especially for large $w_0/w_1$. }
%     \label{fig:w0_w1}
% \end{figure}

\twocolumngrid

\bibliography{reference}

%apsrev4-2.bst 2019-01-14 (MD) hand-edited version of apsrev4-1.bst
%Control: key (0)
%Control: author (8) initials jnrlst
%Control: editor formatted (1) identically to author
%Control: production of article title (0) allowed
%Control: page (0) single
%Control: year (1) truncated
%Control: production of eprint (0) enabled
\begin{thebibliography}{60}%
\makeatletter
\providecommand \@ifxundefined [1]{%
 \@ifx{#1\undefined}
}%
\providecommand \@ifnum [1]{%
 \ifnum #1\expandafter \@firstoftwo
 \else \expandafter \@secondoftwo
 \fi
}%
\providecommand \@ifx [1]{%
 \ifx #1\expandafter \@firstoftwo
 \else \expandafter \@secondoftwo
 \fi
}%
\providecommand \natexlab [1]{#1}%
\providecommand \enquote  [1]{``#1''}%
\providecommand \bibnamefont  [1]{#1}%
\providecommand \bibfnamefont [1]{#1}%
\providecommand \citenamefont [1]{#1}%
\providecommand \href@noop [0]{\@secondoftwo}%
\providecommand \href [0]{\begingroup \@sanitize@url \@href}%
\providecommand \@href[1]{\@@startlink{#1}\@@href}%
\providecommand \@@href[1]{\endgroup#1\@@endlink}%
\providecommand \@sanitize@url [0]{\catcode `\\12\catcode `\$12\catcode
  `\&12\catcode `\#12\catcode `\^12\catcode `\_12\catcode `\%12\relax}%
\providecommand \@@startlink[1]{}%
\providecommand \@@endlink[0]{}%
\providecommand \url  [0]{\begingroup\@sanitize@url \@url }%
\providecommand \@url [1]{\endgroup\@href {#1}{\urlprefix }}%
\providecommand \urlprefix  [0]{URL }%
\providecommand \Eprint [0]{\href }%
\providecommand \doibase [0]{https://doi.org/}%
\providecommand \selectlanguage [0]{\@gobble}%
\providecommand \bibinfo  [0]{\@secondoftwo}%
\providecommand \bibfield  [0]{\@secondoftwo}%
\providecommand \translation [1]{[#1]}%
\providecommand \BibitemOpen [0]{}%
\providecommand \bibitemStop [0]{}%
\providecommand \bibitemNoStop [0]{.\EOS\space}%
\providecommand \EOS [0]{\spacefactor3000\relax}%
\providecommand \BibitemShut  [1]{\csname bibitem#1\endcsname}%
\let\auto@bib@innerbib\@empty
%</preamble>
\bibitem [{\citenamefont {Wolf}(2011)}]{wolf2011principles}%
  \BibitemOpen
  \bibfield  {author} {\bibinfo {author} {\bibfnamefont {E.~L.}\ \bibnamefont
  {Wolf}},\ }\href@noop {} {\emph {\bibinfo {title} {Principles of electron
  tunneling spectroscopy}}},\ Vol.\ \bibinfo {volume} {152}\ (\bibinfo
  {publisher} {OUP Oxford},\ \bibinfo {year} {2011})\BibitemShut {NoStop}%
\bibitem [{\citenamefont {Eisenstein}\ \emph {et~al.}(1991)\citenamefont
  {Eisenstein}, \citenamefont {Gramila}, \citenamefont {Pfeiffer},\ and\
  \citenamefont {West}}]{eisenstein1991probing}%
  \BibitemOpen
  \bibfield  {author} {\bibinfo {author} {\bibfnamefont {J.~P.}\ \bibnamefont
  {Eisenstein}}, \bibinfo {author} {\bibfnamefont {T.~J.}\ \bibnamefont
  {Gramila}}, \bibinfo {author} {\bibfnamefont {L.~N.}\ \bibnamefont
  {Pfeiffer}},\ and\ \bibinfo {author} {\bibfnamefont {K.~W.}\ \bibnamefont
  {West}},\ }\bibfield  {title} {\bibinfo {title} {Probing a two-dimensional
  fermi surface by tunneling},\ }\href
  {https://doi.org/10.1103/PhysRevB.44.6511} {\bibfield  {journal} {\bibinfo
  {journal} {Phys. Rev. B}\ }\textbf {\bibinfo {volume} {44}},\ \bibinfo
  {pages} {6511} (\bibinfo {year} {1991})}\BibitemShut {NoStop}%
\bibitem [{\citenamefont {Murphy}\ \emph {et~al.}(1995)\citenamefont {Murphy},
  \citenamefont {Eisenstein}, \citenamefont {Pfeiffer},\ and\ \citenamefont
  {West}}]{murphy1995lifetime}%
  \BibitemOpen
  \bibfield  {author} {\bibinfo {author} {\bibfnamefont {S.~Q.}\ \bibnamefont
  {Murphy}}, \bibinfo {author} {\bibfnamefont {J.~P.}\ \bibnamefont
  {Eisenstein}}, \bibinfo {author} {\bibfnamefont {L.~N.}\ \bibnamefont
  {Pfeiffer}},\ and\ \bibinfo {author} {\bibfnamefont {K.~W.}\ \bibnamefont
  {West}},\ }\bibfield  {title} {\bibinfo {title} {Lifetime of two-dimensional
  electrons measured by tunneling spectroscopy},\ }\href
  {https://doi.org/10.1103/PhysRevB.52.14825} {\bibfield  {journal} {\bibinfo
  {journal} {Phys. Rev. B}\ }\textbf {\bibinfo {volume} {52}},\ \bibinfo
  {pages} {14825} (\bibinfo {year} {1995})}\BibitemShut {NoStop}%
\bibitem [{\citenamefont {Jang}\ \emph {et~al.}(2017)\citenamefont {Jang},
  \citenamefont {Yoo}, \citenamefont {Pfeiffer}, \citenamefont {West},
  \citenamefont {Baldwin},\ and\ \citenamefont {Ashoori}}]{jang2017full}%
  \BibitemOpen
  \bibfield  {author} {\bibinfo {author} {\bibfnamefont {J.}~\bibnamefont
  {Jang}}, \bibinfo {author} {\bibfnamefont {H.~M.}\ \bibnamefont {Yoo}},
  \bibinfo {author} {\bibfnamefont {L.}~\bibnamefont {Pfeiffer}}, \bibinfo
  {author} {\bibfnamefont {K.}~\bibnamefont {West}}, \bibinfo {author}
  {\bibfnamefont {K.}~\bibnamefont {Baldwin}},\ and\ \bibinfo {author}
  {\bibfnamefont {R.~C.}\ \bibnamefont {Ashoori}},\ }\bibfield  {title}
  {\bibinfo {title} {Full momentum-and energy-resolved spectral function of a
  2d electronic system},\ }\href@noop {} {\bibfield  {journal} {\bibinfo
  {journal} {Science}\ }\textbf {\bibinfo {volume} {358}},\ \bibinfo {pages}
  {901} (\bibinfo {year} {2017})}\BibitemShut {NoStop}%
\bibitem [{\citenamefont {Inbar}\ \emph {et~al.}(2023)\citenamefont {Inbar},
  \citenamefont {Birkbeck}, \citenamefont {Xiao}, \citenamefont {Taniguchi},
  \citenamefont {Watanabe}, \citenamefont {Yan}, \citenamefont {Oreg},
  \citenamefont {Stern}, \citenamefont {Berg},\ and\ \citenamefont
  {Ilani}}]{inbar2023quantum}%
  \BibitemOpen
  \bibfield  {author} {\bibinfo {author} {\bibfnamefont {A.}~\bibnamefont
  {Inbar}}, \bibinfo {author} {\bibfnamefont {J.}~\bibnamefont {Birkbeck}},
  \bibinfo {author} {\bibfnamefont {J.}~\bibnamefont {Xiao}}, \bibinfo {author}
  {\bibfnamefont {T.}~\bibnamefont {Taniguchi}}, \bibinfo {author}
  {\bibfnamefont {K.}~\bibnamefont {Watanabe}}, \bibinfo {author}
  {\bibfnamefont {B.}~\bibnamefont {Yan}}, \bibinfo {author} {\bibfnamefont
  {Y.}~\bibnamefont {Oreg}}, \bibinfo {author} {\bibfnamefont {A.}~\bibnamefont
  {Stern}}, \bibinfo {author} {\bibfnamefont {E.}~\bibnamefont {Berg}},\ and\
  \bibinfo {author} {\bibfnamefont {S.}~\bibnamefont {Ilani}},\ }\bibfield
  {title} {\bibinfo {title} {The quantum twisting microscope},\ }\href@noop {}
  {\bibfield  {journal} {\bibinfo  {journal} {Nature}\ }\textbf {\bibinfo
  {volume} {614}},\ \bibinfo {pages} {682} (\bibinfo {year}
  {2023})}\BibitemShut {NoStop}%
\bibitem [{\citenamefont {Birkbeck}\ \emph {et~al.}(2024)\citenamefont
  {Birkbeck}, \citenamefont {Xiao}, \citenamefont {Inbar}, \citenamefont
  {Taniguchi}, \citenamefont {Watanabe}, \citenamefont {Berg}, \citenamefont
  {Glazman}, \citenamefont {Guinea}, \citenamefont {von Oppen},\ and\
  \citenamefont {Ilani}}]{birkbeck2024measuring}%
  \BibitemOpen
  \bibfield  {author} {\bibinfo {author} {\bibfnamefont {J.}~\bibnamefont
  {Birkbeck}}, \bibinfo {author} {\bibfnamefont {J.}~\bibnamefont {Xiao}},
  \bibinfo {author} {\bibfnamefont {A.}~\bibnamefont {Inbar}}, \bibinfo
  {author} {\bibfnamefont {T.}~\bibnamefont {Taniguchi}}, \bibinfo {author}
  {\bibfnamefont {K.}~\bibnamefont {Watanabe}}, \bibinfo {author}
  {\bibfnamefont {E.}~\bibnamefont {Berg}}, \bibinfo {author} {\bibfnamefont
  {L.}~\bibnamefont {Glazman}}, \bibinfo {author} {\bibfnamefont
  {F.}~\bibnamefont {Guinea}}, \bibinfo {author} {\bibfnamefont
  {F.}~\bibnamefont {von Oppen}},\ and\ \bibinfo {author} {\bibfnamefont
  {S.}~\bibnamefont {Ilani}},\ }\href {https://arxiv.org/abs/2407.13404}
  {\bibinfo {title} {Measuring phonon dispersion and electron-phason coupling
  in twisted bilayer graphene with a cryogenic quantum twisting microscope}}
  (\bibinfo {year} {2024}),\ \Eprint {https://arxiv.org/abs/2407.13404}
  {arXiv:2407.13404 [cond-mat.mes-hall]} \BibitemShut {NoStop}%
\bibitem [{\citenamefont {Xiao}\ \emph {et~al.}(2024)\citenamefont {Xiao},
  \citenamefont {Berg}, \citenamefont {Glazman}, \citenamefont {Guinea},
  \citenamefont {Ilani},\ and\ \citenamefont {von Oppen}}]{xiao2024theory}%
  \BibitemOpen
  \bibfield  {author} {\bibinfo {author} {\bibfnamefont {J.}~\bibnamefont
  {Xiao}}, \bibinfo {author} {\bibfnamefont {E.}~\bibnamefont {Berg}}, \bibinfo
  {author} {\bibfnamefont {L.~I.}\ \bibnamefont {Glazman}}, \bibinfo {author}
  {\bibfnamefont {F.}~\bibnamefont {Guinea}}, \bibinfo {author} {\bibfnamefont
  {S.}~\bibnamefont {Ilani}},\ and\ \bibinfo {author} {\bibfnamefont
  {F.}~\bibnamefont {von Oppen}},\ }\href {https://arxiv.org/abs/2407.12092}
  {\bibinfo {title} {Theory of phonon spectroscopy with the quantum twisting
  microscope}} (\bibinfo {year} {2024}),\ \Eprint
  {https://arxiv.org/abs/2407.12092} {arXiv:2407.12092 [cond-mat.mes-hall]}
  \BibitemShut {NoStop}%
\bibitem [{\citenamefont {Xiao}\ \emph {et~al.}(2023)\citenamefont {Xiao},
  \citenamefont {Vituri},\ and\ \citenamefont {Berg}}]{xiao2023probing}%
  \BibitemOpen
  \bibfield  {author} {\bibinfo {author} {\bibfnamefont {J.}~\bibnamefont
  {Xiao}}, \bibinfo {author} {\bibfnamefont {Y.}~\bibnamefont {Vituri}},\ and\
  \bibinfo {author} {\bibfnamefont {E.}~\bibnamefont {Berg}},\ }\bibfield
  {title} {\bibinfo {title} {Probing the order parameter symmetry of
  two-dimensional superconductors by twisted josephson interferometry},\ }\href
  {https://doi.org/10.1103/PhysRevB.108.094520} {\bibfield  {journal} {\bibinfo
   {journal} {Phys. Rev. B}\ }\textbf {\bibinfo {volume} {108}},\ \bibinfo
  {pages} {094520} (\bibinfo {year} {2023})}\BibitemShut {NoStop}%
\bibitem [{\citenamefont {Peri}\ \emph {et~al.}(2024)\citenamefont {Peri},
  \citenamefont {Ilani}, \citenamefont {Lee},\ and\ \citenamefont
  {Refael}}]{peri2024probing}%
  \BibitemOpen
  \bibfield  {author} {\bibinfo {author} {\bibfnamefont {V.}~\bibnamefont
  {Peri}}, \bibinfo {author} {\bibfnamefont {S.}~\bibnamefont {Ilani}},
  \bibinfo {author} {\bibfnamefont {P.~A.}\ \bibnamefont {Lee}},\ and\ \bibinfo
  {author} {\bibfnamefont {G.}~\bibnamefont {Refael}},\ }\bibfield  {title}
  {\bibinfo {title} {Probing quantum spin liquids with a quantum twisting
  microscope},\ }\href {https://doi.org/10.1103/PhysRevB.109.035127} {\bibfield
   {journal} {\bibinfo  {journal} {Phys. Rev. B}\ }\textbf {\bibinfo {volume}
  {109}},\ \bibinfo {pages} {035127} (\bibinfo {year} {2024})}\BibitemShut
  {NoStop}%
\bibitem [{\citenamefont {Pichler}\ \emph {et~al.}(2024)\citenamefont
  {Pichler}, \citenamefont {Kadow}, \citenamefont {Kuhlenkamp},\ and\
  \citenamefont {Knap}}]{pichler2024probing}%
  \BibitemOpen
  \bibfield  {author} {\bibinfo {author} {\bibfnamefont {F.}~\bibnamefont
  {Pichler}}, \bibinfo {author} {\bibfnamefont {W.}~\bibnamefont {Kadow}},
  \bibinfo {author} {\bibfnamefont {C.}~\bibnamefont {Kuhlenkamp}},\ and\
  \bibinfo {author} {\bibfnamefont {M.}~\bibnamefont {Knap}},\ }\bibfield
  {title} {\bibinfo {title} {Probing magnetism in moir\'e heterostructures with
  quantum twisting microscopes},\ }\href
  {https://doi.org/10.1103/PhysRevB.110.045116} {\bibfield  {journal} {\bibinfo
   {journal} {Phys. Rev. B}\ }\textbf {\bibinfo {volume} {110}},\ \bibinfo
  {pages} {045116} (\bibinfo {year} {2024})}\BibitemShut {NoStop}%
\bibitem [{\citenamefont {Cao}\ \emph {et~al.}(2018{\natexlab{a}})\citenamefont
  {Cao}, \citenamefont {Fatemi}, \citenamefont {Fang}, \citenamefont
  {Watanabe}, \citenamefont {Taniguchi}, \citenamefont {Kaxiras},\ and\
  \citenamefont {Jarillo-Herrero}}]{cao2018unconventional}%
  \BibitemOpen
  \bibfield  {author} {\bibinfo {author} {\bibfnamefont {Y.}~\bibnamefont
  {Cao}}, \bibinfo {author} {\bibfnamefont {V.}~\bibnamefont {Fatemi}},
  \bibinfo {author} {\bibfnamefont {S.}~\bibnamefont {Fang}}, \bibinfo {author}
  {\bibfnamefont {K.}~\bibnamefont {Watanabe}}, \bibinfo {author}
  {\bibfnamefont {T.}~\bibnamefont {Taniguchi}}, \bibinfo {author}
  {\bibfnamefont {E.}~\bibnamefont {Kaxiras}},\ and\ \bibinfo {author}
  {\bibfnamefont {P.}~\bibnamefont {Jarillo-Herrero}},\ }\bibfield  {title}
  {\bibinfo {title} {Unconventional superconductivity in magic-angle graphene
  superlattices},\ }\href@noop {} {\bibfield  {journal} {\bibinfo  {journal}
  {Nature}\ }\textbf {\bibinfo {volume} {556}},\ \bibinfo {pages} {43}
  (\bibinfo {year} {2018}{\natexlab{a}})}\BibitemShut {NoStop}%
\bibitem [{\citenamefont {Cao}\ \emph {et~al.}(2018{\natexlab{b}})\citenamefont
  {Cao}, \citenamefont {Fatemi}, \citenamefont {Demir}, \citenamefont {Fang},
  \citenamefont {Tomarken}, \citenamefont {Luo}, \citenamefont
  {Sanchez-Yamagishi}, \citenamefont {Watanabe}, \citenamefont {Taniguchi},
  \citenamefont {Kaxiras} \emph {et~al.}}]{cao2018correlated}%
  \BibitemOpen
  \bibfield  {author} {\bibinfo {author} {\bibfnamefont {Y.}~\bibnamefont
  {Cao}}, \bibinfo {author} {\bibfnamefont {V.}~\bibnamefont {Fatemi}},
  \bibinfo {author} {\bibfnamefont {A.}~\bibnamefont {Demir}}, \bibinfo
  {author} {\bibfnamefont {S.}~\bibnamefont {Fang}}, \bibinfo {author}
  {\bibfnamefont {S.~L.}\ \bibnamefont {Tomarken}}, \bibinfo {author}
  {\bibfnamefont {J.~Y.}\ \bibnamefont {Luo}}, \bibinfo {author} {\bibfnamefont
  {J.~D.}\ \bibnamefont {Sanchez-Yamagishi}}, \bibinfo {author} {\bibfnamefont
  {K.}~\bibnamefont {Watanabe}}, \bibinfo {author} {\bibfnamefont
  {T.}~\bibnamefont {Taniguchi}}, \bibinfo {author} {\bibfnamefont
  {E.}~\bibnamefont {Kaxiras}}, \emph {et~al.},\ }\bibfield  {title} {\bibinfo
  {title} {Correlated insulator behaviour at half-filling in magic-angle
  graphene superlattices},\ }\href@noop {} {\bibfield  {journal} {\bibinfo
  {journal} {Nature}\ }\textbf {\bibinfo {volume} {556}},\ \bibinfo {pages}
  {80} (\bibinfo {year} {2018}{\natexlab{b}})}\BibitemShut {NoStop}%
\bibitem [{\citenamefont {Zondiner}\ \emph {et~al.}(2020)\citenamefont
  {Zondiner}, \citenamefont {Rozen}, \citenamefont {Rodan-Legrain},
  \citenamefont {Cao}, \citenamefont {Queiroz}, \citenamefont {Taniguchi},
  \citenamefont {Watanabe}, \citenamefont {Oreg}, \citenamefont {von Oppen},
  \citenamefont {Stern} \emph {et~al.}}]{zondiner2020cascade}%
  \BibitemOpen
  \bibfield  {author} {\bibinfo {author} {\bibfnamefont {U.}~\bibnamefont
  {Zondiner}}, \bibinfo {author} {\bibfnamefont {A.}~\bibnamefont {Rozen}},
  \bibinfo {author} {\bibfnamefont {D.}~\bibnamefont {Rodan-Legrain}}, \bibinfo
  {author} {\bibfnamefont {Y.}~\bibnamefont {Cao}}, \bibinfo {author}
  {\bibfnamefont {R.}~\bibnamefont {Queiroz}}, \bibinfo {author} {\bibfnamefont
  {T.}~\bibnamefont {Taniguchi}}, \bibinfo {author} {\bibfnamefont
  {K.}~\bibnamefont {Watanabe}}, \bibinfo {author} {\bibfnamefont
  {Y.}~\bibnamefont {Oreg}}, \bibinfo {author} {\bibfnamefont {F.}~\bibnamefont
  {von Oppen}}, \bibinfo {author} {\bibfnamefont {A.}~\bibnamefont {Stern}},
  \emph {et~al.},\ }\bibfield  {title} {\bibinfo {title} {Cascade of phase
  transitions and dirac revivals in magic-angle graphene},\ }\href@noop {}
  {\bibfield  {journal} {\bibinfo  {journal} {Nature}\ }\textbf {\bibinfo
  {volume} {582}},\ \bibinfo {pages} {203} (\bibinfo {year}
  {2020})}\BibitemShut {NoStop}%
\bibitem [{\citenamefont {Wong}\ \emph {et~al.}(2020)\citenamefont {Wong},
  \citenamefont {Nuckolls}, \citenamefont {Oh}, \citenamefont {Lian},
  \citenamefont {Xie}, \citenamefont {Jeon}, \citenamefont {Watanabe},
  \citenamefont {Taniguchi}, \citenamefont {Bernevig},\ and\ \citenamefont
  {Yazdani}}]{wong2020cascade}%
  \BibitemOpen
  \bibfield  {author} {\bibinfo {author} {\bibfnamefont {D.}~\bibnamefont
  {Wong}}, \bibinfo {author} {\bibfnamefont {K.~P.}\ \bibnamefont {Nuckolls}},
  \bibinfo {author} {\bibfnamefont {M.}~\bibnamefont {Oh}}, \bibinfo {author}
  {\bibfnamefont {B.}~\bibnamefont {Lian}}, \bibinfo {author} {\bibfnamefont
  {Y.}~\bibnamefont {Xie}}, \bibinfo {author} {\bibfnamefont {S.}~\bibnamefont
  {Jeon}}, \bibinfo {author} {\bibfnamefont {K.}~\bibnamefont {Watanabe}},
  \bibinfo {author} {\bibfnamefont {T.}~\bibnamefont {Taniguchi}}, \bibinfo
  {author} {\bibfnamefont {B.~A.}\ \bibnamefont {Bernevig}},\ and\ \bibinfo
  {author} {\bibfnamefont {A.}~\bibnamefont {Yazdani}},\ }\bibfield  {title}
  {\bibinfo {title} {Cascade of electronic transitions in magic-angle twisted
  bilayer graphene},\ }\href@noop {} {\bibfield  {journal} {\bibinfo  {journal}
  {Nature}\ }\textbf {\bibinfo {volume} {582}},\ \bibinfo {pages} {198}
  (\bibinfo {year} {2020})}\BibitemShut {NoStop}%
\bibitem [{\citenamefont {Nuckolls}\ \emph {et~al.}(2023)\citenamefont
  {Nuckolls}, \citenamefont {Lee}, \citenamefont {Oh}, \citenamefont {Wong},
  \citenamefont {Soejima}, \citenamefont {Hong}, \citenamefont
  {C{\u{a}}lug{\u{a}}ru}, \citenamefont {Herzog-Arbeitman}, \citenamefont
  {Bernevig}, \citenamefont {Watanabe} \emph {et~al.}}]{nuckolls2023quantum}%
  \BibitemOpen
  \bibfield  {author} {\bibinfo {author} {\bibfnamefont {K.~P.}\ \bibnamefont
  {Nuckolls}}, \bibinfo {author} {\bibfnamefont {R.~L.}\ \bibnamefont {Lee}},
  \bibinfo {author} {\bibfnamefont {M.}~\bibnamefont {Oh}}, \bibinfo {author}
  {\bibfnamefont {D.}~\bibnamefont {Wong}}, \bibinfo {author} {\bibfnamefont
  {T.}~\bibnamefont {Soejima}}, \bibinfo {author} {\bibfnamefont {J.~P.}\
  \bibnamefont {Hong}}, \bibinfo {author} {\bibfnamefont {D.}~\bibnamefont
  {C{\u{a}}lug{\u{a}}ru}}, \bibinfo {author} {\bibfnamefont {J.}~\bibnamefont
  {Herzog-Arbeitman}}, \bibinfo {author} {\bibfnamefont {B.~A.}\ \bibnamefont
  {Bernevig}}, \bibinfo {author} {\bibfnamefont {K.}~\bibnamefont {Watanabe}},
  \emph {et~al.},\ }\bibfield  {title} {\bibinfo {title} {Quantum textures of
  the many-body wavefunctions in magic-angle graphene},\ }\href@noop {}
  {\bibfield  {journal} {\bibinfo  {journal} {Nature}\ }\textbf {\bibinfo
  {volume} {620}},\ \bibinfo {pages} {525} (\bibinfo {year}
  {2023})}\BibitemShut {NoStop}%
\bibitem [{\citenamefont {Kim}\ \emph {et~al.}(2023)\citenamefont {Kim},
  \citenamefont {Choi}, \citenamefont {Lantagne-Hurtubise}, \citenamefont
  {Lewandowski}, \citenamefont {Thomson}, \citenamefont {Kong}, \citenamefont
  {Zhou}, \citenamefont {Baum}, \citenamefont {Zhang}, \citenamefont {Holleis}
  \emph {et~al.}}]{kim2023imaging}%
  \BibitemOpen
  \bibfield  {author} {\bibinfo {author} {\bibfnamefont {H.}~\bibnamefont
  {Kim}}, \bibinfo {author} {\bibfnamefont {Y.}~\bibnamefont {Choi}}, \bibinfo
  {author} {\bibfnamefont {{\'E}.}~\bibnamefont {Lantagne-Hurtubise}}, \bibinfo
  {author} {\bibfnamefont {C.}~\bibnamefont {Lewandowski}}, \bibinfo {author}
  {\bibfnamefont {A.}~\bibnamefont {Thomson}}, \bibinfo {author} {\bibfnamefont
  {L.}~\bibnamefont {Kong}}, \bibinfo {author} {\bibfnamefont {H.}~\bibnamefont
  {Zhou}}, \bibinfo {author} {\bibfnamefont {E.}~\bibnamefont {Baum}}, \bibinfo
  {author} {\bibfnamefont {Y.}~\bibnamefont {Zhang}}, \bibinfo {author}
  {\bibfnamefont {L.}~\bibnamefont {Holleis}}, \emph {et~al.},\ }\bibfield
  {title} {\bibinfo {title} {Imaging inter-valley coherent order in magic-angle
  twisted trilayer graphene},\ }\href@noop {} {\bibfield  {journal} {\bibinfo
  {journal} {Nature}\ }\textbf {\bibinfo {volume} {623}},\ \bibinfo {pages}
  {942} (\bibinfo {year} {2023})}\BibitemShut {NoStop}%
\bibitem [{\citenamefont {Oh}\ \emph {et~al.}(2021)\citenamefont {Oh},
  \citenamefont {Nuckolls}, \citenamefont {Wong}, \citenamefont {Lee},
  \citenamefont {Liu}, \citenamefont {Watanabe}, \citenamefont {Taniguchi},\
  and\ \citenamefont {Yazdani}}]{oh2021evidence}%
  \BibitemOpen
  \bibfield  {author} {\bibinfo {author} {\bibfnamefont {M.}~\bibnamefont
  {Oh}}, \bibinfo {author} {\bibfnamefont {K.~P.}\ \bibnamefont {Nuckolls}},
  \bibinfo {author} {\bibfnamefont {D.}~\bibnamefont {Wong}}, \bibinfo {author}
  {\bibfnamefont {R.~L.}\ \bibnamefont {Lee}}, \bibinfo {author} {\bibfnamefont
  {X.}~\bibnamefont {Liu}}, \bibinfo {author} {\bibfnamefont {K.}~\bibnamefont
  {Watanabe}}, \bibinfo {author} {\bibfnamefont {T.}~\bibnamefont
  {Taniguchi}},\ and\ \bibinfo {author} {\bibfnamefont {A.}~\bibnamefont
  {Yazdani}},\ }\bibfield  {title} {\bibinfo {title} {Evidence for
  unconventional superconductivity in twisted bilayer graphene},\ }\href@noop
  {} {\bibfield  {journal} {\bibinfo  {journal} {Nature}\ }\textbf {\bibinfo
  {volume} {600}},\ \bibinfo {pages} {240} (\bibinfo {year}
  {2021})}\BibitemShut {NoStop}%
\bibitem [{\citenamefont {Kim}\ \emph {et~al.}(2022)\citenamefont {Kim},
  \citenamefont {Choi}, \citenamefont {Lewandowski}, \citenamefont {Thomson},
  \citenamefont {Zhang}, \citenamefont {Polski}, \citenamefont {Watanabe},
  \citenamefont {Taniguchi}, \citenamefont {Alicea},\ and\ \citenamefont
  {Nadj-Perge}}]{kim2022evidence}%
  \BibitemOpen
  \bibfield  {author} {\bibinfo {author} {\bibfnamefont {H.}~\bibnamefont
  {Kim}}, \bibinfo {author} {\bibfnamefont {Y.}~\bibnamefont {Choi}}, \bibinfo
  {author} {\bibfnamefont {C.}~\bibnamefont {Lewandowski}}, \bibinfo {author}
  {\bibfnamefont {A.}~\bibnamefont {Thomson}}, \bibinfo {author} {\bibfnamefont
  {Y.}~\bibnamefont {Zhang}}, \bibinfo {author} {\bibfnamefont
  {R.}~\bibnamefont {Polski}}, \bibinfo {author} {\bibfnamefont
  {K.}~\bibnamefont {Watanabe}}, \bibinfo {author} {\bibfnamefont
  {T.}~\bibnamefont {Taniguchi}}, \bibinfo {author} {\bibfnamefont
  {J.}~\bibnamefont {Alicea}},\ and\ \bibinfo {author} {\bibfnamefont
  {S.}~\bibnamefont {Nadj-Perge}},\ }\bibfield  {title} {\bibinfo {title}
  {Evidence for unconventional superconductivity in twisted trilayer
  graphene},\ }\href@noop {} {\bibfield  {journal} {\bibinfo  {journal}
  {Nature}\ }\textbf {\bibinfo {volume} {606}},\ \bibinfo {pages} {494}
  (\bibinfo {year} {2022})}\BibitemShut {NoStop}%
\bibitem [{\citenamefont {Tanaka}\ \emph {et~al.}(2024)\citenamefont {Tanaka},
  \citenamefont {j.~Wang}, \citenamefont {Dinh}, \citenamefont {Rodan-Legrain},
  \citenamefont {Zaman}, \citenamefont {Hays}, \citenamefont {Kannan},
  \citenamefont {Almanakly}, \citenamefont {Kim}, \citenamefont {Niedzielski},
  \citenamefont {Serniak}, \citenamefont {Schwartz}, \citenamefont {Watanabe},
  \citenamefont {Taniguchi}, \citenamefont {Grover}, \citenamefont {Orlando},
  \citenamefont {Gustavsson}, \citenamefont {Jarillo-Herrero},\ and\
  \citenamefont {Oliver}}]{tanaka2024kinetic}%
  \BibitemOpen
  \bibfield  {author} {\bibinfo {author} {\bibfnamefont {M.}~\bibnamefont
  {Tanaka}}, \bibinfo {author} {\bibfnamefont {J.~I.}\ \bibnamefont {j.~Wang}},
  \bibinfo {author} {\bibfnamefont {T.~H.}\ \bibnamefont {Dinh}}, \bibinfo
  {author} {\bibfnamefont {D.}~\bibnamefont {Rodan-Legrain}}, \bibinfo {author}
  {\bibfnamefont {S.}~\bibnamefont {Zaman}}, \bibinfo {author} {\bibfnamefont
  {M.}~\bibnamefont {Hays}}, \bibinfo {author} {\bibfnamefont {B.}~\bibnamefont
  {Kannan}}, \bibinfo {author} {\bibfnamefont {A.}~\bibnamefont {Almanakly}},
  \bibinfo {author} {\bibfnamefont {D.~K.}\ \bibnamefont {Kim}}, \bibinfo
  {author} {\bibfnamefont {B.~M.}\ \bibnamefont {Niedzielski}}, \bibinfo
  {author} {\bibfnamefont {K.}~\bibnamefont {Serniak}}, \bibinfo {author}
  {\bibfnamefont {M.~E.}\ \bibnamefont {Schwartz}}, \bibinfo {author}
  {\bibfnamefont {K.}~\bibnamefont {Watanabe}}, \bibinfo {author}
  {\bibfnamefont {T.}~\bibnamefont {Taniguchi}}, \bibinfo {author}
  {\bibfnamefont {J.~A.}\ \bibnamefont {Grover}}, \bibinfo {author}
  {\bibfnamefont {T.~P.}\ \bibnamefont {Orlando}}, \bibinfo {author}
  {\bibfnamefont {S.}~\bibnamefont {Gustavsson}}, \bibinfo {author}
  {\bibfnamefont {P.}~\bibnamefont {Jarillo-Herrero}},\ and\ \bibinfo {author}
  {\bibfnamefont {W.~D.}\ \bibnamefont {Oliver}},\ }\href
  {https://arxiv.org/abs/2406.13740} {\bibinfo {title} {Kinetic inductance,
  quantum geometry, and superconductivity in magic-angle twisted bilayer
  graphene}} (\bibinfo {year} {2024}),\ \Eprint
  {https://arxiv.org/abs/2406.13740} {arXiv:2406.13740 [cond-mat.supr-con]}
  \BibitemShut {NoStop}%
\bibitem [{\citenamefont {Banerjee}\ \emph {et~al.}(2024)\citenamefont
  {Banerjee}, \citenamefont {Hao}, \citenamefont {Kreidel}, \citenamefont
  {Ledwith}, \citenamefont {Phinney}, \citenamefont {Park}, \citenamefont
  {Zimmerman}, \citenamefont {Watanabe}, \citenamefont {Taniguchi},
  \citenamefont {Westervelt}, \citenamefont {Jarillo-Herrero}, \citenamefont
  {Volkov}, \citenamefont {Vishwanath}, \citenamefont {Fong},\ and\
  \citenamefont {Kim}}]{banerjee2024superfluid}%
  \BibitemOpen
  \bibfield  {author} {\bibinfo {author} {\bibfnamefont {A.}~\bibnamefont
  {Banerjee}}, \bibinfo {author} {\bibfnamefont {Z.}~\bibnamefont {Hao}},
  \bibinfo {author} {\bibfnamefont {M.}~\bibnamefont {Kreidel}}, \bibinfo
  {author} {\bibfnamefont {P.}~\bibnamefont {Ledwith}}, \bibinfo {author}
  {\bibfnamefont {I.}~\bibnamefont {Phinney}}, \bibinfo {author} {\bibfnamefont
  {J.~M.}\ \bibnamefont {Park}}, \bibinfo {author} {\bibfnamefont {A.~M.}\
  \bibnamefont {Zimmerman}}, \bibinfo {author} {\bibfnamefont {K.}~\bibnamefont
  {Watanabe}}, \bibinfo {author} {\bibfnamefont {T.}~\bibnamefont {Taniguchi}},
  \bibinfo {author} {\bibfnamefont {R.~M.}\ \bibnamefont {Westervelt}},
  \bibinfo {author} {\bibfnamefont {P.}~\bibnamefont {Jarillo-Herrero}},
  \bibinfo {author} {\bibfnamefont {P.~A.}\ \bibnamefont {Volkov}}, \bibinfo
  {author} {\bibfnamefont {A.}~\bibnamefont {Vishwanath}}, \bibinfo {author}
  {\bibfnamefont {K.~C.}\ \bibnamefont {Fong}},\ and\ \bibinfo {author}
  {\bibfnamefont {P.}~\bibnamefont {Kim}},\ }\href
  {https://arxiv.org/abs/2406.13742} {\bibinfo {title} {Superfluid stiffness of
  twisted multilayer graphene superconductors}} (\bibinfo {year} {2024}),\
  \Eprint {https://arxiv.org/abs/2406.13742} {arXiv:2406.13742
  [cond-mat.supr-con]} \BibitemShut {NoStop}%
\bibitem [{\citenamefont {Choi}\ \emph {et~al.}(2021)\citenamefont {Choi},
  \citenamefont {Kim}, \citenamefont {Lewandowski}, \citenamefont {Peng},
  \citenamefont {Thomson}, \citenamefont {Polski}, \citenamefont {Zhang},
  \citenamefont {Watanabe}, \citenamefont {Taniguchi}, \citenamefont {Alicea}
  \emph {et~al.}}]{choi2021interaction}%
  \BibitemOpen
  \bibfield  {author} {\bibinfo {author} {\bibfnamefont {Y.}~\bibnamefont
  {Choi}}, \bibinfo {author} {\bibfnamefont {H.}~\bibnamefont {Kim}}, \bibinfo
  {author} {\bibfnamefont {C.}~\bibnamefont {Lewandowski}}, \bibinfo {author}
  {\bibfnamefont {Y.}~\bibnamefont {Peng}}, \bibinfo {author} {\bibfnamefont
  {A.}~\bibnamefont {Thomson}}, \bibinfo {author} {\bibfnamefont
  {R.}~\bibnamefont {Polski}}, \bibinfo {author} {\bibfnamefont
  {Y.}~\bibnamefont {Zhang}}, \bibinfo {author} {\bibfnamefont
  {K.}~\bibnamefont {Watanabe}}, \bibinfo {author} {\bibfnamefont
  {T.}~\bibnamefont {Taniguchi}}, \bibinfo {author} {\bibfnamefont
  {J.}~\bibnamefont {Alicea}}, \emph {et~al.},\ }\bibfield  {title} {\bibinfo
  {title} {Interaction-driven band flattening and correlated phases in twisted
  bilayer graphene},\ }\href@noop {} {\bibfield  {journal} {\bibinfo  {journal}
  {Nature Physics}\ }\textbf {\bibinfo {volume} {17}},\ \bibinfo {pages} {1375}
  (\bibinfo {year} {2021})}\BibitemShut {NoStop}%
\bibitem [{\citenamefont {Bocarsly}\ \emph {et~al.}(2024)\citenamefont
  {Bocarsly}, \citenamefont {Roy}, \citenamefont {Bhardwaj}, \citenamefont
  {Uzan}, \citenamefont {Ledwith}, \citenamefont {Shavit}, \citenamefont
  {Banu}, \citenamefont {Zhou}, \citenamefont {Myasoedov}, \citenamefont
  {Watanabe}, \citenamefont {Taniguchi}, \citenamefont {Oreg}, \citenamefont
  {Parker}, \citenamefont {Ronen},\ and\ \citenamefont
  {Zeldov}}]{bocarsly2024imaging}%
  \BibitemOpen
  \bibfield  {author} {\bibinfo {author} {\bibfnamefont {M.}~\bibnamefont
  {Bocarsly}}, \bibinfo {author} {\bibfnamefont {I.}~\bibnamefont {Roy}},
  \bibinfo {author} {\bibfnamefont {V.}~\bibnamefont {Bhardwaj}}, \bibinfo
  {author} {\bibfnamefont {M.}~\bibnamefont {Uzan}}, \bibinfo {author}
  {\bibfnamefont {P.}~\bibnamefont {Ledwith}}, \bibinfo {author} {\bibfnamefont
  {G.}~\bibnamefont {Shavit}}, \bibinfo {author} {\bibfnamefont
  {N.}~\bibnamefont {Banu}}, \bibinfo {author} {\bibfnamefont {Y.}~\bibnamefont
  {Zhou}}, \bibinfo {author} {\bibfnamefont {Y.}~\bibnamefont {Myasoedov}},
  \bibinfo {author} {\bibfnamefont {K.}~\bibnamefont {Watanabe}}, \bibinfo
  {author} {\bibfnamefont {T.}~\bibnamefont {Taniguchi}}, \bibinfo {author}
  {\bibfnamefont {Y.}~\bibnamefont {Oreg}}, \bibinfo {author} {\bibfnamefont
  {D.}~\bibnamefont {Parker}}, \bibinfo {author} {\bibfnamefont
  {Y.}~\bibnamefont {Ronen}},\ and\ \bibinfo {author} {\bibfnamefont
  {E.}~\bibnamefont {Zeldov}},\ }\href {https://arxiv.org/abs/2407.10675}
  {\bibinfo {title} {Imaging coulomb interactions and migrating dirac cones in
  twisted graphene by local quantum oscillations}} (\bibinfo {year} {2024}),\
  \Eprint {https://arxiv.org/abs/2407.10675} {arXiv:2407.10675
  [cond-mat.mes-hall]} \BibitemShut {NoStop}%
\bibitem [{\citenamefont {Li}\ \emph {et~al.}(2024{\natexlab{a}})\citenamefont
  {Li}, \citenamefont {Zhang}, \citenamefont {Wang}, \citenamefont {Chen},
  \citenamefont {Bao}, \citenamefont {Liu}, \citenamefont {Lin}, \citenamefont
  {Zhang}, \citenamefont {Zhang}, \citenamefont {Watanabe} \emph
  {et~al.}}]{li2024evolution}%
  \BibitemOpen
  \bibfield  {author} {\bibinfo {author} {\bibfnamefont {Q.}~\bibnamefont
  {Li}}, \bibinfo {author} {\bibfnamefont {H.}~\bibnamefont {Zhang}}, \bibinfo
  {author} {\bibfnamefont {Y.}~\bibnamefont {Wang}}, \bibinfo {author}
  {\bibfnamefont {W.}~\bibnamefont {Chen}}, \bibinfo {author} {\bibfnamefont
  {C.}~\bibnamefont {Bao}}, \bibinfo {author} {\bibfnamefont {Q.}~\bibnamefont
  {Liu}}, \bibinfo {author} {\bibfnamefont {T.}~\bibnamefont {Lin}}, \bibinfo
  {author} {\bibfnamefont {S.}~\bibnamefont {Zhang}}, \bibinfo {author}
  {\bibfnamefont {H.}~\bibnamefont {Zhang}}, \bibinfo {author} {\bibfnamefont
  {K.}~\bibnamefont {Watanabe}}, \emph {et~al.},\ }\bibfield  {title} {\bibinfo
  {title} {Evolution of the flat band and the role of lattice relaxations in
  twisted bilayer graphene},\ }\href@noop {} {\bibfield  {journal} {\bibinfo
  {journal} {Nature Materials}\ ,\ \bibinfo {pages} {1}} (\bibinfo {year}
  {2024}{\natexlab{a}})}\BibitemShut {NoStop}%
\bibitem [{\citenamefont {Chen}\ \emph {et~al.}(2023)\citenamefont {Chen},
  \citenamefont {Nuckolls}, \citenamefont {Ding}, \citenamefont {Miao},
  \citenamefont {Wong}, \citenamefont {Oh}, \citenamefont {Lee}, \citenamefont
  {He}, \citenamefont {Peng}, \citenamefont {Pei}, \citenamefont {Li},
  \citenamefont {Zhang}, \citenamefont {Liu}, \citenamefont {Liu},
  \citenamefont {Jozwiak}, \citenamefont {Bostwick}, \citenamefont {Rotenberg},
  \citenamefont {Li}, \citenamefont {Han}, \citenamefont {Pan}, \citenamefont
  {Dai}, \citenamefont {Liu}, \citenamefont {Bernevig}, \citenamefont {Wang},
  \citenamefont {Yazdani},\ and\ \citenamefont {Chen}}]{chen2023strong}%
  \BibitemOpen
  \bibfield  {author} {\bibinfo {author} {\bibfnamefont {C.}~\bibnamefont
  {Chen}}, \bibinfo {author} {\bibfnamefont {K.~P.}\ \bibnamefont {Nuckolls}},
  \bibinfo {author} {\bibfnamefont {S.}~\bibnamefont {Ding}}, \bibinfo {author}
  {\bibfnamefont {W.}~\bibnamefont {Miao}}, \bibinfo {author} {\bibfnamefont
  {D.}~\bibnamefont {Wong}}, \bibinfo {author} {\bibfnamefont {M.}~\bibnamefont
  {Oh}}, \bibinfo {author} {\bibfnamefont {R.~L.}\ \bibnamefont {Lee}},
  \bibinfo {author} {\bibfnamefont {S.}~\bibnamefont {He}}, \bibinfo {author}
  {\bibfnamefont {C.}~\bibnamefont {Peng}}, \bibinfo {author} {\bibfnamefont
  {D.}~\bibnamefont {Pei}}, \bibinfo {author} {\bibfnamefont {Y.}~\bibnamefont
  {Li}}, \bibinfo {author} {\bibfnamefont {S.}~\bibnamefont {Zhang}}, \bibinfo
  {author} {\bibfnamefont {J.}~\bibnamefont {Liu}}, \bibinfo {author}
  {\bibfnamefont {Z.}~\bibnamefont {Liu}}, \bibinfo {author} {\bibfnamefont
  {C.}~\bibnamefont {Jozwiak}}, \bibinfo {author} {\bibfnamefont
  {A.}~\bibnamefont {Bostwick}}, \bibinfo {author} {\bibfnamefont
  {E.}~\bibnamefont {Rotenberg}}, \bibinfo {author} {\bibfnamefont
  {C.}~\bibnamefont {Li}}, \bibinfo {author} {\bibfnamefont {X.}~\bibnamefont
  {Han}}, \bibinfo {author} {\bibfnamefont {D.}~\bibnamefont {Pan}}, \bibinfo
  {author} {\bibfnamefont {X.}~\bibnamefont {Dai}}, \bibinfo {author}
  {\bibfnamefont {C.}~\bibnamefont {Liu}}, \bibinfo {author} {\bibfnamefont
  {B.~A.}\ \bibnamefont {Bernevig}}, \bibinfo {author} {\bibfnamefont
  {Y.}~\bibnamefont {Wang}}, \bibinfo {author} {\bibfnamefont {A.}~\bibnamefont
  {Yazdani}},\ and\ \bibinfo {author} {\bibfnamefont {Y.}~\bibnamefont
  {Chen}},\ }\href@noop {} {\bibinfo {title} {Strong inter-valley
  electron-phonon coupling in magic-angle twisted bilayer graphene}} (\bibinfo
  {year} {2023}),\ \Eprint {https://arxiv.org/abs/2303.14903} {arXiv:2303.14903
  [cond-mat.mes-hall]} \BibitemShut {NoStop}%
\bibitem [{\citenamefont {Lisi}\ \emph {et~al.}(2021)\citenamefont {Lisi},
  \citenamefont {Lu}, \citenamefont {Benschop}, \citenamefont {de~Jong},
  \citenamefont {Stepanov}, \citenamefont {Duran}, \citenamefont {Margot},
  \citenamefont {Cucchi}, \citenamefont {Cappelli}, \citenamefont {Hunter}
  \emph {et~al.}}]{lisi2021observation}%
  \BibitemOpen
  \bibfield  {author} {\bibinfo {author} {\bibfnamefont {S.}~\bibnamefont
  {Lisi}}, \bibinfo {author} {\bibfnamefont {X.}~\bibnamefont {Lu}}, \bibinfo
  {author} {\bibfnamefont {T.}~\bibnamefont {Benschop}}, \bibinfo {author}
  {\bibfnamefont {T.~A.}\ \bibnamefont {de~Jong}}, \bibinfo {author}
  {\bibfnamefont {P.}~\bibnamefont {Stepanov}}, \bibinfo {author}
  {\bibfnamefont {J.~R.}\ \bibnamefont {Duran}}, \bibinfo {author}
  {\bibfnamefont {F.}~\bibnamefont {Margot}}, \bibinfo {author} {\bibfnamefont
  {I.}~\bibnamefont {Cucchi}}, \bibinfo {author} {\bibfnamefont
  {E.}~\bibnamefont {Cappelli}}, \bibinfo {author} {\bibfnamefont
  {A.}~\bibnamefont {Hunter}}, \emph {et~al.},\ }\bibfield  {title} {\bibinfo
  {title} {Observation of flat bands in twisted bilayer graphene},\ }\href@noop
  {} {\bibfield  {journal} {\bibinfo  {journal} {Nature Physics}\ }\textbf
  {\bibinfo {volume} {17}},\ \bibinfo {pages} {189} (\bibinfo {year}
  {2021})}\BibitemShut {NoStop}%
\bibitem [{\citenamefont {Bistritzer}\ and\ \citenamefont
  {MacDonald}(2010)}]{bistritzer2010transport}%
  \BibitemOpen
  \bibfield  {author} {\bibinfo {author} {\bibfnamefont {R.}~\bibnamefont
  {Bistritzer}}\ and\ \bibinfo {author} {\bibfnamefont {A.~H.}\ \bibnamefont
  {MacDonald}},\ }\bibfield  {title} {\bibinfo {title} {Transport between
  twisted graphene layers},\ }\href
  {https://doi.org/10.1103/PhysRevB.81.245412} {\bibfield  {journal} {\bibinfo
  {journal} {Phys. Rev. B}\ }\textbf {\bibinfo {volume} {81}},\ \bibinfo
  {pages} {245412} (\bibinfo {year} {2010})}\BibitemShut {NoStop}%
\bibitem [{Note1()}]{Note1}%
  \BibitemOpen
  \bibinfo {note} {In Eq.~\protect \eqref {eq:d2IdV2_dp}, $d\phi /dV =-e$ if
  $\mu _{T,S}$ are fixed}\BibitemShut {NoStop}%
\bibitem [{\citenamefont {Bistritzer}\ and\ \citenamefont
  {MacDonald}(2011)}]{bistritzer2011moire}%
  \BibitemOpen
  \bibfield  {author} {\bibinfo {author} {\bibfnamefont {R.}~\bibnamefont
  {Bistritzer}}\ and\ \bibinfo {author} {\bibfnamefont {A.~H.}\ \bibnamefont
  {MacDonald}},\ }\bibfield  {title} {\bibinfo {title} {Moir{\'e} bands in
  twisted double-layer graphene},\ }\href@noop {} {\bibfield  {journal}
  {\bibinfo  {journal} {Proceedings of the National Academy of Sciences}\
  }\textbf {\bibinfo {volume} {108}},\ \bibinfo {pages} {12233} (\bibinfo
  {year} {2011})}\BibitemShut {NoStop}%
\bibitem [{\citenamefont {Balents}(2019)}]{balents2019general}%
  \BibitemOpen
  \bibfield  {author} {\bibinfo {author} {\bibfnamefont {L.}~\bibnamefont
  {Balents}},\ }\bibfield  {title} {\bibinfo {title} {General continuum model
  for twisted bilayer graphene and arbitrary smooth deformations},\ }\href@noop
  {} {\bibfield  {journal} {\bibinfo  {journal} {SciPost Physics}\ }\textbf
  {\bibinfo {volume} {7}},\ \bibinfo {pages} {048} (\bibinfo {year}
  {2019})}\BibitemShut {NoStop}%
\bibitem [{\citenamefont {Scheer}\ \emph {et~al.}(2022)\citenamefont {Scheer},
  \citenamefont {Gu},\ and\ \citenamefont {Lian}}]{scheer2022magic}%
  \BibitemOpen
  \bibfield  {author} {\bibinfo {author} {\bibfnamefont {M.~G.}\ \bibnamefont
  {Scheer}}, \bibinfo {author} {\bibfnamefont {K.}~\bibnamefont {Gu}},\ and\
  \bibinfo {author} {\bibfnamefont {B.}~\bibnamefont {Lian}},\ }\bibfield
  {title} {\bibinfo {title} {Magic angles in twisted bilayer graphene near
  commensuration: Towards a hypermagic regime},\ }\href
  {https://doi.org/10.1103/PhysRevB.106.115418} {\bibfield  {journal} {\bibinfo
   {journal} {Phys. Rev. B}\ }\textbf {\bibinfo {volume} {106}},\ \bibinfo
  {pages} {115418} (\bibinfo {year} {2022})}\BibitemShut {NoStop}%
\bibitem [{\citenamefont {Vafek}\ and\ \citenamefont
  {Kang}(2023)}]{vafek2023continuum}%
  \BibitemOpen
  \bibfield  {author} {\bibinfo {author} {\bibfnamefont {O.}~\bibnamefont
  {Vafek}}\ and\ \bibinfo {author} {\bibfnamefont {J.}~\bibnamefont {Kang}},\
  }\bibfield  {title} {\bibinfo {title} {Continuum effective hamiltonian for
  graphene bilayers for an arbitrary smooth lattice deformation from
  microscopic theories},\ }\href {https://doi.org/10.1103/PhysRevB.107.075123}
  {\bibfield  {journal} {\bibinfo  {journal} {Phys. Rev. B}\ }\textbf {\bibinfo
  {volume} {107}},\ \bibinfo {pages} {075123} (\bibinfo {year}
  {2023})}\BibitemShut {NoStop}%
\bibitem [{\citenamefont {Kang}\ and\ \citenamefont
  {Vafek}(2023)}]{kang2023pesudomagnetic}%
  \BibitemOpen
  \bibfield  {author} {\bibinfo {author} {\bibfnamefont {J.}~\bibnamefont
  {Kang}}\ and\ \bibinfo {author} {\bibfnamefont {O.}~\bibnamefont {Vafek}},\
  }\bibfield  {title} {\bibinfo {title} {Pseudomagnetic fields, particle-hole
  asymmetry, and microscopic effective continuum hamiltonians of twisted
  bilayer graphene},\ }\href {https://doi.org/10.1103/PhysRevB.107.075408}
  {\bibfield  {journal} {\bibinfo  {journal} {Phys. Rev. B}\ }\textbf {\bibinfo
  {volume} {107}},\ \bibinfo {pages} {075408} (\bibinfo {year}
  {2023})}\BibitemShut {NoStop}%
\bibitem [{\citenamefont {Li}\ \emph {et~al.}(2024{\natexlab{b}})\citenamefont
  {Li}, \citenamefont {Kumar}, \citenamefont {Stepanov}, \citenamefont
  {Pantaleón}, \citenamefont {Zhan}, \citenamefont {Agarwal}, \citenamefont
  {Bercher}, \citenamefont {Barrier}, \citenamefont {Watanabe}, \citenamefont
  {Taniguchi}, \citenamefont {Kuzmenko}, \citenamefont {Guinea}, \citenamefont
  {Torre},\ and\ \citenamefont {Koppens}}]{li2024infrare}%
  \BibitemOpen
  \bibfield  {author} {\bibinfo {author} {\bibfnamefont {G.}~\bibnamefont
  {Li}}, \bibinfo {author} {\bibfnamefont {R.~K.}\ \bibnamefont {Kumar}},
  \bibinfo {author} {\bibfnamefont {P.}~\bibnamefont {Stepanov}}, \bibinfo
  {author} {\bibfnamefont {P.~A.}\ \bibnamefont {Pantaleón}}, \bibinfo
  {author} {\bibfnamefont {Z.}~\bibnamefont {Zhan}}, \bibinfo {author}
  {\bibfnamefont {H.}~\bibnamefont {Agarwal}}, \bibinfo {author} {\bibfnamefont
  {A.}~\bibnamefont {Bercher}}, \bibinfo {author} {\bibfnamefont
  {J.}~\bibnamefont {Barrier}}, \bibinfo {author} {\bibfnamefont
  {K.}~\bibnamefont {Watanabe}}, \bibinfo {author} {\bibfnamefont
  {T.}~\bibnamefont {Taniguchi}}, \bibinfo {author} {\bibfnamefont {A.~B.}\
  \bibnamefont {Kuzmenko}}, \bibinfo {author} {\bibfnamefont {F.}~\bibnamefont
  {Guinea}}, \bibinfo {author} {\bibfnamefont {I.}~\bibnamefont {Torre}},\ and\
  \bibinfo {author} {\bibfnamefont {F.~H.~L.}\ \bibnamefont {Koppens}},\ }\href
  {https://arxiv.org/abs/2404.05716} {\bibinfo {title} {Infrared spectroscopy
  for diagnosing superlattice minibands in magic-angle twisted bilayer
  graphene}} (\bibinfo {year} {2024}{\natexlab{b}}),\ \Eprint
  {https://arxiv.org/abs/2404.05716} {arXiv:2404.05716 [cond-mat.mes-hall]}
  \BibitemShut {NoStop}%
\bibitem [{\citenamefont {Song}\ \emph {et~al.}(2019)\citenamefont {Song},
  \citenamefont {Wang}, \citenamefont {Shi}, \citenamefont {Li}, \citenamefont
  {Fang},\ and\ \citenamefont {Bernevig}}]{song2019all}%
  \BibitemOpen
  \bibfield  {author} {\bibinfo {author} {\bibfnamefont {Z.}~\bibnamefont
  {Song}}, \bibinfo {author} {\bibfnamefont {Z.}~\bibnamefont {Wang}}, \bibinfo
  {author} {\bibfnamefont {W.}~\bibnamefont {Shi}}, \bibinfo {author}
  {\bibfnamefont {G.}~\bibnamefont {Li}}, \bibinfo {author} {\bibfnamefont
  {C.}~\bibnamefont {Fang}},\ and\ \bibinfo {author} {\bibfnamefont {B.~A.}\
  \bibnamefont {Bernevig}},\ }\bibfield  {title} {\bibinfo {title} {All magic
  angles in twisted bilayer graphene are topological},\ }\href
  {https://doi.org/10.1103/PhysRevLett.123.036401} {\bibfield  {journal}
  {\bibinfo  {journal} {Phys. Rev. Lett.}\ }\textbf {\bibinfo {volume} {123}},\
  \bibinfo {pages} {036401} (\bibinfo {year} {2019})}\BibitemShut {NoStop}%
\bibitem [{\citenamefont {Bernevig}\ \emph
  {et~al.}(2021{\natexlab{a}})\citenamefont {Bernevig}, \citenamefont {Song},
  \citenamefont {Regnault},\ and\ \citenamefont {Lian}}]{bernevig2021twisted}%
  \BibitemOpen
  \bibfield  {author} {\bibinfo {author} {\bibfnamefont {B.~A.}\ \bibnamefont
  {Bernevig}}, \bibinfo {author} {\bibfnamefont {Z.-D.}\ \bibnamefont {Song}},
  \bibinfo {author} {\bibfnamefont {N.}~\bibnamefont {Regnault}},\ and\
  \bibinfo {author} {\bibfnamefont {B.}~\bibnamefont {Lian}},\ }\bibfield
  {title} {\bibinfo {title} {Twisted bilayer graphene. i. matrix elements,
  approximations, perturbation theory, and a
  $k\ifmmode\cdot\else\textperiodcentered\fi{}p$ two-band model},\ }\href
  {https://doi.org/10.1103/PhysRevB.103.205411} {\bibfield  {journal} {\bibinfo
   {journal} {Phys. Rev. B}\ }\textbf {\bibinfo {volume} {103}},\ \bibinfo
  {pages} {205411} (\bibinfo {year} {2021}{\natexlab{a}})}\BibitemShut
  {NoStop}%
\bibitem [{\citenamefont {Kerelsky}\ \emph {et~al.}(2019)\citenamefont
  {Kerelsky}, \citenamefont {McGilly}, \citenamefont {Kennes}, \citenamefont
  {Xian}, \citenamefont {Yankowitz}, \citenamefont {Chen}, \citenamefont
  {Watanabe}, \citenamefont {Taniguchi}, \citenamefont {Hone}, \citenamefont
  {Dean} \emph {et~al.}}]{kerelsky2019maximized}%
  \BibitemOpen
  \bibfield  {author} {\bibinfo {author} {\bibfnamefont {A.}~\bibnamefont
  {Kerelsky}}, \bibinfo {author} {\bibfnamefont {L.~J.}\ \bibnamefont
  {McGilly}}, \bibinfo {author} {\bibfnamefont {D.~M.}\ \bibnamefont {Kennes}},
  \bibinfo {author} {\bibfnamefont {L.}~\bibnamefont {Xian}}, \bibinfo {author}
  {\bibfnamefont {M.}~\bibnamefont {Yankowitz}}, \bibinfo {author}
  {\bibfnamefont {S.}~\bibnamefont {Chen}}, \bibinfo {author} {\bibfnamefont
  {K.}~\bibnamefont {Watanabe}}, \bibinfo {author} {\bibfnamefont
  {T.}~\bibnamefont {Taniguchi}}, \bibinfo {author} {\bibfnamefont
  {J.}~\bibnamefont {Hone}}, \bibinfo {author} {\bibfnamefont {C.}~\bibnamefont
  {Dean}}, \emph {et~al.},\ }\bibfield  {title} {\bibinfo {title} {Maximized
  electron interactions at the magic angle in twisted bilayer graphene},\
  }\href@noop {} {\bibfield  {journal} {\bibinfo  {journal} {Nature}\ }\textbf
  {\bibinfo {volume} {572}},\ \bibinfo {pages} {95} (\bibinfo {year}
  {2019})}\BibitemShut {NoStop}%
\bibitem [{\citenamefont {Choi}\ \emph {et~al.}(2019)\citenamefont {Choi},
  \citenamefont {Kemmer}, \citenamefont {Peng}, \citenamefont {Thomson},
  \citenamefont {Arora}, \citenamefont {Polski}, \citenamefont {Zhang},
  \citenamefont {Ren}, \citenamefont {Alicea}, \citenamefont {Refael} \emph
  {et~al.}}]{choi2019electronic}%
  \BibitemOpen
  \bibfield  {author} {\bibinfo {author} {\bibfnamefont {Y.}~\bibnamefont
  {Choi}}, \bibinfo {author} {\bibfnamefont {J.}~\bibnamefont {Kemmer}},
  \bibinfo {author} {\bibfnamefont {Y.}~\bibnamefont {Peng}}, \bibinfo {author}
  {\bibfnamefont {A.}~\bibnamefont {Thomson}}, \bibinfo {author} {\bibfnamefont
  {H.}~\bibnamefont {Arora}}, \bibinfo {author} {\bibfnamefont
  {R.}~\bibnamefont {Polski}}, \bibinfo {author} {\bibfnamefont
  {Y.}~\bibnamefont {Zhang}}, \bibinfo {author} {\bibfnamefont
  {H.}~\bibnamefont {Ren}}, \bibinfo {author} {\bibfnamefont {J.}~\bibnamefont
  {Alicea}}, \bibinfo {author} {\bibfnamefont {G.}~\bibnamefont {Refael}},
  \emph {et~al.},\ }\bibfield  {title} {\bibinfo {title} {Electronic
  correlations in twisted bilayer graphene near the magic angle},\ }\href@noop
  {} {\bibfield  {journal} {\bibinfo  {journal} {Nature physics}\ }\textbf
  {\bibinfo {volume} {15}},\ \bibinfo {pages} {1174} (\bibinfo {year}
  {2019})}\BibitemShut {NoStop}%
\bibitem [{\citenamefont {Xie}\ \emph {et~al.}(2019)\citenamefont {Xie},
  \citenamefont {Lian}, \citenamefont {J{\"a}ck}, \citenamefont {Liu},
  \citenamefont {Chiu}, \citenamefont {Watanabe}, \citenamefont {Taniguchi},
  \citenamefont {Bernevig},\ and\ \citenamefont
  {Yazdani}}]{xie2019spectroscopic}%
  \BibitemOpen
  \bibfield  {author} {\bibinfo {author} {\bibfnamefont {Y.}~\bibnamefont
  {Xie}}, \bibinfo {author} {\bibfnamefont {B.}~\bibnamefont {Lian}}, \bibinfo
  {author} {\bibfnamefont {B.}~\bibnamefont {J{\"a}ck}}, \bibinfo {author}
  {\bibfnamefont {X.}~\bibnamefont {Liu}}, \bibinfo {author} {\bibfnamefont
  {C.-L.}\ \bibnamefont {Chiu}}, \bibinfo {author} {\bibfnamefont
  {K.}~\bibnamefont {Watanabe}}, \bibinfo {author} {\bibfnamefont
  {T.}~\bibnamefont {Taniguchi}}, \bibinfo {author} {\bibfnamefont {B.~A.}\
  \bibnamefont {Bernevig}},\ and\ \bibinfo {author} {\bibfnamefont
  {A.}~\bibnamefont {Yazdani}},\ }\bibfield  {title} {\bibinfo {title}
  {Spectroscopic signatures of many-body correlations in magic-angle twisted
  bilayer graphene},\ }\href@noop {} {\bibfield  {journal} {\bibinfo  {journal}
  {Nature}\ }\textbf {\bibinfo {volume} {572}},\ \bibinfo {pages} {101}
  (\bibinfo {year} {2019})}\BibitemShut {NoStop}%
\bibitem [{\citenamefont {Bi}\ \emph {et~al.}(2019)\citenamefont {Bi},
  \citenamefont {Yuan},\ and\ \citenamefont {Fu}}]{bi2019designing}%
  \BibitemOpen
  \bibfield  {author} {\bibinfo {author} {\bibfnamefont {Z.}~\bibnamefont
  {Bi}}, \bibinfo {author} {\bibfnamefont {N.~F.~Q.}\ \bibnamefont {Yuan}},\
  and\ \bibinfo {author} {\bibfnamefont {L.}~\bibnamefont {Fu}},\ }\bibfield
  {title} {\bibinfo {title} {Designing flat bands by strain},\ }\href
  {https://doi.org/10.1103/PhysRevB.100.035448} {\bibfield  {journal} {\bibinfo
   {journal} {Phys. Rev. B}\ }\textbf {\bibinfo {volume} {100}},\ \bibinfo
  {pages} {035448} (\bibinfo {year} {2019})}\BibitemShut {NoStop}%
\bibitem [{\citenamefont {Tang}\ \emph {et~al.}(2024)\citenamefont {Tang},
  \citenamefont {Wang}, \citenamefont {Ni}, \citenamefont {Mrejen},
  \citenamefont {Watanabe}, \citenamefont {Taniguchi}, \citenamefont {Fang},
  \citenamefont {Jarillo-Herrero},\ and\ \citenamefont {Yao}}]{tang2024chip}%
  \BibitemOpen
  \bibfield  {author} {\bibinfo {author} {\bibfnamefont {H.}~\bibnamefont
  {Tang}}, \bibinfo {author} {\bibfnamefont {Y.}~\bibnamefont {Wang}}, \bibinfo
  {author} {\bibfnamefont {X.}~\bibnamefont {Ni}}, \bibinfo {author}
  {\bibfnamefont {M.}~\bibnamefont {Mrejen}}, \bibinfo {author} {\bibfnamefont
  {K.}~\bibnamefont {Watanabe}}, \bibinfo {author} {\bibfnamefont
  {T.}~\bibnamefont {Taniguchi}}, \bibinfo {author} {\bibfnamefont {N.~X.}\
  \bibnamefont {Fang}}, \bibinfo {author} {\bibfnamefont {P.}~\bibnamefont
  {Jarillo-Herrero}},\ and\ \bibinfo {author} {\bibfnamefont {A.}~\bibnamefont
  {Yao}},\ }\bibfield  {title} {\bibinfo {title} {On-chip
  multi-degree-of-freedom control of two-dimensional materials},\ }\href
  {https://doi.org/10.1038/s41586-024-07826-x} {\bibfield  {journal} {\bibinfo
  {journal} {Nature}\ }\textbf {\bibinfo {volume} {632}},\ \bibinfo {pages}
  {1038} (\bibinfo {year} {2024})}\BibitemShut {NoStop}%
\bibitem [{\citenamefont {Polini}\ \emph {et~al.}(2008)\citenamefont {Polini},
  \citenamefont {Asgari}, \citenamefont {Borghi}, \citenamefont {Barlas},
  \citenamefont {Pereg-Barnea},\ and\ \citenamefont
  {MacDonald}}]{polini2008plasmons}%
  \BibitemOpen
  \bibfield  {author} {\bibinfo {author} {\bibfnamefont {M.}~\bibnamefont
  {Polini}}, \bibinfo {author} {\bibfnamefont {R.}~\bibnamefont {Asgari}},
  \bibinfo {author} {\bibfnamefont {G.}~\bibnamefont {Borghi}}, \bibinfo
  {author} {\bibfnamefont {Y.}~\bibnamefont {Barlas}}, \bibinfo {author}
  {\bibfnamefont {T.}~\bibnamefont {Pereg-Barnea}},\ and\ \bibinfo {author}
  {\bibfnamefont {A.~H.}\ \bibnamefont {MacDonald}},\ }\bibfield  {title}
  {\bibinfo {title} {Plasmons and the spectral function of graphene},\ }\href
  {https://doi.org/10.1103/PhysRevB.77.081411} {\bibfield  {journal} {\bibinfo
  {journal} {Phys. Rev. B}\ }\textbf {\bibinfo {volume} {77}},\ \bibinfo
  {pages} {081411} (\bibinfo {year} {2008})}\BibitemShut {NoStop}%
\bibitem [{\citenamefont {Hwang}\ and\ \citenamefont
  {Das~Sarma}(2008)}]{hwang2008quasiparticle}%
  \BibitemOpen
  \bibfield  {author} {\bibinfo {author} {\bibfnamefont {E.~H.}\ \bibnamefont
  {Hwang}}\ and\ \bibinfo {author} {\bibfnamefont {S.}~\bibnamefont
  {Das~Sarma}},\ }\bibfield  {title} {\bibinfo {title} {Quasiparticle spectral
  function in doped graphene: Electron-electron interaction effects in arpes},\
  }\href {https://doi.org/10.1103/PhysRevB.77.081412} {\bibfield  {journal}
  {\bibinfo  {journal} {Phys. Rev. B}\ }\textbf {\bibinfo {volume} {77}},\
  \bibinfo {pages} {081412} (\bibinfo {year} {2008})}\BibitemShut {NoStop}%
\bibitem [{\citenamefont {Rhodes}\ \emph {et~al.}(2019)\citenamefont {Rhodes},
  \citenamefont {Chae}, \citenamefont {Ribeiro-Palau},\ and\ \citenamefont
  {Hone}}]{rhodes2019disorder}%
  \BibitemOpen
  \bibfield  {author} {\bibinfo {author} {\bibfnamefont {D.}~\bibnamefont
  {Rhodes}}, \bibinfo {author} {\bibfnamefont {S.~H.}\ \bibnamefont {Chae}},
  \bibinfo {author} {\bibfnamefont {R.}~\bibnamefont {Ribeiro-Palau}},\ and\
  \bibinfo {author} {\bibfnamefont {J.}~\bibnamefont {Hone}},\ }\bibfield
  {title} {\bibinfo {title} {Disorder in van der waals heterostructures of 2d
  materials},\ }\href@noop {} {\bibfield  {journal} {\bibinfo  {journal}
  {Nature materials}\ }\textbf {\bibinfo {volume} {18}},\ \bibinfo {pages}
  {541} (\bibinfo {year} {2019})}\BibitemShut {NoStop}%
\bibitem [{\citenamefont {Shallcross}\ \emph {et~al.}(2008)\citenamefont
  {Shallcross}, \citenamefont {Sharma},\ and\ \citenamefont
  {Pankratov}}]{shallcross2008quantum}%
  \BibitemOpen
  \bibfield  {author} {\bibinfo {author} {\bibfnamefont {S.}~\bibnamefont
  {Shallcross}}, \bibinfo {author} {\bibfnamefont {S.}~\bibnamefont {Sharma}},\
  and\ \bibinfo {author} {\bibfnamefont {O.~A.}\ \bibnamefont {Pankratov}},\
  }\bibfield  {title} {\bibinfo {title} {Quantum interference at the twist
  boundary in graphene},\ }\href
  {https://doi.org/10.1103/PhysRevLett.101.056803} {\bibfield  {journal}
  {\bibinfo  {journal} {Phys. Rev. Lett.}\ }\textbf {\bibinfo {volume} {101}},\
  \bibinfo {pages} {056803} (\bibinfo {year} {2008})}\BibitemShut {NoStop}%
\bibitem [{\citenamefont {Jiang}\ \emph {et~al.}(2022)\citenamefont {Jiang},
  \citenamefont {Zhao},\ and\ \citenamefont {Zhang}}]{jiang2022tunable}%
  \BibitemOpen
  \bibfield  {author} {\bibinfo {author} {\bibfnamefont {X.-C.}\ \bibnamefont
  {Jiang}}, \bibinfo {author} {\bibfnamefont {Y.-Y.}\ \bibnamefont {Zhao}},\
  and\ \bibinfo {author} {\bibfnamefont {Y.-Z.}\ \bibnamefont {Zhang}},\
  }\bibfield  {title} {\bibinfo {title} {Tunable band gap in twisted bilayer
  graphene},\ }\href {https://doi.org/10.1103/PhysRevB.105.115106} {\bibfield
  {journal} {\bibinfo  {journal} {Phys. Rev. B}\ }\textbf {\bibinfo {volume}
  {105}},\ \bibinfo {pages} {115106} (\bibinfo {year} {2022})}\BibitemShut
  {NoStop}%
\bibitem [{\citenamefont {Talkington}\ and\ \citenamefont
  {Mele}(2023)}]{talktington2023terahertz}%
  \BibitemOpen
  \bibfield  {author} {\bibinfo {author} {\bibfnamefont {S.}~\bibnamefont
  {Talkington}}\ and\ \bibinfo {author} {\bibfnamefont {E.~J.}\ \bibnamefont
  {Mele}},\ }\bibfield  {title} {\bibinfo {title} {Terahertz circular dichroism
  in commensurate twisted bilayer graphene},\ }\href
  {https://doi.org/10.1103/PhysRevB.108.085421} {\bibfield  {journal} {\bibinfo
   {journal} {Phys. Rev. B}\ }\textbf {\bibinfo {volume} {108}},\ \bibinfo
  {pages} {085421} (\bibinfo {year} {2023})}\BibitemShut {NoStop}%
\bibitem [{\citenamefont {Guinea}\ and\ \citenamefont
  {Walet}(2018)}]{guinea2018electrostatic}%
  \BibitemOpen
  \bibfield  {author} {\bibinfo {author} {\bibfnamefont {F.}~\bibnamefont
  {Guinea}}\ and\ \bibinfo {author} {\bibfnamefont {N.~R.}\ \bibnamefont
  {Walet}},\ }\bibfield  {title} {\bibinfo {title} {Electrostatic effects, band
  distortions, and superconductivity in twisted graphene bilayers},\
  }\href@noop {} {\bibfield  {journal} {\bibinfo  {journal} {Proceedings of the
  National Academy of Sciences}\ }\textbf {\bibinfo {volume} {115}},\ \bibinfo
  {pages} {13174} (\bibinfo {year} {2018})}\BibitemShut {NoStop}%
\bibitem [{\citenamefont {Zhu}\ \emph {et~al.}(2024)\citenamefont {Zhu},
  \citenamefont {Torre}, \citenamefont {Polini},\ and\ \citenamefont
  {MacDonald}}]{zhu2024gw}%
  \BibitemOpen
  \bibfield  {author} {\bibinfo {author} {\bibfnamefont {J.}~\bibnamefont
  {Zhu}}, \bibinfo {author} {\bibfnamefont {I.}~\bibnamefont {Torre}}, \bibinfo
  {author} {\bibfnamefont {M.}~\bibnamefont {Polini}},\ and\ \bibinfo {author}
  {\bibfnamefont {A.~H.}\ \bibnamefont {MacDonald}},\ }\bibfield  {title}
  {\bibinfo {title} {Gw theory of magic-angle twisted bilayer graphene},\
  }\href@noop {} {\bibfield  {journal} {\bibinfo  {journal} {arXiv preprint
  arXiv:2401.02872}\ } (\bibinfo {year} {2024})}\BibitemShut {NoStop}%
\bibitem [{\citenamefont {Ezzi}\ \emph {et~al.}(2024)\citenamefont {Ezzi},
  \citenamefont {Peng}, \citenamefont {Liu}, \citenamefont {Chao},
  \citenamefont {Pallewela}, \citenamefont {Foo},\ and\ \citenamefont
  {Adam}}]{ezzi2024self}%
  \BibitemOpen
  \bibfield  {author} {\bibinfo {author} {\bibfnamefont {M.~M.~A.}\
  \bibnamefont {Ezzi}}, \bibinfo {author} {\bibfnamefont {L.}~\bibnamefont
  {Peng}}, \bibinfo {author} {\bibfnamefont {Z.}~\bibnamefont {Liu}}, \bibinfo
  {author} {\bibfnamefont {J.~H.~Z.}\ \bibnamefont {Chao}}, \bibinfo {author}
  {\bibfnamefont {G.~N.}\ \bibnamefont {Pallewela}}, \bibinfo {author}
  {\bibfnamefont {D.}~\bibnamefont {Foo}},\ and\ \bibinfo {author}
  {\bibfnamefont {S.}~\bibnamefont {Adam}},\ }\href
  {https://arxiv.org/abs/2404.17638} {\bibinfo {title} {A self-consistent
  hartree theory for lattice-relaxed magic-angle twisted bilayer graphene}}
  (\bibinfo {year} {2024}),\ \Eprint {https://arxiv.org/abs/2404.17638}
  {arXiv:2404.17638 [cond-mat.str-el]} \BibitemShut {NoStop}%
\bibitem [{\citenamefont {Sharpe}\ \emph {et~al.}(2019)\citenamefont {Sharpe},
  \citenamefont {Fox}, \citenamefont {Barnard}, \citenamefont {Finney},
  \citenamefont {Watanabe}, \citenamefont {Taniguchi}, \citenamefont
  {Kastner},\ and\ \citenamefont {Goldhaber-Gordon}}]{sharpe2019emergent}%
  \BibitemOpen
  \bibfield  {author} {\bibinfo {author} {\bibfnamefont {A.~L.}\ \bibnamefont
  {Sharpe}}, \bibinfo {author} {\bibfnamefont {E.~J.}\ \bibnamefont {Fox}},
  \bibinfo {author} {\bibfnamefont {A.~W.}\ \bibnamefont {Barnard}}, \bibinfo
  {author} {\bibfnamefont {J.}~\bibnamefont {Finney}}, \bibinfo {author}
  {\bibfnamefont {K.}~\bibnamefont {Watanabe}}, \bibinfo {author}
  {\bibfnamefont {T.}~\bibnamefont {Taniguchi}}, \bibinfo {author}
  {\bibfnamefont {M.}~\bibnamefont {Kastner}},\ and\ \bibinfo {author}
  {\bibfnamefont {D.}~\bibnamefont {Goldhaber-Gordon}},\ }\bibfield  {title}
  {\bibinfo {title} {Emergent ferromagnetism near three-quarters filling in
  twisted bilayer graphene},\ }\href@noop {} {\bibfield  {journal} {\bibinfo
  {journal} {Science}\ }\textbf {\bibinfo {volume} {365}},\ \bibinfo {pages}
  {605} (\bibinfo {year} {2019})}\BibitemShut {NoStop}%
\bibitem [{\citenamefont {Serlin}\ \emph {et~al.}(2020)\citenamefont {Serlin},
  \citenamefont {Tschirhart}, \citenamefont {Polshyn}, \citenamefont {Zhang},
  \citenamefont {Zhu}, \citenamefont {Watanabe}, \citenamefont {Taniguchi},
  \citenamefont {Balents},\ and\ \citenamefont {Young}}]{serlin2020intrinsic}%
  \BibitemOpen
  \bibfield  {author} {\bibinfo {author} {\bibfnamefont {M.}~\bibnamefont
  {Serlin}}, \bibinfo {author} {\bibfnamefont {C.}~\bibnamefont {Tschirhart}},
  \bibinfo {author} {\bibfnamefont {H.}~\bibnamefont {Polshyn}}, \bibinfo
  {author} {\bibfnamefont {Y.}~\bibnamefont {Zhang}}, \bibinfo {author}
  {\bibfnamefont {J.}~\bibnamefont {Zhu}}, \bibinfo {author} {\bibfnamefont
  {K.}~\bibnamefont {Watanabe}}, \bibinfo {author} {\bibfnamefont
  {T.}~\bibnamefont {Taniguchi}}, \bibinfo {author} {\bibfnamefont
  {L.}~\bibnamefont {Balents}},\ and\ \bibinfo {author} {\bibfnamefont
  {A.}~\bibnamefont {Young}},\ }\bibfield  {title} {\bibinfo {title} {Intrinsic
  quantized anomalous hall effect in a moir{\'e} heterostructure},\ }\href@noop
  {} {\bibfield  {journal} {\bibinfo  {journal} {Science}\ }\textbf {\bibinfo
  {volume} {367}},\ \bibinfo {pages} {900} (\bibinfo {year}
  {2020})}\BibitemShut {NoStop}%
\bibitem [{\citenamefont {Zhou}\ \emph {et~al.}(2021)\citenamefont {Zhou},
  \citenamefont {Xie}, \citenamefont {Ghazaryan}, \citenamefont {Holder},
  \citenamefont {Ehrets}, \citenamefont {Spanton}, \citenamefont {Taniguchi},
  \citenamefont {Watanabe}, \citenamefont {Berg}, \citenamefont {Serbyn} \emph
  {et~al.}}]{zhou2021half}%
  \BibitemOpen
  \bibfield  {author} {\bibinfo {author} {\bibfnamefont {H.}~\bibnamefont
  {Zhou}}, \bibinfo {author} {\bibfnamefont {T.}~\bibnamefont {Xie}}, \bibinfo
  {author} {\bibfnamefont {A.}~\bibnamefont {Ghazaryan}}, \bibinfo {author}
  {\bibfnamefont {T.}~\bibnamefont {Holder}}, \bibinfo {author} {\bibfnamefont
  {J.~R.}\ \bibnamefont {Ehrets}}, \bibinfo {author} {\bibfnamefont {E.~M.}\
  \bibnamefont {Spanton}}, \bibinfo {author} {\bibfnamefont {T.}~\bibnamefont
  {Taniguchi}}, \bibinfo {author} {\bibfnamefont {K.}~\bibnamefont {Watanabe}},
  \bibinfo {author} {\bibfnamefont {E.}~\bibnamefont {Berg}}, \bibinfo {author}
  {\bibfnamefont {M.}~\bibnamefont {Serbyn}}, \emph {et~al.},\ }\bibfield
  {title} {\bibinfo {title} {Half-and quarter-metals in rhombohedral trilayer
  graphene},\ }\href@noop {} {\bibfield  {journal} {\bibinfo  {journal}
  {Nature}\ }\textbf {\bibinfo {volume} {598}},\ \bibinfo {pages} {429}
  (\bibinfo {year} {2021})}\BibitemShut {NoStop}%
\bibitem [{\citenamefont {Wolf}\ \emph {et~al.}(2024)\citenamefont {Wolf},
  \citenamefont {Wei}, \citenamefont {Zhou},\ and\ \citenamefont
  {Huang}}]{wolf2024magnetism}%
  \BibitemOpen
  \bibfield  {author} {\bibinfo {author} {\bibfnamefont {T.}~\bibnamefont
  {Wolf}}, \bibinfo {author} {\bibfnamefont {N.}~\bibnamefont {Wei}}, \bibinfo
  {author} {\bibfnamefont {H.}~\bibnamefont {Zhou}},\ and\ \bibinfo {author}
  {\bibfnamefont {C.}~\bibnamefont {Huang}},\ }\href
  {https://arxiv.org/abs/2408.15884} {\bibinfo {title} {Magnetism in the dilute
  electron gas of rhombohedral multilayer graphene}} (\bibinfo {year} {2024}),\
  \Eprint {https://arxiv.org/abs/2408.15884} {arXiv:2408.15884
  [cond-mat.str-el]} \BibitemShut {NoStop}%
\bibitem [{\citenamefont {Xie}\ \emph {et~al.}(2024)\citenamefont {Xie},
  \citenamefont {Xu}, \citenamefont {Dong}, \citenamefont {Cui}, \citenamefont
  {Ou}, \citenamefont {Erdi}, \citenamefont {Watanabe}, \citenamefont
  {Taniguchi}, \citenamefont {Tongay}, \citenamefont {Levitov} \emph
  {et~al.}}]{xie2024long}%
  \BibitemOpen
  \bibfield  {author} {\bibinfo {author} {\bibfnamefont {T.}~\bibnamefont
  {Xie}}, \bibinfo {author} {\bibfnamefont {S.}~\bibnamefont {Xu}}, \bibinfo
  {author} {\bibfnamefont {Z.}~\bibnamefont {Dong}}, \bibinfo {author}
  {\bibfnamefont {Z.}~\bibnamefont {Cui}}, \bibinfo {author} {\bibfnamefont
  {Y.}~\bibnamefont {Ou}}, \bibinfo {author} {\bibfnamefont {M.}~\bibnamefont
  {Erdi}}, \bibinfo {author} {\bibfnamefont {K.}~\bibnamefont {Watanabe}},
  \bibinfo {author} {\bibfnamefont {T.}~\bibnamefont {Taniguchi}}, \bibinfo
  {author} {\bibfnamefont {S.~A.}\ \bibnamefont {Tongay}}, \bibinfo {author}
  {\bibfnamefont {L.~S.}\ \bibnamefont {Levitov}}, \emph {et~al.},\ }\bibfield
  {title} {\bibinfo {title} {Long-lived isospin excitations in magic-angle
  twisted bilayer graphene},\ }\href@noop {} {\bibfield  {journal} {\bibinfo
  {journal} {Nature}\ ,\ \bibinfo {pages} {1}} (\bibinfo {year}
  {2024})}\BibitemShut {NoStop}%
\bibitem [{\citenamefont {Lewandowski}\ and\ \citenamefont
  {Levitov}(2019)}]{lewandowski2019intrinsically}%
  \BibitemOpen
  \bibfield  {author} {\bibinfo {author} {\bibfnamefont {C.}~\bibnamefont
  {Lewandowski}}\ and\ \bibinfo {author} {\bibfnamefont {L.}~\bibnamefont
  {Levitov}},\ }\bibfield  {title} {\bibinfo {title} {Intrinsically undamped
  plasmon modes in narrow electron bands},\ }\href@noop {} {\bibfield
  {journal} {\bibinfo  {journal} {Proceedings of the National Academy of
  Sciences}\ }\textbf {\bibinfo {volume} {116}},\ \bibinfo {pages} {20869}
  (\bibinfo {year} {2019})}\BibitemShut {NoStop}%
\bibitem [{\citenamefont {Kumar}\ \emph {et~al.}(2021)\citenamefont {Kumar},
  \citenamefont {Xie},\ and\ \citenamefont {MacDonald}}]{kumar2021lattice}%
  \BibitemOpen
  \bibfield  {author} {\bibinfo {author} {\bibfnamefont {A.}~\bibnamefont
  {Kumar}}, \bibinfo {author} {\bibfnamefont {M.}~\bibnamefont {Xie}},\ and\
  \bibinfo {author} {\bibfnamefont {A.~H.}\ \bibnamefont {MacDonald}},\
  }\bibfield  {title} {\bibinfo {title} {Lattice collective modes from a
  continuum model of magic-angle twisted bilayer graphene},\ }\href
  {https://doi.org/10.1103/PhysRevB.104.035119} {\bibfield  {journal} {\bibinfo
   {journal} {Phys. Rev. B}\ }\textbf {\bibinfo {volume} {104}},\ \bibinfo
  {pages} {035119} (\bibinfo {year} {2021})}\BibitemShut {NoStop}%
\bibitem [{\citenamefont {Bernevig}\ \emph
  {et~al.}(2021{\natexlab{b}})\citenamefont {Bernevig}, \citenamefont {Lian},
  \citenamefont {Cowsik}, \citenamefont {Xie}, \citenamefont {Regnault},\ and\
  \citenamefont {Song}}]{bernevig2021twistedv}%
  \BibitemOpen
  \bibfield  {author} {\bibinfo {author} {\bibfnamefont {B.~A.}\ \bibnamefont
  {Bernevig}}, \bibinfo {author} {\bibfnamefont {B.}~\bibnamefont {Lian}},
  \bibinfo {author} {\bibfnamefont {A.}~\bibnamefont {Cowsik}}, \bibinfo
  {author} {\bibfnamefont {F.}~\bibnamefont {Xie}}, \bibinfo {author}
  {\bibfnamefont {N.}~\bibnamefont {Regnault}},\ and\ \bibinfo {author}
  {\bibfnamefont {Z.-D.}\ \bibnamefont {Song}},\ }\bibfield  {title} {\bibinfo
  {title} {Twisted bilayer graphene. v. exact analytic many-body excitations in
  coulomb hamiltonians: Charge gap, goldstone modes, and absence of cooper
  pairing},\ }\href {https://doi.org/10.1103/PhysRevB.103.205415} {\bibfield
  {journal} {\bibinfo  {journal} {Phys. Rev. B}\ }\textbf {\bibinfo {volume}
  {103}},\ \bibinfo {pages} {205415} (\bibinfo {year}
  {2021}{\natexlab{b}})}\BibitemShut {NoStop}%
\bibitem [{\citenamefont {Khalaf}\ \emph {et~al.}(2020)\citenamefont {Khalaf},
  \citenamefont {Bultinck}, \citenamefont {Vishwanath},\ and\ \citenamefont
  {Zaletel}}]{khalaf2020soft}%
  \BibitemOpen
  \bibfield  {author} {\bibinfo {author} {\bibfnamefont {E.}~\bibnamefont
  {Khalaf}}, \bibinfo {author} {\bibfnamefont {N.}~\bibnamefont {Bultinck}},
  \bibinfo {author} {\bibfnamefont {A.}~\bibnamefont {Vishwanath}},\ and\
  \bibinfo {author} {\bibfnamefont {M.~P.}\ \bibnamefont {Zaletel}},\ }\href
  {https://arxiv.org/abs/2009.14827} {\bibinfo {title} {Soft modes in magic
  angle twisted bilayer graphene}} (\bibinfo {year} {2020}),\ \Eprint
  {https://arxiv.org/abs/2009.14827} {arXiv:2009.14827 [cond-mat.str-el]}
  \BibitemShut {NoStop}%
\bibitem [{\citenamefont {Moon}\ and\ \citenamefont
  {Koshino}(2013)}]{moon2013optical}%
  \BibitemOpen
  \bibfield  {author} {\bibinfo {author} {\bibfnamefont {P.}~\bibnamefont
  {Moon}}\ and\ \bibinfo {author} {\bibfnamefont {M.}~\bibnamefont {Koshino}},\
  }\bibfield  {title} {\bibinfo {title} {Optical absorption in twisted bilayer
  graphene},\ }\href {https://doi.org/10.1103/PhysRevB.87.205404} {\bibfield
  {journal} {\bibinfo  {journal} {Phys. Rev. B}\ }\textbf {\bibinfo {volume}
  {87}},\ \bibinfo {pages} {205404} (\bibinfo {year} {2013})}\BibitemShut
  {NoStop}%
\bibitem [{\citenamefont {Devakul}\ \emph {et~al.}(2023)\citenamefont
  {Devakul}, \citenamefont {Ledwith}, \citenamefont {Xia}, \citenamefont {Uri},
  \citenamefont {de~la Barrera}, \citenamefont {Jarillo-Herrero},\ and\
  \citenamefont {Fu}}]{devakul2023magic}%
  \BibitemOpen
  \bibfield  {author} {\bibinfo {author} {\bibfnamefont {T.}~\bibnamefont
  {Devakul}}, \bibinfo {author} {\bibfnamefont {P.~J.}\ \bibnamefont
  {Ledwith}}, \bibinfo {author} {\bibfnamefont {L.-Q.}\ \bibnamefont {Xia}},
  \bibinfo {author} {\bibfnamefont {A.}~\bibnamefont {Uri}}, \bibinfo {author}
  {\bibfnamefont {S.~C.}\ \bibnamefont {de~la Barrera}}, \bibinfo {author}
  {\bibfnamefont {P.}~\bibnamefont {Jarillo-Herrero}},\ and\ \bibinfo {author}
  {\bibfnamefont {L.}~\bibnamefont {Fu}},\ }\bibfield  {title} {\bibinfo
  {title} {Magic-angle helical trilayer graphene},\ }\href@noop {} {\bibfield
  {journal} {\bibinfo  {journal} {Science Advances}\ }\textbf {\bibinfo
  {volume} {9}},\ \bibinfo {pages} {eadi6063} (\bibinfo {year}
  {2023})}\BibitemShut {NoStop}%
\end{thebibliography}%

\end{document}